\def\submissiontype{0}				

\def\showComments{1}				

\newif\ifTightOnSpace				
\TightOnSpacefalse					

\newif\ifEprint					
\newif\ifSubmission				
\newif\ifCameraReady			
\newif\ifSupplementaryMaterial	
\SupplementaryMaterialfalse

\ifnum\submissiontype=0     												 	
	\Eprinttrue
	\Submissionfalse
	\CameraReadyfalse
\else																			
	\Eprintfalse								
	
	\ifnum\submissiontype=1     												
		\Submissiontrue
		\CameraReadyfalse
		
		\SupplementaryMaterialtrue												
	
	\else     																	
		\CameraReadyfalse
		\Submissionfalse
	\fi 
\fi

\RequirePackage{amsmath}  

\ifEprint
\documentclass[a4paper,runningheads,envcountsame,envcountsect]{llncs}
	\usepackage[all]{nowidow}
\else
	\documentclass{llncs} 
\fi

\usepackage[utf8]{inputenc}
\usepackage[T1]{fontenc}

\usepackage{soul}

\usepackage{float} 

\usepackage{tikz}
\usetikzlibrary{positioning}
\usetikzlibrary{calc}
\usetikzlibrary{shapes.geometric}
\usetikzlibrary{matrix,arrows}

\usepackage{nicefrac}
\usepackage{nicodemus}
\usepackage{stmaryrd}

\usepackage{amsfonts,amsmath,amscd,amssymb,array}
\usepackage{algorithm2e,algpseudocode}
\usepackage{graphicx}
\usepackage{xspace} 
\usepackage{ifthen}
\usepackage{braket}
\usepackage{dsfont}
\usepackage{breakcites}

\usepackage{thmtools, thm-restate}

\usepackage{csquotes}

\usepackage{caption}
\usepackage{enumitem}

\usepackage{xcolor}
\definecolor{linkcolor}{rgb}{0.65,0,0}
\definecolor{citecolor}{rgb}{0,0.65,0}
\definecolor{urlcolor}{rgb}{0,0,0.65}
\definecolor{oraclecolor}{rgb}{0.8,0,0}
\definecolor{qkdoraclecolor}{rgb}{0.5,0,0.5}
\definecolor{bookkeepcolor}{rgb}{0,0.6,0}
\usepackage[colorlinks=true, linkcolor=linkcolor, urlcolor=urlcolor, 
citecolor=citecolor,hypertexnames=false, bookmarks=true]{hyperref}
\usepackage{xr} 
\usepackage[capitalise]{cleveref}

\newcounter{pfcase} 
\newcounter{pfsubcase}[pfcase]
\newcounter{pfsubsubcase}[pfsubcase]
\newcounter{pfsubsubsubcase}[pfsubsubcase]

\makeatletter
\if@cref@capitalise
\crefname{pfcase}{Case}{Cases}
\crefname{pfsubcase}{Case}{Cases}
\crefname{pfsubsubcase}{Case}{Cases}
\crefname{pfsubsubsubcase}{Case}{Cases}
\crefname{line}{Line}{Lines}
\else
\crefname{pfcase}{case}{cases}
\crefname{pfsubcase}{case}{cases}
\crefname{pfsubsubcase}{case}{cases}
\crefname{pfsubsubsubcase}{case}{cases}
\crefname{line}{line}{lines}
\fi
\makeatother

\ifSupplementaryMaterial
	\ifTightOnSpace
		\crefname{appendix}{Suppl. Mat.}{Supplementary Material} 
	\else
		\crefname{appendix}{Supplementary Material}{Supplementary Material} 
	\fi
\fi

\usepackage[framemethod=default]{mdframed}

\usepackage{mathtools}

\usepackage{ctable} 
\makeatletter
\renewcommand{\tnote}[2][a]{%
  \ifx\@CTnotespar\@CTtrue%
  \@CTtextsuperscript{\normalfont\textit{#1}}\,#2
  \else%
  \hbox{\@CTtextsuperscript{\normalfont\textit{#1}}}&#2\tabularnewline
  \fi
}
\makeatother

\newcommand{\mathsc}[1]{\text{\textsc{#1}}}

\usepackage{listings}
\usepackage{varwidth}

\newenvironment{nicodemus}[1][\thenicolinenr]{
	\begin{enumerate}[
		topsep=0ex,
		label=\nicolinenrformat\PaddingUp*,
		ref=\nicorefprefix\PaddingUp*,
		align=right,
		leftmargin=0em,
		itemindent=!,
		labelindent=0em,
		labelwidth=\nicolinenrwidth,
		labelsep=\nicolinenrsep,
		listparindent=\parindent,
		noitemsep,
		]%
		\setcounter{enumi}{#1}%
		\addtocounter{enumi}{-1}%
	}{%
	\end{enumerate}%
	\addtocounter{enumi}{1}%
	\setcounter{nicolinenr}{\theenumi}%
}			
\newcommand{\attackname}{Dependent-Key\xspace}

\newcommand{\todoitem}[4][]{%
  \ifnum\showComments=1{%
    \ifthenelse{\equal{#1}{done}}{%
      {\textcolor{lightgray}{[#2 (done): #4]}}%
    }{%
      {\textcolor{#3}{[#2: #4]}}%
    }%
  }\fi%
}

\def\subheading#1{\medskip\noindent{\boldmath\textbf{#1}}~\ignorespaces}
\newcommand{\heading}[1]{{\vspace{1ex}\noindent\sc{#1}}}

\newcommand{\uni}{\ensuremath{\leftarrow_\$}}										
\newcommand{\bool}[1]{\ensuremath{\llbracket #1\rrbracket}\xspace}		

\newcommand{\Adv}{\ensuremath{\mathrm{Adv}}\xspace}						

\newcommand{\gcom}[1]{\hfill $\sslash$#1}								
\newcommand{\pcfor}{\textbf{for }}
\newcommand{\pcif}{\textbf{if }}
\newcommand{\pcelse}{\textbf{else }}

\newcommand{\pcand}{\textbf{ and }}
\newcommand{\pcor}{\textbf{ or }}

\newcommand{\true}{\textbf{true}}
\newcommand{\false}{\textbf{false}}
\newcommand{\pcreturn}{\textbf{return }}
\newcommand{\pcabort}{\textbf{abort}}

\newcommand{\concat}{\mathbin{\|}}

\newcommand{\arwrite}[1]{{\color{bookkeepcolor}\ensuremath{\textup{\texttt{#1}}}}\xspace} 
\newcommand{\varwrite}[1]{\ensuremath{\mathit{#1}}\xspace} 

\newcommand{\algowrite}[1]{\textsf{#1}\xspace} 
\newcommand{\notionwrite}[1]{\ensuremath{\mathsf{#1}}\xspace} 
\newcommand{\orwrite}[1]{\textsc{#1}\xspace} 

\newcommand{\MAC}{\algowrite{MAC}}
\newcommand{\PQCMAC}{\ensuremath{\MAC^{(pqc)}}}
\newcommand{\QKDMAC}{\ensuremath{\MAC^{(qkd)}}}

\newcommand{\MKG}{\algowrite{MKG}}
\newcommand{\MVerify}{\ensuremath{\algowrite{MVerify}}}
\newcommand{\Oracle}{{\color{oraclecolor}\mathcal{O}}}

\newcommand{\PRF}{\algowrite{PRF}}

\newcommand{\KDF}{\algowrite{KDF}}		

\newcommand{\pk}{\ensuremath{\mathit{pk}}}
\newcommand{\sk}{\ensuremath{\mathit{sk}}}

\newcommand{\KEM}{\algowrite{KEM}}
\newcommand{\KeyGen}{\algowrite{KeyGen}}

\newcommand{\Encaps}{\algowrite{Encaps}}
\newcommand{\Decaps}{\algowrite{Decaps}}

\newcommand{\KEMeph}{\algowrite{KEM}_\varwrite{eph}}
\newcommand{\KeyGenEph}{\algowrite{KeyGen}_\varwrite{eph}}
\newcommand{\EncapsEph}{\algowrite{Encaps}_\varwrite{eph}}
\newcommand{\DecapsEph}{\algowrite{Decaps}_\varwrite{eph}}
\newcommand{\KEMstat}{\algowrite{KEM}_\varwrite{stat}}
\newcommand{\KeyGenStat}{\algowrite{KeyGen}_\varwrite{stat}}
\newcommand{\EncapsStat}{\algowrite{Encaps}_\varwrite{stat}}
\newcommand{\DecapsStat}{\algowrite{Decaps}_\varwrite{stat}}

\newcommand{\muKg}{\mu(\KeyGen)}
\newcommand{\muKgEph}{\mu(\KeyGenEph)}

\newcommand{\muEnc}{\mu(\Encaps)}
\newcommand{\muEncEph}{\mu(\EncapsEph)}
\newcommand{\muEncStat}{\mu(\EncapsStat)}
\newcommand{\muSec}{\mu(\algowrite{Secret})}
\newcommand{\muSecStat}{\mu(\algowrite{Secret}_\varwrite{stat})}

\newcommand{\deltaeph}{\delta_\varwrite{eph}}
\newcommand{\deltastat}{\delta_\varwrite{stat}}

\newcommand{\IND}{\notionwrite{IND}}

\newcommand{\CPA}{\notionwrite{CPA}}
\newcommand{\CCA}{\notionwrite{CCA}}
\newcommand{\INDCPA}{\IND{}\text{-}\CPA}
\newcommand{\INDCCA}{\IND{}\text{-}\CCA}
\newcommand{\INDOCCA}{\IND{}\text{-}\notionwrite{1}{}\CCA}
\newcommand{\oracleDecaps}{\color{oraclecolor}{\orwrite{Dec}}}

\newcommand{\secAA}{\notionwrite{AA}}
\newcommand{\StAA}{\notionwrite{StAA}}

\newcommand{\SUFCMA}{\notionwrite{sEUF}{}\text{-}\notionwrite{CMA}}
\newcommand{\OTSUFCMA}{\notionwrite{OT}{}\text{-}\SUFCMA}
\newcommand{\ROTSUFCMA}{\notionwrite{Robust}{}\text{-}\notionwrite{OT}{}\text{-}\SUFCMA}

\newcommand{\INDStAAPQC}{\IND{}\text{-}\StAA{}\text{-}\notionwrite{PQC}}
\newcommand{\INDAAPQC}{\IND{}\text{-}\secAA{}\text{-}\notionwrite{PQC}}
\newcommand{\INDAAQKD}{\IND{}\text{-}\secAA{}\text{-}\notionwrite{QKD}}

\newcommand{\party}{\algowrite{P}} 				
\newcommand{\Init}{\algowrite{Init}}			
\newcommand{\SendMOne}{\algowrite{SendM1}}
\newcommand{\SendMTwo}{\algowrite{SendM2}}

\newcommand{\TrivialQKD}{{\color{bookkeepcolor}\algowrite{TrivialQKD}}}
\newcommand{\TrivialPQC}{{\color{bookkeepcolor}\algowrite{TrivialPQC}}}

\newcommand{\orEst}{{\color{oraclecolor}\orwrite{EST}}\xspace}
\newcommand{\orTest}{{\color{oraclecolor}\orwrite{TEST}}\xspace}

\newcommand{\orSend}{{\color{oraclecolor}\orwrite{SEND-*}}\xspace}	
\newcommand{\orSendInit}{{\color{oraclecolor}\orwrite{SEND-INIT}}\xspace}	
\newcommand{\orSendMOne}{{\color{oraclecolor}\orwrite{SEND-M1}}\xspace}	
\newcommand{\orSendMTwo}{{\color{oraclecolor}\orwrite{SEND-M2}}\xspace}	

\newcommand{\orRevState}{{\color{oraclecolor}\orwrite{REVEAL-STATE}}\xspace}
\newcommand{\orCorrupt}{{\color{oraclecolor}\orwrite{CORRUPT}}\xspace}
\newcommand{\orReveal}{{\color{oraclecolor}\orwrite{REVEAL}}\xspace}
\newcommand{\orQKDSet}{{\color{oraclecolor}\orwrite{QKD-OVERRIDE}}\xspace}
\newcommand{\orQKDGet}{{\color{oraclecolor}\orwrite{QKD-LEAK}}\xspace}
\newcommand{\orQKDSids}{{\color{oraclecolor}\orwrite{QKD-KEY-HOLDERS}}\xspace}

\newcommand{\numberParties}{\ensuremath{\varwrite{n_{\texttt{p\!\!\:t}}}}\xspace} 	
\newcommand{\numberSessions}{\ensuremath{\varwrite{n_{\texttt{s}}}}\xspace}

\newcommand{\sessionID}{{\color{bookkeepcolor}\varwrite{sID}}} 		
\newcommand{\ctr}{\varwrite{ctr}}				
\newcommand{\sid}{\sessionID} 
\newcommand{\sidinit}{\sid_{\text{init}}}
\newcommand{\sidresp}{\sid_{\text{resp}}}
\newcommand{\sidctr}{\ensuremath{\sid_\ctr}} 		
\newcommand{\sidmatch}{\ensuremath{\overline{\sid}}} 		

\newcommand{\peer}{\arwrite{peer}}				
\newcommand{\owner}{\arwrite{owner}}				
\newcommand{\role}{\arwrite{role}}				
\newcommand{\roleInit}{{\color{bookkeepcolor}\mathsc{Initiator}}}		
\newcommand{\roleResp}{{\color{bookkeepcolor}\mathsc{Responder}}}	

\newcommand{\sessionkeyArray}{\arwrite{sesKey}}	
\newcommand{\sessionkey}{\varwrite{sesKey}}		

\newcommand{\sent}{\arwrite{sent}}			
\newcommand{\received}{\arwrite{received}}		
\newcommand{\key}{\arwrite{qKey}}  
\newcommand{\flag}{\arwrite{qStatus}}  
\newcommand{\kdfIn}{\arwrite{kdfIn}}  

\newcommand{\honest}{{\color{bookkeepcolor}\mathsc{Honest}}} 
\newcommand{\corrupt}{{\color{bookkeepcolor}\mathsc{Corrupt}}} 
\newcommand{\leaked}{{\color{bookkeepcolor}\mathsc{Revealed}}} 
\newcommand{\Accept}{{\color{bookkeepcolor}\mathsc{Accept}}} 
\newcommand{\Reject}{{\color{bookkeepcolor}\mathsc{Reject}}} 

\makeatletter
\newcommand{\msgdst}{\@ifstar{\ensuremath{p_\varwrite{out}}}{\ensuremath{\vec{p_\varwrite{out}}}}} 
\makeatother

\newcommand{\state}{\arwrite{state}}					

\newcommand{\keyID}{\varwrite{kID}}				
\newcommand{\kid}{\keyID} 		
\newcommand{\kidctr}{\ensuremath{\keyID_\ctr}} 		
\newcommand{\kidstar}{{\color{bookkeepcolor}\ensuremath{\kid^*}}} 		
\newcommand{\kidUsed}{\arwrite{kidUsed}} 		
\newcommand{\qsent}{\arwrite{qSentSid}}
\newcommand{\qrecv}{\arwrite{qRecvSid}}

\newcommand{\Attack}{{\color{bookkeepcolor}\algowrite{AttackPQC}}}
\newcommand{\FindMatches}{{\color{bookkeepcolor}\algowrite{FindMatches}}}
\newcommand{\Completed}{{\color{bookkeepcolor}\algowrite{Completed}}}

\newcommand{\corrupted}{\arwrite{corrupted}}		
\newcommand{\revealedState}{\arwrite{revState}}	
\newcommand{\revealed}{\arwrite{revealed}}		
\newcommand{\matchingSessions}{{\color{bookkeepcolor}\ensuremath{\varwrite{matches}}}}
\newcommand{\kdfMatch}{{\color{bookkeepcolor}\ensuremath{\varwrite{kdfMatch}}}}
\newcommand{\complete}{{\color{bookkeepcolor}\ensuremath{\varwrite{complete}}}}

\newcommand{\qkdInit}{{\color{bookkeepcolor}\ensuremath{\algowrite{QkdInit}}}}
\newcommand{\enckey}{{\color{qkdoraclecolor}\orwrite{QKD-GET-KEY}}}
\newcommand{\deckey}{{\color{qkdoraclecolor}\orwrite{QKD-GET-KEY-WITH-ID}}}
\newcommand{\kpqc}{\ensuremath{k_\varwrite{pqc}}}
\newcommand{\kpqcm}{\ensuremath{k_{\varwrite{pqc},m}}}
\newcommand{\kpqcs}{\ensuremath{k_{\varwrite{pqc},s}}}
\newcommand{\kqkd}{\ensuremath{k_\varwrite{qkd}}}
\newcommand{\kqkdm}{\ensuremath{k_{\varwrite{qkd},m}}}
\newcommand{\kqkds}{\ensuremath{k_{\varwrite{qkd},s}}}

\newcommand{\ksess}{\ensuremath{k_\varwrite{sess}}}
\newcommand{\kalice}{\ensuremath{k_a}}
\newcommand{\kbob}{\ensuremath{k_b}}
\newcommand{\keph}{\ensuremath{k_e}}
\newcommand{\calice}{\ensuremath{c_a}}
\newcommand{\cbob}{\ensuremath{c_b}}
\newcommand{\ceph}{\ensuremath{c_e}}

\newcommand{\game}[1]{\ensuremath{G_{#1}}}
\newcommand{\gameb}[1]{\ensuremath{G_{#1}}}
\newcommand{\gamenst}[1]{\ensuremath{G_{#1}^{\neg\text{st}}}}
\newcommand{\gamenstb}[1]{\ensuremath{G_{#1}^{{\neg\text{st}}}}}
\newcommand{\gamenski}[1]{\ensuremath{G_{#1}^{{\neg sk(i)}}}}
\newcommand{\gamenm}[1]{\ensuremath{G_{#1}^{\neg\text{ms}}}}
\newcommand{\gamenmi}[1]{\ensuremath{G_{#1}^{\neg\text{ms},\text{i}}}}
\newcommand{\gamenmik}[1]{\ensuremath{G_{#1}^{\neg\text{ms},\text{i},\text{k}}}}
\newcommand{\gamenmink}[1]{\ensuremath{G_{#1}^{\neg\text{ms},\text{i},\neg\text{k}}}}
\newcommand{\gamenmr}[1]{\ensuremath{G_{#1}^{\neg\text{ms},\text{r}}}}
\newcommand{\gamenmrk}[1]{\ensuremath{G_{#1}^{\neg\text{ms},\text{r},\text{k}}}}
\newcommand{\gamenmrnk}[1]{\ensuremath{G_{#1}^{\neg\text{ms},\text{r},\neg\text{k}}}}
\newcommand{\gamenqz}[1]{\ensuremath{G_{#1}^{\text{qkd}}}}
\newcommand{\gamenqr}[1]{\ensuremath{G_{#1}^{\text{qkd},\text{r}}}}
\newcommand{\gamenqnt}[1]{\ensuremath{G_{#1}^{\text{qkd},\text{r},\neg\text{mt}}}}
\newcommand{\gamenqt}[1]{\ensuremath{G_{#1}^{\text{qkd},\text{r},\text{mt}}}}

\newcommand{\sinit}{\ensuremath{s_\varwrite{init}}}
\newcommand{\sresp}{\ensuremath{s_\varwrite{resp}}}
\newcommand{\tinit}{\ensuremath{t_\varwrite{qRecv}}}
\newcommand{\tresp}{\ensuremath{t_\varwrite{qSent}}}

\newcommand{\sidqsend}{\ensuremath{\sid_\varwrite{qs}}}
\newcommand{\sqsend}{\ensuremath{\sid_\varwrite{qSent}}}
\newcommand{\sqrec}{\ensuremath{\sid_\varwrite{qRecv}}}

\newcommand{\leneph}{\ensuremath{\ell_\texttt{eph}}}
\newcommand{\lenstat}{\ensuremath{\ell_\texttt{stat}}}
\newcommand{\lenpqc}{\ensuremath{\ell_\texttt{pqc}}}

\newcommand{\lenqkd}{\ensuremath{\ell_\texttt{qkd}}}

\newcommand{\lensess}{\ensuremath{\ell_\texttt{sess}}}

\newcommand{\deph}{\ensuremath{\delta_\varwrite{eph}}}
\newcommand{\dstat}{\ensuremath{\delta_\varwrite{stat}}}

\titlerunning{QKD Oracles for AKE}	
\title{QKD Oracles for Authenticated Key Exchange}

\ifSubmission
	\author{\vspace{-0.5in}}
	\institute{}
\else
	\author{
		Kathrin Hövelmanns\inst{1}
		\and
		Daan Planken\inst{2,3}
		\and
		Christian Schaffner\inst{2,3}
		\and
    Sebastian R. Verschoor\inst{2,3}
	}
	\institute{
		Eindhoven University of Technology, The Netherlands\\
		\and
		University of Amsterdam, The Netherlands\\
		\and
		QuSoft, Amsterdam, The Netherlands\\
		\email{qkdOraclesForAKE@hoevelmanns.net}
	}
	\authorrunning{Hövelmanns, Planken, Schaffner, Verschoor}
\fi

\def\acknowledgmenttext{
KH was supported by an NWO VENI grant (Project No. VI.Veni.222.397).
DP and SV were supported by the Dutch National Growth Fund (NGF), as part of the Quantum Delta NL programme. CS is (partially) supported by gravitation project Challenges in Cyber Security (CiCS) with file number 024.006.037 which is financed by the Dutch Research Council (NWO).
}

\begin{document}

	


\maketitle

\begin{abstract} 
	Authenticated Key Exchange (AKE) establishes shared (`symmetric') cryptographic keys which are essential for secure online communication. 
	AKE protocols can be constructed from public-key cryptography like Key Encapsulation Mechanisms (KEMs). 
	Another approach is to use Quantum Key Distribution (QKD) to establish a symmetric key, which uses quantum communication.
	\emph{Combining} post-quantum AKE and QKD \emph{appropriately} may provide security against quantum attacks even if only one of the two approaches turns out to be secure.

	We provide an extensive review of existing security analyses for combined AKE and their formal security models,
	and identify some gaps in their treatment of QKD key IDs. In particular, improper handling of QKD key IDs leads to \attackname attacks on AKE.

	As our main conceptual contribution, we model QKD as an oracle that closely resembles the standard ETSI 014 QKD interface.
  	We demonstrate the usability of our QKD oracle for cryptographic security analyses by integrating it into a prominent security model for AKE, called CK$^+$ model,
  	thereby obtaining a security model for combined AKE that catches \attackname attacks.
  	In this model, we formally prove security of a new protocol that combines QKD with a triple-KEM handshake.
  	This is the first provably secure hybrid protocol that maintains information-theoretic security of QKD.
  
\end{abstract}

\ifSubmission \else
	\begingroup
	\makeatletter
	\renewcommand\thefootnote{}\footnote{%
  \relax \acknowledgmenttext
Date: September 15, 2025}
  \addtocounter{footnote}{-1}%
	\endgroup
\fi


\tikzstyle{entity} = [
	rectangle,
	minimum width=30mm,
	minimum height=10mm,
	text centered,
	text width=30mm,
	draw=black,
	fill=white
]
\tikzstyle{op} = [rectangle, draw=black, fill=white, align=left]
\tikzstyle{msg} = [anchor=south, fill=white, fill opacity=0.9, text opacity=1]

\section{Introduction} \label{sec:intro}
	Today's secure online communication heavily relies on symmetric cryptography.
One way to establish shared symmetric \emph{session keys} is to use Authenticated Key Exchange (AKE) protocols, which are often based on public-key cryptography. AKE guarantees that session keys remain unknown to anyone but the two intended parties. By this guarantee, unauthorized parties cannot obtain these session keys, and thus the honest parties' communication is protected by symmetric cryptography.
It is, however, not completely straightforward how these AKE security guarantees should be formalized --
over the past three decades, different security definitions~\cite{C:BelRog93,EC:CanKra01,C:Krawczyk05,PROVSEC:LaMLauMit07} have been given,
all coming with subtle differences and advantages/disadvantages.

\subheading{Post-quantum AKE.}
Quantum computers threaten the security of many widely deployed AKE protocols,
as Shor's quantum algorithm~\cite{FOCS:Shor94} breaks the assumptions underlying the current public-key schemes.
One countermeasure is to replace the current cryptographic algorithms with ones that can be executed on \emph{classical} (non-quantum) hardware, but that resist attacks by an adversary equipped with a quantum computer.
This type of cryptography is called Post-Quantum Cryptography (PQC)~\cite{BL17}.
Recent efforts have led to the development of new digital signatures and Key Encapsulation Mechanisms (KEMs),
cryptographic building blocks that can be used as replacements in existing cryptographic protocols like AKE.
There has been much scientific debate on which concrete signatures and KEMs should be chosen for deployment,
with the ongoing effort accumulating in several (forthcoming) standards~\cite{AAC+22,ABC+25}.

One concern with novel post-quantum cryptography (and public-key algorithms in general) is that their security relies on the conjectured hardness of certain computational problems.
Accordingly, future technological or cryptanalytic advances may turn out to invalidate their security.
This is unavoidable because it is impossible to have information-theoretic secure public-key cryptography --
an unbounded attacker can always look up the public key and then run a brute-force search for the private key.
To hedge against potential future breaks, it is thus widely recommended~\cite{BSIhybrid,ANSSIhybrid,SCN+24} to deploy \emph{hybrid} solutions:
hybrid solutions combine multiple algorithms such that the combination remains secure even if all but one of its components are broken.
For KEMs, recent research~\cite{PKC:GiaHeuPoe18,PQCRYPTO:BBFGS19} proved that a natural KEM combiner approach -- adding several KEM keys together via an exclusive-or (XOR) -- 
is not sufficient to achieve practical (\emph{chosen-ciphertext}) security,
and provided non-trivial alternatives that actually meet this security goal.

\subheading{QKD.}
Another approach to preventing quantum attacks is to use Quantum Key Distribution (QKD)~\cite{BEN84,Ekert91}.
QKD is an interactive protocol which allows Alice to establish a cryptographic key with Bob
by encoding information into quantum states that are sent over a quantum channel.
QKD provides security guarantees that do not rely on computational hardness assumptions.
This is also called information-theoretic security (ITS).
However, for QKD to have ITS, the classical communication in a QKD protocol must be authenticated with ITS one-time Message Authentication Codes (MACs),
which is commonly achieved by using pre-shared symmetric keys.
If these pre-shared keys were directly used for symmetric cryptography, they would be used up quickly,
while QKD allows to extend these pre-shared keys -- something that is impossible with classical communication alone.
In other words, the ITS property of QKD eliminates the threat of future cryptanalytic advances.

On the other hand, QKD requires careful handling of quantum information which makes implementations technologically challenging. As a result, QKD is currently limited in range (to a few hundred kilometers) and in key rates by the quantum communication hardware, and the service might not always be available. 
Some of these limitations are mitigated in practice by having QKD running continuously in the background, producing keys that are stored in a buffer for later use. Users can request keys  via ETSI 014~\cite{etsi014} or a similar interface. The keys are then provided to the user together with an identifier.
To ensure that both users retrieve the same key from their devices, the users need to communicate that identifier as part of their protocol.
Notably, the key identifier is the only value that binds the QKD key to the session. As we will demonstrate, it is essential to include these identifiers in the higher-level protocol --
we describe (in \cref{sec:mix-and-match}) an attack on protocols that exchange the identifiers out-of-bound.
Crucially, that inclusion should not make use of computationally secure primitives (such as cryptographic hash functions),
as application of computational primitives to a QKD key would void information-theoretic security,
the major selling point of QKD.

While several QKD standardization processes are ongoing~\cite{Dev+22,Saez2024}, a widely accepted standard for QKD protocols is still lacking. In order to certify QKD deployment in practice, we expect it to be mandatory to combine their key material with post-quantum or current public-key cryptography in a hybrid AKE protocol.

\medskip
Given the importance of AKE, and the uncertainties surrounding post-quantum approaches and QKD, is it natural to ask:

\medskip
\emph{What is the appropriate way of combining post-quantum cryptography and QKD to achieve a hybrid protocol that remains secure if either the post-quantum cryptography or QKD is secure?}

\subsection{Our contributions}
\begin{itemize}
\item Our main conceptual contribution is a tool that allows to formally analyze the security provided by QKD in cryptographic contexts. Concretely, we propose \textbf{QKD oracles} whose functionality closely resemble the standard ETSI 014 QKD interface~\cite{etsi014}, but they can easily be adapted to other QKD device interfaces as well, see \cref{fig:qkd} in \cref{sec:secModel}.
The oracles allow an honest party to retrieve a QKD key (implemented as a uniformly random bitstring)
accompanied by a unique identifier. By querying the oracle with that same identifier, the honest peer can retrieve the same key.
This interface forces protocols to explicitly communicate the key identifier.
We model compromised parties or other attacks on QKD by allowing the adversary to query the oracle, to learn or even change the key.
\item These oracles can be integrated into AKE models, as we demonstrate by incorporating them into the CK$^+$ security model for AKE~\cite{C:Krawczyk05}, 
resulting in a \textbf{novel AKE security model} that captures \attackname attacks, see \cref{sec:secModel}.
We build upon the work of H\"ovelmanns, Kiltz, Sch\"age, and Unruh~\cite{PKC:HKSU20},
who specify the CK$^+$ security model in pseudo-code to ensure there is as little ambiguity as possible.
\item We propose a \textbf{new hybrid AKE protocol} that combines QKD with a Triple-KEM PQC protocol, see \cref{fig:protocol} in \cref{sec:protocol}.
To cryptographically bind the sessions together,
we compute two one-time MACs:
one with (part of) the QKD key and one with (part of) the PQC key.
The session key is given by the XOR of the remainder of both keys.
In \cref{sec:security-proof}, we prove that this protocol is secure in our CK$^+$-like model,
even if one of its components fails.
If QKD is secure, we obtain information-theoretic security,
while the security of the PQC protocol is relative to the security of the KEMs and is given in the Quantum Random Oracle Model (QROM).
\item As a minor contribution,
we demonstrate the importance of explicitly communicating QKD key identifiers
by giving an \textbf{explicit \attackname attack} against protocols that consider this out-of-bounds, see \cref{sec:mix-and-match}.
We also provide an extensive review of existing literature in \cref{sec:related_work},
and we highlight an issue in the security proof of ITS KEM combiners.
\end{itemize}

\subheading{Outline}
Our article is structured as follows. 
In \Cref{sec:related_work}, we provide an extensive review of related work on KEM combiners and QKD integration.
\Cref{sec:Prels} provides the necessary background on AKE, PQC and QKD.
\Cref{sec:protocol} presents our hybrid AKE protocol combining QKD with Triple-KEM PQC (see \cref{fig:protocol}), 
and \cref{sec:mix-and-match} explains potential \attackname attacks when QKD key IDs are not handled properly.
\Cref{sec:secModel} defines our AKE security model including the formalization of our new QKD oracles in \cref{fig:qkd}.
\Cref{sec:security-proof} provides the security proof for our protocol.
\Cref{sec:conclusion} concludes the paper and discusses future work.


\subsection{Related work} 
\label{sec:related_work}

\subheading{Original work on QKD in an Authenticated Key Exchange (AKE) framework.}
To model QKD security in an AKE-like fashion, Mosca, Stebila, and Ustaoğlu~\cite{PQCRYPTO:MosSteUst13} enhanced an AKE security model, called eCK model,
with an oracle (called SendQ oracle) that models quantum communication.
The authors concluded the work with the question how end-to-end protocols could best combine QKD and classical (non-quantum) AKE keys.

\subheading{KEM Combiners as `hybrid AKE' (QKD-AKE) combiners?}
Our end-to-end protocol was designed to combine a QKD exchange with an AKE (a KEM handshake) in a way that preserves security even if one of the components fails.
This goal strongly resembles the goal underlying recent KEM combiners~\cite{PKC:GiaHeuPoe18, PQCRYPTO:BBFGS19}.
There already exist various `hybrid AKE' constructions that combine QKD with classical cryptography,
which we will discuss and compare below.
What many approaches have in common is that they simply use QKD as input to existing KEM combiners.
We want to immediately stress, however, that this raises two independent problems:

\ifTightOnSpace	\else \begin{enumerate} \fi
	\ifTightOnSpace	\else \item \fi
		Firstly, \emph{QKD is not a KEM} --- KEMs \emph{and their security} are inherently based on the fact that they use an asymmetric key pair.
		QKD inherently lacks the input requirements of a KEM combiner.
		As we will detail in~\cref{ssec:xtm-combiner}, recent KEM combiners do not mesh well with QKD and
		finding a suitable QKD-AKE combiner is less straightforward than one might initially think.
	\ifTightOnSpace	\else \item \fi
		Secondly, the output of a KEM combiner is \emph{still a KEM} --
		even if it was `proven secure' (as a \emph{KEM}),
		it doesn't automatically yield a secure \emph{protocol} for \emph{Authenticated} Key Exchange.
		The resulting (combined, \emph{authenticated}) protocol still needs to be analyzed in a proper AKE-like security model.
		To that end, QKD needs to be modeled properly via some abstraction such as, e.g., our oracle (see~\cref{sec:secModel}),
		or a shared array (as done in HAKE~\cite{PQCRYPTO:DowHanPat20}, see below).
\ifTightOnSpace	\else \end{enumerate} \fi

We summarize the reviewed literature in~\cref{tab:comparison}.
The given constructions vary widely concerning the availability of a technical security analysis,
with the level ranging from non-existent to full formal proofs in a suitable security model.
We list the type of cryptographic construction that is being given (`primitive'),
the security definition against which it is analyzed (if any),
whether the analysis is given as a formal proof,
whether or not the combiner includes the QKD key ID somewhere (which proves important due to the attack described in~\cref{sec:mix-and-match}),
and whether or not the combiner retains information-theoretic security (ITS).
We substantiate the table by detailing our findings below.

\ctable[
caption={Comparison of hybrid constructions that integrate QKD.},
label={tab:comparison},
pos= ht
]{lcccccc}{
  \tnote[a]{Despite the work suggesting that the construction is chosen-ciphertext secure (see below).}
  \tnote[b]{Status unclear: proof contains inconsistencies (see below).}
  \tnote[c]{Despite what is stated in the work (see below).}
  \tnote[d]{Status unclear: incomplete security definition (see below).}
}{
  \toprule
  & Primitive & Security Notion & Proof~ & Key ID & ITS \\
  \midrule
  OpenQKD~\cite{TLLM18} & PSK/XOR & None & No & No & Maybe \\
  X.1714~\cite{X1714} & Key combiner & None & No & No & Maybe \\
  3-key combiner~\cite{RDM+24} & dual PRF & dual PRF\tmark[a] & ?\tmark[b] & Yes & No \\
  QR TLS 1.3~\cite{GAO+23} & AKE & None & No & Yes & No \\
  ART~\cite{ART24} & AKE & chosen-ciphertext{} & No & Yes & No \\
  Muckle~\cite{PQCRYPTO:DowHanPat20} & AKE & HAKE & Yes & No & No & \\
  Muckle+~\cite{MucklePlus} & AKE & HAKE & Yes & No & No \\
  Muckle++~\cite{GPH+24} & AKE & HAKE & ?\tmark[b] & Yes & No\tmark[c] \\
  Muckle\#~\cite{BSP+24} & AKE & HAKE & ?\tmark[d] & No & No \\
  VMuckle~\cite{BBB+25a} & AKE & HAKE & Yes & No & No \\
  This work & AKE & CK$^+$ (with QKD oracle) & Yes & Yes & Yes \\
  \bottomrule
}

{\subheading{OpenQKD.}
	Tysowski, Ling, L\"utkenhaus, and Mosca~\cite{TLLM18}
	suggest to use the QKD key directly as a pre-shared key (PSK) in existing protocols
	when ITS is not required.
	Alternatively, they give an example for a XOR key combination,
	and substantiate compliance of the XOR combiner with the NIST recommendations by
	referring to the NIST PQC project FAQ.
	By now, however, XORing keys is no longer consistent with NIST recommendations --
	the respective FAQ answer has since been clarified.
	It now refers to SP 800-56C~\cite{sp800-56c} for a generic composite key establishment scheme
	that concatenates the keys and then hashes them with a cryptographic hash function.
	More importantly, however, we would advise against simply XORing the keys since this is vulnerable to \attackname attacks
	(see~\cref{sec:mix-and-match}).
}

{\subheading{ITU-T X.1714.}
	recommends~\cite{X1714} to combine QKD with other methods using ``key combiners''.
	By this they mean a function that takes multiple keys as input and outputs another key.
	We note that their function does not take KEM ciphertexts (or other context) as input.
	The document provides no formal security definition.
}
 
{\subheading{PRF-based combiners.}
	A dual Pseudo-Random Function (PRF) takes two inputs and provides a pseudorandom output if either input is pseudorandom.
	Aviram, Dowling, Komargodski, Paterson, Ronen, and Yogev~\cite{EPRINT:ADKPRY22}
	give a (proven secure) dual PRF construction (called KeyCombine).
	Bindel, Brendel, Fischlin, Goncalves and Stebila~\cite{PQCRYPTO:BBFGS19}
	show that such dual PRFs can be used for KEM combiners,
	by providing a (proven secure) KEM combiner (called Dual-PRF Combiner)
	that first applies a dual PRF to the KEM keys,
	and then uses the result as the key for a PRF computed over the ciphertexts.
	The authors warn that ``the naive approach of directly using a dual PRF'' does not preserve practical (chosen-ciphertext) security.
	
	In~\cite{EPRINT:ADKPRY22}, the authors also gave a generalized construction (called multi-key KeyCombine) with more than two keys as input,
	and accordingly adapted the security definition.
	A variant of the multi-key KeyCombine construction, called 3-key combiner,
	was given by Ricci, Dobias, Malina, Hajny and Jedlicka in~\cite{RDM+24}.
	The authors argue chosen-ciphertext security of the 3-key combiner by stating
	that it achieves the security notion for multi-input PRFs
	and then referring to the proof of the dual PRF \emph{KEM combiner}.
	We now argue, however, that there are several independent issues with this.
	First, in this context, the combiner yields an interactive protocol, not a KEM,
	and chosen-ciphertext security will thus not be the relevant security property in this context.
	Assuming it were, we secondly note that the work uses the multi-input PRF `naively'
	in the sense described by~\cite{PQCRYPTO:BBFGS19},
	and as~\cite{PQCRYPTO:BBFGS19} warned, this does not achieve chosen-ciphertext security.
	Thirdly, the combiner expects KEM outputs as its input,
	whereas the construction in~\cite{RDM+24} uses QKD and Kyber.AKE~\cite{BDK+18} as inputs, none of which are KEMs.
	In conclusion, we believe that chosen-ciphertext security neither holds nor is it the relevant security property.
	
}

{\subheading{QR TLS 1.3.}
	In the quantum-resistant TLS 1.3 protocol by Rubio García, Cano Aguilera, Vegas Olmos, Tafur Monroy, and Rommel~\cite{GAO+23},
	the authors take the QKD key and concatenate it
	with the existing Input Keying Material of the TLS 1.3 key schedule.
	(The QKD key ID is included in the ServerHello message.)
	The protocol is described as ``achieving enhanced security against quantum adversaries'',
	but the work does not provide a corresponding security definition or (informal) security analysis to clarify this description.
}

{\subheading{ART.}
	Acquina, Rommel and Tafur Monroy~\cite{ART24} construct a protocol
	that runs two ephemeral KEMs in parallel with QKD
	and then apply a KEM combiner resembling one from~\cite{PKC:GiaHeuPoe18}.
	All protocol messages are both signed (with EdDSA and with Dilithium) and tagged (with Poly1305, using part of the QKD key).
	Security of the protocol is not treated explicitly -- no formal security definition or proof are given.
	We see two issues with the construction, resembling those of the 3-key combiner:
	Firstly, the resulting protocol is not a KEM, chosen-ciphertext security of the combiner thus might not be enough.
	Secondly, QKD is not a KEM, which the combiner would expect as input.
	The authors argue that QKD could be viewed as a KEM where ``public and secret keys play no role'' --
	the QKD step that produces the key and its ID could essentially be viewed
	as the KEM algorithm that produces a key and a ``ciphertext'',
	and the QKD step that takes the key ID and looks up the corresponding QKD key could essentially be viewed
	as the KEM algorithm that recovers shared keys from ciphertexts.
	We would argue, however, that this would not hold up in terms of security --
	QKD
	does not use a public-key pair like a KEM does and is therefore incompatible with the usual security notions for KEMs,
	including chosen-ciphertext security (as required by the combiner).
}

{\subheading{HAKE and Muckle.}
	Dowling, Hansen, and Paterson~\cite{PQCRYPTO:DowHanPat20} also combined classical (non-PQC) cryptography, PQC and QKD,
	accordingly calling the result Hybrid AKE (HAKE).
	They give a formal security definition, by adapting an AKE security model to combined protocols,
	reflecting that each of the three components by itself might be vulnerable.
	QKD is described as being modeled as a shared array of secret bits
	which honest parties can access to obtain their QKD keys,
	using an index that would be functionally equivalent to the QKD key ID (also see~\cref{ssec:qkd}).
	We noticed, however, that parties will always get the correct QKD key without having to specify or communicate any index at all,
	since the model~\cite[Appendix~C, Fig.~5]{EPRINT:DowHanPat20}
	fixes a shared key per protocol stage.
	It thus does not capture attacks where the adversary tricks an honest party into using the wrong key,
	such as the \attackname attack described in~\cref{fig:mix-and-match}.
	
	We also noticed that the security definition exhibits some ambiguity due to ambiguity of the cleanliness predicate,
	which in turn stems from a technical issue regarding how QKD keys are shared between sessions.
	Concretely, the HAKE \algowrite{Create} query initializes a fixed key (per session, per stage) in a variable called \textbf{esk}.
	When the adversary triggers subsequent sessions, they can specify if \textbf{esk} should be copied from an already initialized session at the peer.
	Specifically for protocols like Muckle (see below), this power results in ambiguity:
	the cleanliness predicate (Definition 3: \textsf{clean}$_{q\textsf{HAKE}}$)
    only forbids ``trivial'' leakage of the test session's QKD key, i.e., leakage either through the test session itself or its matching session.
    A security proof in this model does not capture that two parties might run a shared session,
    using different QKD keys (leading to desynchronization that might only be detected at a later stage)
    and it could happen that a QKD key might be copied over to another session that should in principle be independent.
	This issue can be resolved by accordingly adjusting the \algowrite{Create} query
	or by modeling QKD through an oracle instead, as we will describe in~\cref{fig:qkd}.

	In the same work, the authors present Muckle,
	a multi-stage protocol where parties share a pre-shared key (PSK),
	a secret state,
	and have access to QKD.
	At each stage, the session generates three ephemeral keys (with a classical KEM, a PQC KEM, and QKD),
	and authenticates the sent messages with a MAC.
	The ephemeral keys, secret state, protocol transcript and a counter are
	processed in a key schedule (a chain of PRFs),
	which outputs an updated secret state and the session keys.
	Since Muckle derives the keys with a PRF, it can be computationally secure at best.
	Furthermore, it is known~\cite[Theorem~19]{CSF:CohCreGar16} that correct one-round protocols cannot achieve full PCS without being vulnerable to denial-of-service attacks,
	and indeed Muckle can be desynchronized by dropping a single response message.
	Security is proven in the HAKE model.
	We note a very minor issue with the main security theorem \cite[Theorem~1]{PQCRYPTO:DowHanPat20}
	that analyses attacks when the adversary compromised the PQC KEM or QKD (but not both).
	The proof analyses the two cases separately and then adds the resulting terms,
	whereas it could (and should) take their minimum.	  
}

{\subheading{Muckle+.}
	Bruckner, Ramacher, and Striecks~\cite{MucklePlus} later presented Muckle+,
	a Sign-and-MAC style protocol which replaces Muckle's authentication method
	with a new approach: whereas Muckle MAC'ed the sent messages, using a PSK-based key,
	Muckle+ takes (the hash of) a partial transcript of the protocol run, signs it,
	and then MACs it with a key that is derived from the established ephemeral keys and the partial transcript.
	This makes Muckle+ significantly distinct from Muckle and Krawczyk's SIGMA protocol~\cite{C:Krawczyk03}.
	The security analysis in in the HAKE model.
	We remark that the authors state that they need the QKD keys to be distributed via multi-path techniques,
  which is not captured by HAKE (and thus not by the proof).
	
	Technically, Muckle+ contains an (easily fixable) bug, stemming from the treatment of the session key.
	The authors determine~\cite[Section 3.3]{MucklePlus} the to-be-attacked session key as 
	a master secret $\varwrite{MS}$.
	But this creates a vulnerable dependency -- in the next stage of the protocol, the session's internal state is updated to
	$\varwrite{SecState} \gets \PRF(\varwrite{MS}, pub)$,
	where $pub$ is some publicly available information.
	When attacking the key $\varwrite{MS}$, the adversary gets a value $\varwrite{MS}'$ that is either random or equals $\varwrite{MS}$
	and is tasked with determining which it is.
	To do so, the adversary can now 
	exploit that the HAKE model allows to reveal $\varwrite{SecState}$, the next stage's state.
	By checking whether the state fits the rule $\varwrite{SecState} = \PRF(\varwrite{MS}', pub)$,
	the adversary immediately knows if $\varwrite{MS}'$ equals $\varwrite{MS}$.
	This can easily be fixed by setting the session key to $(\varwrite{CATS}, \varwrite{SATS})$,
	which are application traffic secrets from which no following values are derived.
	We reported this issue to the authors and it was resolved in follow-up work (VMuckle, see below).
}

{\subheading{Muckle++.}
	Muckle++ is a protocol 
	by Garms, Paraïso, Hanley, Khalid, Rafferty, Grant, Newman, Shields, Cid, and O'Neill
	that modifies Muckle by (optionally) replacing the \MAC{}s with signatures,
	and by replacing the \KDF with an ITS primitive~\cite{GPH+24}.
	Specifically, the key schedule replaces the \PRF in the key schedule and derivation of the \MAC keys
	with something called a QPRF.
	The authors write:
	``To instantiate the QPRF, a keyed-hash \MAC (HMAC)-based \KDF
	(HKDF) was modified by xor-ing quantum key material to the \PRF output.''
	It is never specified where this ``quantum key material'' comes from,
	nor is it ever provided as input to QPRF in the protocol description.
	More important, however, is that in the security proof [Theorem~3, Supporting Information],
	the authors model the QPRF as a dual-PRF, which can never achieve ITS.
}

{\subheading{Muckle\#.}
	Muckle\# is a protocol
	by Battarbee, Striecks, Perret, Ramacher, and Verhaeghe
	that modifies Muckle+ by replacing the signatures
	used for authentication with static KEMs~\cite{BSP+24}.
	The security analysis appears to be incomplete --
	we were unable to verify the proof due to two issues.
	Firstly, Definition 8 (malicious acceptance) appears to be incomplete, as the sentence contains the phrase ``has not been issued'', but it is unclear what this is referring to.
	It is not stated how this definition is integrated into the HAKE security experiment.
	Secondly, we note that the cleanliness definition is incomplete:
	the passive case lists that no \algowrite{CompromiseQK} query may have been issued for the test session/stage,
	but the proof of that case considers two identical sub-cases that assume no \algowrite{CompromiseSK} query has been made.
	The wording for the cleanliness definition for the active case suggests
	there should be a list of queries that are disallowed, but no queries are listed.
}

{\subheading{VMuckle.}
	VMuckle is a protocol
	by Buruaga, Bugler, Brito, Martin, and Striecks~\cite{BBB+25a},
	that adds cryptographic agility to Muckle+
	for the purpose of authentication in the MACsec protocol:
	parties authenticate either using a pre-shared key or with digital signatures.
  The session key $(\varwrite{CATS}, \varwrite{SATS})$ is then used as the Master Session Key input
	for the MACsec Key Agreement (MKA) protocol.
}

\section{Preliminaries} \label{sec:Prels}
	In this section, we recollect standard notation and background for (post-quantum) cryptography like cryptographic primitives, Authenticated Key Exchange (AKE), and the Quantum Random Oracle Model (QROM), as well as Quantum Key Distribution (QKD).

By $\bool{b}$ we denote the indicator variable of $b$,
which takes value 1 if the boolean statement $b$ is true and 0 if $b$ is false.
A list $[x, y]$ contains two elements, $[\ ]$ is an empty list,
and to append $x$ to list $\ell$, we write $\ell \mathrel{+}= [x]$.
We write $x \uni S$ to denote that $x$ is sampled uniformly random from $S$,
we write $x \gets A()$ if $x$ is the output of a probabilistic algorithm $A$, and
we write $x := A()$ if $x$ is the output of a deterministic algorithm $A$.
We write $A^H$ to indicate that an algorithm $A$ has access to an oracle $H$.
Algorithms are efficient with respect to the security parameter,
with the exception of computationally unbounded adversaries.
While formally this requires each algorithm to receive the security parameter as (unary) input,
we omit this to avoid notational clutter.
\notionwrite{SECURITY\text{-}NOTIONS} are written in capitalized sans-serif font.
Bookkeeping is denoted in {\color{bookkeepcolor}green}, using different fonts for
$\arwrite{associative arrays}$,
${\color{bookkeepcolor}\mathsc{Constants}}$,
and ${\color{bookkeepcolor}\varwrite{variables}}$.
\algowrite{Subroutines} are written in sans-serif font,
and \orwrite{ORACLES} in capitals,
with {\color{oraclecolor}red} for {\color{oraclecolor}\orwrite{ADVERSARY-ORACLES}}
and {\color{qkdoraclecolor}purple} for {\color{qkdoraclecolor}\orwrite{QKD-ORACLES}}.
The symbol $\bot$ represents failure,
for example when indexing an associative array on a key that has not been set yet.
For clarity, we explicitly initialize some variables,
others are implicitly initialized at a zero or empty value.

\subheading{Computational and information-theoretic security.}
Some cryptographic primitives require \emph{computational} security,
meaning their security definition should hold against
quantum-polynomial time (QPT) adversaries.
Other primitives should provide security even against computationally unbounded adversaries,
then we say that the primitive provides \emph{information-theoretic} security (ITS).

\subsection{Cryptographic Primitives - Key Encapsulation Mechanisms}
For a fixed security parameter,
a key encapsulation mechanism $\KEM = (\KeyGen, \Encaps, \Decaps)$ 
consists of three algorithms.
Key generation $\KeyGen()$ outputs a keypair $(\pk, \sk)$.
Encapsulation computes
$(c, k) \gets \Encaps(\pk)$,
where $k \in \{0,1\}^\ell$ is called the key,
$\ell$ is called the key length,
and $c$ is called the encapsulation (or sometimes the ciphertext).
Decapsulation computes
$k' := \Decaps(\sk, c)$,
where $k' \in \{0,1\}^\ell \cup \{\bot\}$:
the symbol $\bot \not\in \{0,1\}^\ell$ indicates a decapsulation failure.

\begin{definition}[correctness]
  A \KEM is \emph{$\delta$-correct} if
  \[
    \Pr[\Decaps(\sk, c) \neq k \mid (\pk, \sk) \gets \KeyGen(); \, (c, k) \gets \Encaps(\pk)] \leq \delta
  \]
  where the probability is taken over the random coins of $\KeyGen$ and $\Encaps$.\label{def:KemCorrectness}
\end{definition}

In our proof we will require a bound on the collision probability of honestly generated \KEM values.
\begin{definition}[KEM collision probabilities]\label{def:MuKg}\label{def:MuEnc}\label{def:MuSec}
	We define the public-key collision probability of a KEM as 
	\[
		\muKg := \Pr[\pk=\pk' \mid (\pk, \sk) \gets \KeyGen(), (\pk',\sk') \gets \KeyGen()] \enspace .
	\]
	Additionally, we define the collision probabilities for its ciphertexts and for its encapsulated secrets as 
	\begin{align*}
		\muEnc &:= \Pr[c=c' \mid (\pk, \sk)\gets \KeyGen(),(c,k), \gets \Encaps(\pk), (c', k') \gets \Encaps(\pk)]\enspace , \\
		\muSec &:= \Pr[k=k' \mid (\pk, \sk)\gets \KeyGen(),(c,k), \gets \Encaps(\pk), (c', k') \gets \Encaps(\pk)] \enspace .
	\end{align*}
\end{definition}

\subheading{Standard security notions for KEMs.}
Next, we recall `passive' and `active' security notions for KEMs,
\textbf{IND}istinguishability
under \textbf{C}hosen-\textbf{P}laintext \textbf{A}ttacks (\INDCPA)
and
under \textbf{C}hosen-\textbf{C}iphertext \textbf{A}ttacks (\INDCCA).
In the \INDCPA game, the game prepares a key pair and a `challenge' encapsulation together with the corresponding secret.
It provides the adversary $A$ with the public key, the encapsulation, and either the corresponding secret or a random one.
This choice is formalized via a 'challenge bit' $b$.
$A$ is tasked with guessing which kind of secret it received, by outputting a guessing bit $b'$.
The advantage of $A$ measures how well $A$ does at distinguishing between the two cases.
The \INDCCA game is exactly the same, except that $A$ can additionally request decapsulations of encapsulations of its own choosing (except for the challenge),
by interacting with a decapsulation oracle.

\begin{definition}[\INDCPA, \INDCCA]
	Let \KEM be a key encapsulation mechanism.
	For each bit $b \in \lbrace 0,1\rbrace$, we define an \INDCPA game $\INDCPA^{\KEM, A}_b$ in~\cref{fig:indcpa-cca},
	as well as a game $\INDCCA^{\KEM, A}_b$.
	We define the \INDCPA and the \INDCCA \emph{advantage} of an adversary $A$ against \KEM as
	\begin{align*}
		\Adv_{\KEM}^{\INDCPA}(A) &:= | \Pr[\INDCPA^{\KEM, A}_0 \Rightarrow 1] - \Pr[\INDCPA^{\KEM, A}_1 \Rightarrow 1] | \enspace , \\
		\Adv_{\KEM}^{\INDCCA}(A) &:= | \Pr[\INDCCA^{\KEM, A}_0 \Rightarrow 1] - \Pr[\INDCCA^{\KEM, A}_1 \Rightarrow 1] | \enspace .		
	\end{align*}
    
    \begin{figure}[ht]\begin{center}\fbox{\small
    	\nicoresetlinenr
    	\begin{minipage}[t]{5cm}	
    		\underline{{\bf Game} $\INDCPA^{\KEM, A}_b$}
    		\begin{nicodemus}
    			\item $(\pk, \sk) \gets \KeyGen()$
    			\item $(c, k_0) \gets \Encaps(\pk)$
    			\item $k_1 \uni \{0,1\}^\ell$
    			\item $b' \gets A(\pk, c, k_b)$
    			\item $\pcreturn b'$
    		\end{nicodemus}
    	\end{minipage}
    	\quad
    	
    	\begin{minipage}[t]{5cm}
    		\underline{\textbf{GAME} $\INDCCA^{\KEM, A}_b$}
    		\begin{nicodemus}
    			\item $(\pk, \sk) \gets \KeyGen()$
    			\item $(c, k_0) \gets \Encaps(\pk)$
    			\item $k_1 \uni \{0,1\}^\ell$
    			\item $b' \gets A^{\oracleDecaps(\cdot)}(\pk, c, k_b)$
    			\item $\pcreturn b'$
    		\end{nicodemus}
    		\ 
    		\\
    		\underline{$\oracleDecaps(c')$}
    		\begin{nicodemus}
    			\item $\pcif c' = c: \pcreturn \bot$
    			\item $\pcreturn \Decaps(\sk, c')$
    		\end{nicodemus}
    	\end{minipage}
    }
	\end{center}
    	\caption{\INDCPA and \INDCCA security games for $\KEM = (\KeyGen, \Encaps, \Decaps)$.}
    	\label{fig:indcpa-cca}
    \end{figure}
    
\end{definition}

\subsection{Cryptographic Primitives - Message Authentication Codes}

A message authentication code (MAC) is a triple $\mathcal{M} = (\MKG,\allowbreak \MAC,\allowbreak \MVerify)$,
where $\MKG$ generates a key $k$,
the tag generation algorithm $\MAC$ takes a key $k$ and message $m$ and
computes a tag $\tau \gets \MAC_k(m)$,
and the (deterministic) verification algorithm $\MVerify$
takes a key $k$, message $m$ and tag $\tau$
and outputs a boolean value.
We require \emph{correctness}:
for each $k \gets \MKG()$ and for each $m$, it must hold that
$\MVerify_k(m, \MAC_k(m)) = \true$.

\subheading{Canonical MAC.}
A \MAC{} is called \emph{canonical} if
$\MKG$ generates a uniformly random bitstring,
tag generation (algorithm $\MAC$) is deterministic, and
$\MVerify$ recomputes the tag and compares it against its input.
In this case, we use \MAC interchangeably to refer to both
the triple $\mathcal{M}$ and the tag generation algorithm.
We assume \MAC{}s are canonical,
because in practice most \MAC{}s are canonical,
and canonical \MAC{} give slightly tighter bounds in our security proof.

\subheading{Message integrity.}
We define message integrity via 
\textbf{O}ne-\textbf{T}ime \textbf{s}trong \textbf{E}xistential 
\textbf{U}n\textbf{F}orgeability under \textbf{C}hosen \textbf{M}essage \textbf{A}ttacks (\OTSUFCMA).
In the \OTSUFCMA game, the attacker $A$ is granted to request a tag for a message of their own choosing,
and subsequently tasked with creating a valid message-tag pair that is different from the obtained one.

\begin{definition}[\OTSUFCMA] \label{defn:adv-otsufcma}
	Let $\mathcal{M}$ be a MAC.
	We define the \OTSUFCMA game against $\mathcal{M}$ as in~\cref{fig:otsufcma},
	and the \OTSUFCMA \emph{advantage} of an adversary $A$ against $\mathcal{M}$ as
	\[
		\Adv_{\mathcal{M}}^{\OTSUFCMA}(A) = \Pr[\OTSUFCMA^{\mathcal{M}, A} \Rightarrow 1] \enspace .
	\]
	
	\begin{figure}[ht]
		\nicoresetlinenr
		\centering
		\fbox{\begin{minipage}{.51\textwidth}
				\underline{\textbf{GAME} $\OTSUFCMA^{\mathcal{M}, A}$}
				\begin{nicodemus}
					\item $k \gets \MKG()$
					\item $m \gets A_0()$
					\item $\tau \gets \MAC_k(m)$
					\item $(m', \tau')  \gets A_1(m, \tau)$
					\item $\pcreturn \bool{(m', \tau') \neq (m, \tau) \pcand \MVerify_k(m', \tau')}$
				\end{nicodemus}
		\end{minipage}}
		\caption{
			\OTSUFCMA game for $\mathcal{M} = (\MKG, \MAC, \MVerify)$.
		}\label{fig:otsufcma}
	\end{figure}	
\end{definition}


\subsection{Authenticated Key Exchange}
Authenticated Key Exchange (AKE) is arguably one of the most important cryptographic building blocks in modern security systems.
Most AKE protocols are based on Public-Key Cryptography (PKC),
such as an ad hoc Diffie-Hellman key exchange.
That exchange is then authenticated, usually either via digital signatures, non-interactive key exchange, or (sometimes) public-key encryption or key encapsulation mechanisms~\cite{STOC:BelCanKra98,ACISP:BCNP08,PKC:FSXY12,ASIACCS:FSXY13,ACISP:AlaBoySte14,PKC:HKSU20}.

\subheading{AKE security models.}
Research on AKE has made tremendous progress concerning the development of more solid theoretical foundations,
by specifying the targeted security guarantees via a formal security model.
Notable models are, amongst others, the Bellare-Rogaway (BR) model~\cite{C:BelRog93},
the Canetti-Krawczyk (CK) model~\cite{EC:CanKra01},
the CK$^+$ model~\cite{C:Krawczyk05}, and
the extended Canetti-Krawczyk (eCK) model~\cite{PROVSEC:LaMLauMit07}.

What they have in common is that they formalize the idea that a key computed between two parties should be suitable for the application of symmetric primitives,
by formalizing that the key should be indistinguishable from random to any attacker.
Where they differ is how they actually model the threat, in the sense that different models make different assumptions about the attacker's capabilities
and about what could go wrong during real-world protocol runs.
This is captured by a tailored set of capabilities that is assumed to be held by the attacker.
In more detail, the model formalizes these capabilities by granting the attacker with a set of `query' oracles.
Each such query oracle allows to reveal a certain type of secret values.
These assumptions have been extended considerably by more recent models, with the differences in definition leading to differences that are cryptographically meaningful.

Another way in which the models differ is the concrete formalization.
At times, these can be rather informal descriptions of how attacker queries are handled, which can allow for misinterpretations.
There do, however, also exist works like~\cite{PKC:HKSU20} that
formalize these descriptions via pseudo-code, an approach we also follow in this work.

To handle the complexity of these models,
proofs (such as ours) typically use game-hopping techniques.
A simple but useful observation
is called the difference lemma~\cite{shoup2004sequences}:
given events $A, B, F$
it holds that
if $A \land \lnot F \Leftrightarrow B \land \lnot F$ (i.e. the two events $A \land \lnot F$ and $B \land \lnot F$ are identical),
then $\left| \Pr[A] - \Pr[B] \right| \leq \Pr[F]$.


\subsection{(Quantum) Random Oracle Model and One-way to Hiding}
The standard way of modeling hash functions in modern cryptography is the Random Oracle Model (ROM).
Random oracles are functions to which all algorithms have oracle access,
with the output for every input being chosen uniformly at random from the codomain of the function.

Formally, the random oracle is defined as follows:
a random oracle $H:\mathcal X \to \mathcal Y$ is an empty table of input and output values.
As soon as a value $x\in \mathcal X$ that is not in the table gets queried,
a uniformly random value $y\in \mathcal Y$ gets sampled, $H(x)$ is set to $y$
and the pair $(x,H(x))$ is added to the table.
When a value $x$ that is already in the table gets queried, the value $H(x)$ from
the table is returned.
Using this lazy-sampling approach, any newly queried value is uniformly random, but if a value is queried again,
the output is consistent with earlier queries.

In this paper, we will be working in the quantum-accessible version of the ROM,
which was introduced as the quantum random oracle model (QROM) in~\cite{AC:BDFLSZ11}.
Following~\cite{AC:BDFLSZ11}, we model the quantum random oracle as follows:
\[\mathcal O^H:\sum_{x,y} \alpha_{x,y} \ket{x, y} \mapsto \sum_{x,y} \alpha_{x,y} \ket{x, y \oplus H(x)}\]

The assumptions for the (Q)ROM model are stronger than any actual hash function implementation could ever meet,
so a proof in the (Q)ROM is not as strong as a proof in the standard model.
It can, however, serve as an indicator that the only possible point of failure is the interplay between hash function and algorithm.

In this paper, we will re-purpose the following QROM Lemma~\cite[Lemma 6.2]{Unr15}, called `One-way to Hiding' in the literature.
Intuitively, this lemma states that even a quantum algorithm interacting with a random oracle cannot distinguish an oracle output $H(x)$ from a random co-domain element $y$ (`hiding')
unless it poses a query with sufficient amplitude on $x$ --
in which case $x$ can be measured and a preimage to $H(x)$ has been found (`one-way').
(We made slight changes for the sake of accessibility, changing only syntax and naming.)
\begin{lemma}[One-way to Hiding]\label{lem:ow2h}
	Consider an oracle algorithm $B$ that makes at most $q_H$ many (possibly quantum) queries to a (quantum-accessible) random oracle $H:\{0,1\}^n\to \{0,1\}^m$.
	Let $C$ be an oracle algorithm that on input $x \in \{0,1\}^n$ does the following:
	pick $i \uni \{1,...,q_H\}$ and $y\uni \{0,1\}^m$, run $B^H(x,y)$ until (just before) the $i$-th query,
	measure the argument of the query input register in the computational basis,
	and output the measurement outcome.
	(When $B$ makes less than $i$ queries, $C$ outputs $\bot\notin \{0,1\}^n$.)
	
	Let
	\begin{align*}
		P_{H(x)} &:= \Pr[b'=1 \mid b'\gets B^H(x,H(x))] \enspace ,\\
		P_{y} &:= \Pr[b'=1 \mid	y\uni \{0,1\}^m, \, b'\gets B^H(x,y)] \enspace ,\\
		P_C &:= \Pr[x=x' \mid x'\gets C^H(x)] \enspace ,
	\end{align*}
	where each probability is also taken over the random oracle choice $H\uni (\{0,1\}^n\to \{0,1\}^m)$
	and the input choice $x\uni \{0,1\}^n$.
	Then 
	\[
		|P_{H(x)} - P_{y}| \le 2q_H\sqrt{P_C} \enspace .
	\]
\end{lemma}


\subsection{Quantum Key Distribution}\label{ssec:qkd}
Ever since the publication of~\cite{BEN84}, quantum key distribution (QKD) has promised
to deliver cryptography that is secure under a weaker assumption than any public-key cryptography.
More precisely, QKD is information-theoretically secure,
meaning that any attempt by an attacker to gain information
about the key will be detected with high probability,
regardless of the (quantum) computational power to which the attacker has access to.

Since then, several protocols and variants have been proposed,
and several manufacturers sell commercially available QKD devices
implementing these protocols.
Many QKD schemes have been proven to be theoretically secure,
however the implementation of a QKD scheme could introduce additional vulnerabilities
which are not covered by a theoretical security proof.
These vulnerabilities could include implementation errors, side-channel attacks,
and bad randomness.
Furthermore, the case needs to be considered where the QKD device itself cannot be trusted,
for example because it is acquired from an untrusted vendor,
or because of a supply-chain attack.

Concretely, the information-theoretic security of such protocols can be defined as follows:
give an unbounded QKD adversary access to an oracle that either outputs the key that was generated during the protocol execution
or a uniformly random one.
The scheme is information-theoretically secure if the advantage the computationally unbounded adversary
has with respect to guessing (based on all the quantum and classical information gathered when attacking the protocol) whether it received an honest key or a uniformly random key
is negligible in the security parameter.

The information-theoretic security of QKD ensures that no amount of computation
will let the adversary compromise security.
However, if we use the QKD key in a primitive that is only computationally secure,
then no ITS guarantees can be provided anymore.
Moreover if QKD is used in a computationally secure primitive,
often there exists an alternative solution built only from classical computational cryptography
that can provide the same security guarantees but with lower system complexity and/or with better scalability~\cite{ACD+25b}.
This applies when we use QKD directly for data encryption,
but also when we use QKD in a combined AKE protocol: 
if the combiner itself is only computationally secure,
there might exist a simpler system using only computationally secure primitives that provides the same security guarantees. 
Therefore, it is essential to argue that ITS security of the combined AKE protocols is preserved in case QKD is secure, as we do in \cref{thm:qkd-security}.

\subsubsection{QKD interface (ETSI 014)}\label{sssec:etsi}
Almost all deployed QKD servers deliver keys to their clients via the
Representational State Transfer (REST)-based
Application Programmer Interface (API)
defined in~\cite{etsi014}.\footnote{
  Other methods for key delivery exist,
  such as the ETSI 004 streaming API~\cite{etsi004} or proprietary interfaces,
  but for our purpose of modelling security these behave similarly.
}
The two most important methods are the ``Get key'' method
and the ``Get key with key IDs'' method.
In the former method, the initiating party inputs the identity of their peer
and receives a pair consisting of secret random bytes (the key)
and a corresponding key ID.
Optionally, the requesting party specifies the size and number of keys they require.
In the latter method, the peer inputs the identity of the initiating party and the key ID
and receives the same key, if this key has been established earlier via the ``Get key'' method.
Both methods require that keys are removed from the server upon delivery
(and can therefore only be extracted once per party).
No restrictions are put on the generation of the key ID.
How the initiator communicates the key ID to their peer is out of scope for the standard,
but it is an essential part of any key-exchange protocol,
as exemplified by the \attackname attack described in \cref{fig:mix-and-match}.

A significant part of the ETSI standard deals with multitenancy
(multiple clients can request QKD keys from one server),
but for our purpose we consider only single tenancy.
This simplifies the identity management required when implementing the standard.
In our description, a party corresponds to a secure site in ETSI terminology,
and a party ID corresponds to either the client ID or the server ID.\footnote{
  In ETSI terminology the client would be the \emph{Secure Application Entity}
  and the server would be the \emph{Key Management Entity}.
}

The above interface is reflected in our QKD oracle model, specified in \cref{fig:qkd}.
An important detail to notice is that users of the QKD keys
do not have access to the QKD protocol execution itself,
or even to the transcript that generated the key.
In fact the delivered key could be a concatenation of (parts of) multiple QKD protocol executions.
Only the key and its ID are available.


\section{End-to-end protocol} \label{sec:protocol}
	On a high level, our end-to-end protocol (in \cref{fig:protocol}) combines a post-quantum (or post-quantum-and-classical `hybrid')
triple \KEM handshake with a concurrent QKD key exchange,
in such a way that the output session key is `secure' (indistinguishable from random, see~\cref{sec:security-proof})
even if either the PQC handshake or QKD breaks.
We will show that the output key is \emph{statistically} indistinguishable if the QKD protocol is secure,
i.e., we obtain ITS security against active attackers that are \emph{computationally unbounded},
see~\cref{thm:qkd-security} for the formal statement.

One might suspect that it is straightforward how to combine a \KEM handshake with a QKD key exchange,
e.g., by applying \KEM combiners such as the ones proposed in~\cite{PKC:GiaHeuPoe18, PQCRYPTO:BBFGS19}.
However, QKD is not a \KEM!
In this section, we first analyze difficulties with applying \KEM combiners to our AKE-QKD scenario (see~\cref{ssec:xtm-combiner}) --
in short, most constructions require computational assumptions (and thus violate ITS security),
and the only known construction that avoids this, called XtM, seems to be hard to instantiate.
The reason is that XtM needs a \MAC that meets strong requirements,
and we found that the existing \MAC instantiations do not meet these requirements.

We conclude that so far, we lack the tools to enable an information-theoretic security proof.
We thus construct an alternative protocol in~\cref{ssec:protocol-nested}.
Our alternative is somewhat similar to XtM,
but it avoids the aforementioned gap (by using nested \MAC{}s to authenticate the transcript).
Lastly, in~\cref{sec:mix-and-match}, we argue that the \MAC is in fact cryptographically relevant
because it prevents the severely undesirable situation in which the attacker can enforce sessions with related keys
(which AKE should definitely avoid).

\subsection{Difficulties using KEM combiners}\label{ssec:xtm-combiner}

\KEM{} combiners take multiple `base' \KEM{}s as input and transform them into a new \KEM{}
that should be secure as long as one of its input (base) \KEM{}s is secure.
We will now discuss why the known constructions are not fully suitable for our purposes.

\subheading{No unconditionally \INDCCA secure constructions in~\cite{PKC:GiaHeuPoe18}.}
The \KEM{} combiners introduced by Giacon, Heuer, and Poettering in~\cite{PKC:GiaHeuPoe18}
use some concrete \emph{`core function'} that transforms
the keys and ciphertexts generated by the base \KEM{}s into a single output key.
One example for such a core function would be XOR-ing the base keys.
While this would provably preserve security against \emph{passive} attackers (\INDCPA security),
it does not preserve security against \emph{active} attackers (\INDCCA security), which is the de-facto standard security goal for KEMs.
This problem is reflected in a \emph{Mix-and-Match attack} described by
Bindel, Brendel, Fischlin, Goncalves, and Stebila in~\cite{PQCRYPTO:BBFGS19}:
given a challenge ciphertext $(c_1^*, c_2^*)$,
an adversary $A$ can recover the challenge key $k_1^* \oplus k_2^*$ as follows: 
create two fresh encapsulations $(c_1, k_1) \leftarrow \Encaps_1$ and  $(c_2, k_2) \leftarrow \Encaps_2$,
make the two decapsulation queries $(c_1, c_2^*)$ and $(c_1^*, c_2)$ to obtain $k_1 \oplus k_2^*$ and $k_1^* \oplus k_2$,
use the responses to recover $k_1^*$ and $k_2^*$.
As we will detail in~\cref{sec:mix-and-match}, this attack informed our \attackname attack
on certain types of combined QKD-AKE protocols and thereby our own protocol design.

We note that~\cite{PKC:GiaHeuPoe18} does contain an \INDCCA-preserving combiner,
but it is based on split-key \PRF{}s and thus would introduce a computational assumption.

\subheading{XOR-then-MAC (XtM) combiner.}
In~\cite{PQCRYPTO:BBFGS19}, the authors also provide a combiner called XOR-then-MAC (XtM) combiner.
To the best of our knowledge, this is the only known combiner that does not require computational assumptions to achieve \INDCCA security.
Given two \KEM{}s $\KEM_1$ and $\KEM_2$, the XtM combiner acts as follows:
For each KEM index $1 \leq i \leq 2$, obtain an encapsulation $(c_i, k_{i}) \gets \Encaps_i(\pk_i)$,
and parse $k_{i}$ into two separate keys: $k_{\varwrite{kem}, i} \concat k_{\varwrite{mac}, i} := k_{i}$.
The XtM combiner's final output key is the XOR of the respective keys' first parts: $k := k_{\varwrite{kem},1} \oplus k_{\varwrite{kem},2}$.
To prevent chosen-ciphertext attacks, the XtM combiner additionally computes a tag $\tau$ over both ciphertexts (which is verified during decapsulation):
$\tau := \MAC_{(k_{\varwrite{mac},1}, k_{\varwrite{mac},2})}(c_1 \concat c_2)$. 
The combiner's final ciphertext is $(c_1, c_2, \tau)$.

\subheading{Difficulties in meeting XtM's requirements.}
To enable the proven \INDCCA statement, the combiner requires a \MAC that 
is robust and one-time strong unforgeable.
This is defined with respect to the game $\ROTSUFCMA$ as given in \cref{fig:robust-seufcma}:
it is like the \OTSUFCMA game with a pair of keys
and where the adversary can both choose one of the two keys
and change it between generation and verification of the tag,
capturing that one of the two keys can be broken.
Concerning computational security, the authors discuss a solution based on HMAC.

\begin{figure}\begin{center}\fbox{\begin{minipage}{.5\textwidth}
    \nicoresetlinenr
    \underline{{\bf Game} $\ROTSUFCMA^{\mathcal{M},A}$}
    \begin{nicodemus}
      \item $(k_1, k_2) \gets \MKG()$
      \item $(m^*, b, k_b^*) \gets A_0()$
      \item $\pcif b = 1$
      \item $\quad k^* := (k_1^*, k_2)$
      \item $\pcelse$
      \item $\quad k^* := (k_1, k_2^*)$
      \item $\tau^* \gets \MAC_{k^*}(m^*)$
    \item $(m', \tau', k_b') \gets A_1^{\orwrite{ver}(\cdot)}(\tau^*)$
      \item $\pcif b = 1$
      \item $\quad k' := (k_1', k_2)$
      \item $\pcelse$
      \item $\quad k' := (k_1, k_2')$
      \item $\pcreturn \bool{(m', \tau') \neq (m^*, \tau^*) \land \MVerify_{k'}(m', \tau'))}$
    \end{nicodemus}
  \end{minipage}}\end{center}
  \caption{Security experiment for robust one-time strongly unforgeable MACs, simplified version of Expt$^{\textsf{X}^\textsf{y}\textsf{Z-OT-sEUF}}_\mathcal{M}(\mathcal{A})$~\cite[Figure~7]{EPRINT:BBFGS18}.
    The full version specifies $\orwrite{ver}(\cdot)$:
    a (possibly quantum) verification oracle where the adversary controls one key $k_b$,
    but this oracle is not relevant for our discussion.}\label{fig:robust-seufcma}
\end{figure}

The authors suggest four constructions for instantiating the \MAC, the first three of which are claimed to be unconditionally secure in~\cite{PQCRYPTO:BBFGS19}.
\begin{enumerate}
  \item \textbf{Concatenating} two \MAC{}s, each computed under one of the keys:
    $\tau = (\tau_1, \tau_2) = (\MAC^{(1)}_{k_1}(m),\allowbreak \MAC^{(2)}_{k_2}(m))$
    where $\MAC^{(1)}$ and $\MAC^{(2)}$ are both strongly secure \MAC{}s.
  \item \textbf{XOR aggregation} of the two \MAC{}s:
  	$\tau = \tau_1 \oplus \tau_2$.
  \item \textbf{Carter-Wegman:} When using the Carter-Wegman paradigm~\cite{WegCar81},
    suppose one uses hashing of the form $am + b$ over some finite field,
    where $k = (a, b)$ is the \MAC key.
    Then one can compute a single \MAC over the key $k = k_1 + k_2 = (a_1 + a_2, b_1 + b_2)$,
    so that $\tau = (a_1 + a_2)m + b_1 + b_2$.
  \item \textbf{Computationally:} Use HMAC to instantiate the MAC directly, or rely on the HKDF paradigm~\cite{C:Krawczyk10}.
\end{enumerate}

We believe the first three constructions are not secure with respect to the \ROTSUFCMA game.
This is due to the fact that an adversary who can change one of the keys can change the tag accordingly,
thus being able to compute a different valid tag on the same message.
Formally, we demonstrate the problem by describing $\ROTSUFCMA$-adversaries on each of the respective constructions.
\begin{enumerate} 
  \item The adversary chooses any message $m^*$ and any key $k^*_2$,
    then gets $\tau^* = (\tau_1, \tau_2)$ where $\tau_2 = \MAC^{(2)}_{k^*_2}(m^*))$.
    The adversary chooses any $k'_2 \neq k^*_2$ and computes
    $\tau'_2 = \MAC^{(2)}_{k'_2}(m^*)$. (In the unlikely case that $\tau_2 = \tau_2'$, the adversary can resample $k_2'$.)
    Now $(m^*, \tau', k'_2)$ with $\tau' = (\tau_1, \tau'_2)$ is a successful forgery,
    since the tags differ and $\tau'$ is valid under key $k'$.
  \item Similar to the above, the adversary chooses any $m^*$, any $k^*_2$, and any $k'_2 \neq k^*_2$.
    Since the adversary knows $\tau_2 = \MAC^{(2)}_{k^*_2}(m^*)$ and $\tau'_2 = \MAC^{(2)}_{k'_2}(m^*)$ and is given $\tau^*$,
    they can create a forgery $\tau' = \tau^* \oplus \tau_2 \oplus \tau'_2$ on message $m' = m^*$.
  \item The adversary chooses any $m^* = m'$, any key $k^*_2 = (a_2, b_2)$,
    and chooses $k'_2 = (a_2, b_2')$ such that $b'_2 \neq b_2$.
    When given $\tau^*$, the adversary creates forgery $\tau' = \tau^* - b_2 + b'_2$.
\end{enumerate}
We found no issue with the fourth construction, but it does make a computational assumption.

\subsection{New combined protocol: protocol with nested MACs}\label{ssec:protocol-nested}

\begin{figure}
  \centering
  \begin{tikzpicture}[
    every node/.style={inner ysep=2pt, minimum width=20mm},
    column 1/.style={anchor=base west},
    column 2/.style={anchor=base},
    column 3/.style={anchor=base west},
  ]
    \matrix (m) [matrix of math nodes] {
\text{Alice } (\sk_A, \pk_B) & & \text{Bob } (\sk_B, \pk_A) \\[2mm]
(\cbob,\kbob)\gets \EncapsStat(\pk_B) & & \\
(\pk_e,\sk_e)\gets \KeyGenEph() & & \\
& \cbob, \pk_e & \\
& & \kbob' \gets \DecapsStat(\sk_B, \cbob)\\
& & (\calice, \kalice) \gets \EncapsStat(\pk_A)\\
& & (\ceph, \keph)\gets \EncapsEph(pk_e) \\
& \calice, \ceph & \\
\kalice' \gets \DecapsStat(\sk_A, \calice) & & \\
\keph' \gets \DecapsStat(\sk_e, \ceph) & & \\
\kpqc := \KDF(\kbob, \kalice', \keph') & & \kpqc := \KDF(\kbob', \kalice, \keph) \\
& & (\kid, \kqkd) \gets \enckey(\lenqkd) & & \\
& \kid & \\
\kqkd \gets \deckey(\kid) \\
t = (\cbob, \pk_e, \calice, \ceph, \kid) & & t = (\cbob, \pk_e, \calice, \ceph, \kid) \\
\kqkdm \concat \kqkds := \kqkd & & \kqkdm \concat \kqkds := \kqkd \\
\kpqcm \concat \kpqcs := \kpqc & & \kpqcm \concat \kpqcs := \kpqc \\
& & \tau_1\gets\QKDMAC_{\kqkdm}(t \concat \party_A \concat \party_B)\\
& & \tau_2\gets\PQCMAC_{\kpqcm}(t \concat \tau_1 \concat \party_A \concat \party_B)\\
& \tau_1, \tau_2 & \\
\pcif \QKDMAC_{\kqkdm}(t \concat \party_A \concat \party_B) \neq \tau_1 & & \\
\quad \pcabort & & \\
\pcif \PQCMAC_{\kpqcm}(t \concat \tau_1 \concat \party_A \concat \party_B) \neq \tau_2 & & \\
\quad \pcabort & & \\
\ksess := \kqkds \oplus \kpqcs & & \ksess := \kqkds \oplus \kpqcs \\
    };
    \draw(m-1-1.south west)--(m-1-1.south east);
    \draw(m-1-3.south west)--(m-1-3.south east);
    \draw[->](m-4-2.south west)--(m-4-2.south east);
    \draw[<-](m-8-2.south west)--(m-8-2.south east);
    \draw[<-](m-13-2.south west)--(m-13-2.south east);
    \draw[<-](m-20-2.south west)--(m-20-2.south east);
\end{tikzpicture}
\caption{
  Our new protocol that combines a triple-KEM handshake and QKD to establish a session key $\ksess$.
  Alice and Bob each have a private key (resp. $\sk_A, \sk_B$)
  and know each other's public key (resp. $\pk_A, \pk_B$)
  and party identifier (resp. $\party_A, \party_B$).
  They also share a direct QKD connection,
  accessible via \enckey{} and \deckey{}
  (see \cref{fig:qkd} for a formalization of these, including session IDs).
  Key lengths (including $\lenqkd$) are public system parameters. 
  Bob can combine his two messages into one,
  resulting in an efficient two-message protocol.
}\label{fig:protocol}
\end{figure}
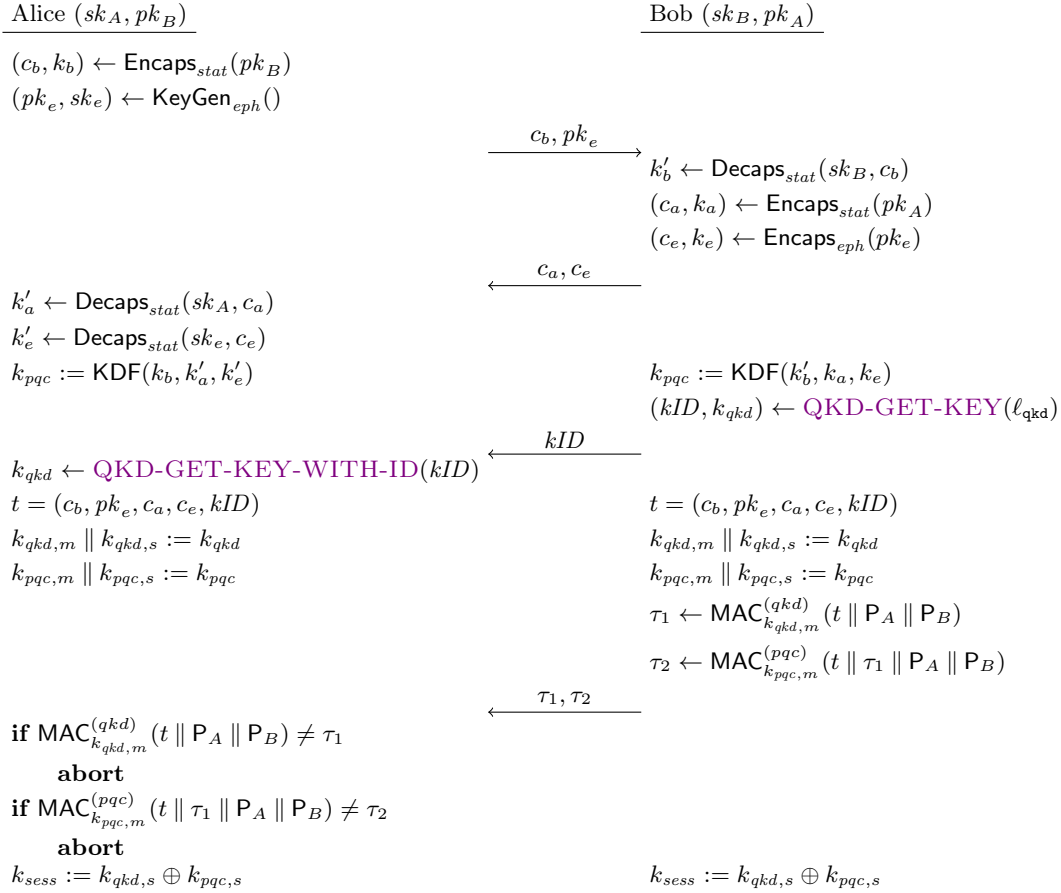

We propose the protocol as shown in \cref{fig:protocol},
which combines QKD with a post-quantum secure triple \KEM AKE.
The triple \KEM protocol can be seen as a simplification of the \textsf{FO}$_{\textsf{AKE}}$ protocol~\cite{PKC:HKSU20}.
Alice and Bob share a QKD link, from which they can get shared QKD key $\kqkd$ using the \enckey{} and \deckey{} methods
(we defer the formalization of these methods to \cref{sec:secModel}).
Triple \KEM takes the encapsulations to three keypairs (two static and one ephemeral)
and combines them using a \KDF to get key $\kpqc$.
The (concatenated) protocol transcript consists of the QKD key ID $\kid$,
the \KEM ciphertexts and the ephemeral \KEM public key.

Bob then splits the QKD key ($\kqkdm \concat \kqkds := \kqkd$)
and the PQC key ($\kpqcm \concat \kpqcs := \kpqc$).
Using the \MAC key $\kqkdm$ he computes a \MAC tag $\tau_1$ over the transcript (and over the party IDs).
Then he uses \MAC key $\kpqcm$ to compute a \MAC tag $\tau_2$ over the transcript \emph{and over $\tau_1$} (and over the party IDs).
He outputs session key $\ksess = \kqkd \oplus \kpqc$.
Alice splits her keys the same way
and if she can verify the tags, she will output the same session key.

\subheading{Forward secrecy.} 
The proof of security for this protocol is deferred to \cref{sec:security-proof},
after we introduced the security model.
We already note, however, that it is known~\cite{C:Krawczyk05} that pure two-message AKE protocols -- like the KEM handshake used in our protocol to establish \kpqc{} --
cannot possibly achieve forward secrecy against active attackers.
That is, if an attacker actively replaces one of the parties and later also steals the long-term key of the party they replaced,
no security could possibly be guaranteed.
(In our case, the attacker could get \kpqc\ as follows: create $(\cbob, \pk_e)$, knowing $\kbob$ and $\sk_e$,
steal $\sk_A$ to get $\kalice$, and use $\sk_e$ to get $\keph$.)
Our combined protocol bypasses this result if QKD is secure and the attacker cannot access \kqkd, see~\cref{thm:qkd-security}.
In the case that QKD is insecure, we still achieve weak forward secrecy:
honestly executed sessions for which the long-term keys leak only later should still remain secure.

\subsection{Motivation of our protocol: \attackname attack}\label{sec:mix-and-match}

We will now argue that the \MAC tags -- which cryptographically bind the two handshakes -- 
are in fact cryptographically relevant,
in the sense that they are necessary to prevent the severely undesirable situation in which attackers can establish sessions with keys that are related to one another.
(We will capture this below via a class of attacks we will call \emph{\attackname} attacks.)
Crucially, AKE guarantees that an established session key remains unpredictable even if some other session keys leaked --
session keys often end up in memory or logs, and AKE should prevent that leakage of one key endangers others,
which we now show is not satisfied if the \MAC tags are omitted.

\subheading{Attack scenario.} In more detail, the requirement of session key unpredictability even under leakage of session keys
was already formalized in the first formal AKE security model~\cite{C:BelRog93}.
It was modeled by allowing the attacker (Mallory) to perform `\orReveal' queries that leak session keys of their choosing.
This has two `trivial' exceptions: (1) the session key itself cannot be leaked directly,
and (2) if Mallory attacks one of Alice's session keys,
and they learns that key by forcing Bob to leak it on his side, then no security could possibly still be guaranteed.
Mallory is thus forbidden from performing a \orReveal query on the `matching' session that happens on Bob's side.

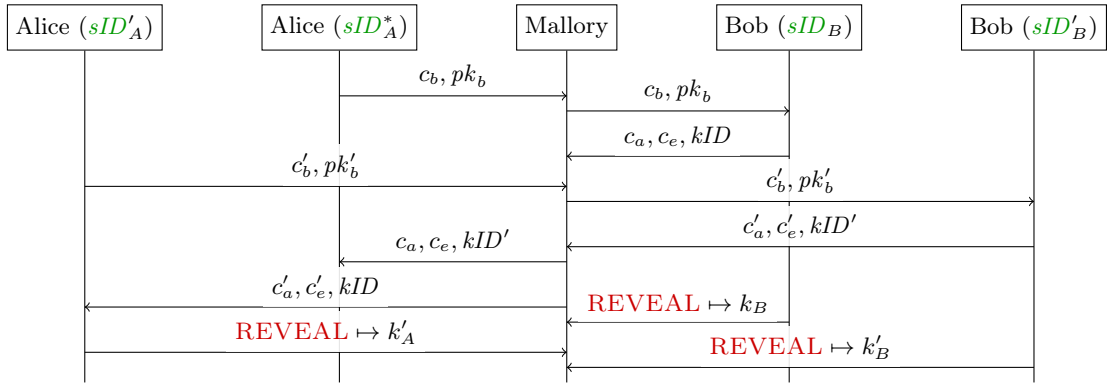
\begin{figure}
  \begin{center}
  \begin{tikzpicture}[node distance=6mm and 13mm]
    \pgfdeclarelayer{vertical}
    \pgfsetlayers{vertical, main}

    \node (ap) [draw=black, minimum height=6mm] {Alice ($\sid_A'$)};
    \node (a)  [draw=black, minimum height=6mm, right=of ap] {Alice ($\sid^*_A$)};
    \node (m)  [draw=black, minimum height=6mm, right=of a] {Mallory};
    \node (d)  [draw=black, minimum height=6mm, right=of m] {Bob ($\sid_B$)};
    \node (dp) [draw=black, minimum height=6mm, right=of d] {Bob ($\sid_B'$)};

    \coordinate[below=of a] (a1);
    \coordinate (m1) at (m |- a1);
    \draw[->] (a1) -- node [msg] {$\cbob, \pk_b$} (m1);

    \coordinate[below=2mm of m1] (m2);
    \coordinate (d2) at (d |- m2);
    \draw[->] (m2) -- node [msg] {$\cbob, \pk_b$} (d2);

    \coordinate[below=of m2] (m3);
    \coordinate (d3) at (d |- m3);
    \draw[<-] (m3) -- node [msg] {$\calice, \ceph, \kid$} (d3);

    \coordinate[below=4mm of m3] (m4);
    \coordinate (a4) at (ap |- m4);
    \draw[->] (a4) -- node [msg] {$\cbob', \pk_b'$} (m4);

    \coordinate[below=2mm of m4] (m5);
    \coordinate (d5) at (dp |- m5);
    \draw[->] (m5) -- node [msg] {$\cbob', \pk_b'$} (d5);

    \coordinate[below=of m5] (m6);
    \coordinate (d6) at (dp |- m6);
    \draw[<-] (m6) -- node [msg] {$\calice', \ceph', \kid'$} (d6);

    \coordinate[below=2mm of m6] (m7);
    \coordinate (a7) at (a |- m7);
    \draw[<-] (a7) -- node [msg] {$\calice, \ceph, \kid'$} (m7);

    \coordinate[below=of m7] (m8);
    \coordinate (a8) at (ap |- m8);
    \draw[<-] (a8) -- node [msg] {$\calice', \ceph', \kid$} (m8);

    \coordinate[below=2mm of m8] (m9);
    \coordinate (d9) at (d |- m9);
    \draw[<-] (m9) -- node [msg] {$\orReveal \mapsto k_B$} (d9);

    \coordinate[below=4mm of m9] (m10);
    \coordinate (a10) at (ap |- m10);
    \draw[->] (a10) -- node [msg] {$\orReveal \mapsto k_A'$} (m10);

    \coordinate[below=2mm of m10] (m11);
    \coordinate (d11) at (dp |- m11);
    \draw[<-] (m11) -- node [msg] {$\orReveal \mapsto k_B'$} (d11);

    \coordinate[below=2mm of m11] (end);

    \begin{pgfonlayer}{vertical}
      \foreach \ent in {ap, a, m, d, dp} {
        \draw (\ent) -- (\ent |- end);
      }
    \end{pgfonlayer}
  \end{tikzpicture}
  \end{center}
  \caption{\attackname attack \emph{on the protocol without \MAC{}s}.
    Mallory sits on the wire, aiming to attack test session $\sid^*_A$.
    She modifies the two session's messages from Bob to Alice on transit, swapping the key IDs that were communicated by Bob.
    She is allowed to request leakage (`\orReveal') from the (non-matching) sessions $\protect\sid_A'$, $\protect\sid_B$ and $\protect\sid_B'$,
    thereby learning the session keys $k_A' = \kpqc' \oplus \kqkd'$,
    $k_B = \kpqc \oplus \kqkd'$,
    and $k_B' = \kpqc' \oplus \kqkd$,
    respectively.
    She can now easily compute the to-be-attacked session key
    $k^*_A = \kpqc \oplus \kqkd = k_A' \oplus k_B \oplus k_B'$.}%
  \label{fig:mix-and-match}
\end{figure}

\subheading{Attack on protocol without MACs.}
Now consider the same protocol, but without \MAC tags.
In other words, the construction essentially becomes the XOR combiner, applied to two key exchanges:
the session key $\ksess = \kpqc \oplus \kqkd$ is output
immediately after Bob sent/Alice received Bob's ciphertext message $(\calice, \ceph)$ and the key ID.
In~\cref{fig:mix-and-match}, we depict a \attackname attack on that protocol:
The adversary, Mallory, sits on the wire between Alice and Bob, modifying Bob's outgoing messages on transit.
Concretely, Mallory swaps the two key IDs Bob intended to use for the two sessions,
thereby decoupling Alice's input messages from Bob's output messages.
Effectively, the first session on Alice's side is thus no longer  `matched' to any session on Bob's side.
As a result, \emph{all} session keys are fair game to be leaked, except Alice's to-be-attacked key itself.
Mallory is thus allowed to learn Alice's other session key as well as both of Bob's session keys.
Suitably XORing these three keys allows them to learn Alice's to-be-attacked key, thereby completely breaking security.

Note that there exist many variants of the attack if Mallory should happen to exhibit additional powers --
for example, if Mallory manages to
learn Alice's long term key, then they can run session $\sid_B'$ directly with Bob himself and so they do not need session $\sid_A'$ to learn $k_A'$.
Similarly if Mallory breaks PQC, then they do not need $\sid_B$ to learn $\kpqc$,
and if they break QKD then they don't need $\sid_B'$ to learn $\kqkd$.

\subheading{Preventing the attack via MAC tags.}
The exchange of MACs in our protocol (see~\cref{fig:protocol}) prevents this kind of \attackname attacks --
swapping the key IDs would now lead to differing transcripts on the two sides due to the differing tags,
and thus, Alice would reject.
We conclude that it is crucial to include the QKD key ID into the transcripts,
thereby binding the used QKD key to the AKE session.
If key IDs should not be available,
e.g. due to QKD keys being delivered via some other interface,
then the binding should occur via some other associated value.
This could be the session ID or post-processing transcript associated to
the QKD session that generated the used key.


\section{QKD Oracles for End-to-end security} \label{sec:secModel}
	Following the game-based literature on Authenticated Key Exchange (see, e.g., \cite{C:BelRog93, EC:CanKra01, EC:CanKra02, ACISP:BCNP08, C:Krawczyk05, PROVSEC:LaMLauMit07, ACISP:BCNP08, PKC:FSXY12, PKC:HKSU20}), we model 
\begin{itemize}
	\item security as key indistinguishability, via a game in which the attacker
    chooses a `test' session whose key \sessionkey it aims to distinguish from random. To that end, the game provides an oracle \orTest{} that the attacker can query for a session of its own choosing, and that returns either that session's key \sessionkey or a uniformly random key. The attacker finishes by outputting a `guessing bit' to indicate whether it believes the key to be real or random.
	\item  adversarial interactions with (potentially many) honest parties by providing the attacker with black-box access to various `Send' oracles that execute the protocol for parties of the attacker's choosing. This choice reflects that the attacker is granted full control over the network -- they could forward honest parties' output faithfully as well as modify the data in transit.
  While we could technically do with a single Send query,
  for clarity we make the queries explicit for the individual messages of the protocol. 
  \item adversarial interactions to extract secret data from the model,
  by which we model various attacks that may leak information to the adversary.
  Distinguishing the session key from random becomes easy when too much secret information has been revealed,
for example when the adversary directly learned the test session key through a \orReveal{} query.
  In that case no protocol can be expected to provide any security,
  so we classify such attacks as \emph{trivial}
  and we define security to mean that all non-trivial attacks succeed with at most negligible probability.
\end{itemize}

We detail the setup of the game's execution environment in \cref{subsec:ExecutionEnvironment} below,
after highlighting some key aspects of our model.

\heading{Differences to previous QKD-related models.}
A major difference to the HAKE model is that we model the QKD functionality in a different way---unlike works in the HAKE framework~\cite{PQCRYPTO:DowHanPat20,MucklePlus,GPH+24,BSP+24}, we model the QKD link layer with an ideal functionality that can be accessed via oracle calls, see \cref{fig:qkd}.
With this model, we provide the adversary with explicit methods to both learn \emph{and modify} QKD key material, reflecting the capabilities of an adversary that can compromise QKD.
We explicitly model how key identifiers ($\kid$s) are used when interacting with real-world QKD devices~\cite{etsi014}.
Unlike previous models, this allows us to capture \attackname attacks in our model,
which can lead to loss of security as we demonstrate in \cref{sec:mix-and-match}.

\heading{Capturing weaknesses via oracles.}
Given that we want to investigate hybrid QKD-PKC protocols and to express security even if one of their components fail, our model will provide the attacker with oracles that reveal secret process data (see \cref{subsec:AdvInteraction}).
We briefly summarize how typical attacks/weaknesses are captured by this model:
\begin{enumerate}
	\item Man-in-the-middle attacks on end-to-end messages are reflected in the Send oracles.
  \item Weak forward secrecy is captured through the \orCorrupt{} query:
    even if the adversary reveals both long-term keys,
    the key of an honestly executed test session should be indistinguishable from random.
  \item Malicious parties are captured via the \orCorrupt{} query:
    after corruption, the adversary can impersonate (the PQC part of) the corrupted party.
  \item Key Compromise Impersonation (KCI) is captured via the \orCorrupt{} query:
    an adversary that corrupts a party can impersonate another party \emph{to} the corrupted party,
    but then the test session key should still be indistinguishable from random.
  \item Maximal EXposure (MEX) is captured via \orCorrupt{} and \orRevState{} queries:
    when the test session contains
    at least an uncompromised long-term key or a non-revealed ephemeral key
    for both the initiator and responder,
    then the test session key should still be indistinguishable from random.
  \item Broken QKD implementations are modelled by allowing adversarial access to the idealized QKD functionality,
    allowing both read-outs via \orQKDGet{} and overwrites via \orQKDSet{}.
  \item \attackname attacks are captured through the \orReveal{} query:
    session keys of other sessions (except for a matching session) can be requested to be revealed,
    and even then, the test session key should still be indistinguishable from random.
\end{enumerate}

\subsection{Execution environment}\label{subsec:ExecutionEnvironment}

The adversary drives the execution of the protocol.
After generation of the long-term public keys and initialization of the QKD oracle,
the adversary can interact with the model via oracle queries.
Each session is uniquely identified by a session identifier $\sid$.
To ensure each session has a unique $\sid$,
a global counter $\sidctr$ is incremented whenever a new session is established
via the \orEst{} query.
In our model the $\sid$ is merely a unique identifier for disambiguating sessions,
and it is not accessible by honest parties,
unlike some other models where a (partial) local transcript represents the session ID.

\heading{Standard game variables.} 
We consider a network of \numberParties many nodes $\party_1, \dots , \party_\numberParties$ that possibly have many sessions at once.
Party IDs are represented by an index in $\{1, \dots \numberParties\}$.
Each session runs the protocol with access to the involved parties' pre-established QKD/PKC material,
while also having its own set of session-specific local variables:
\begin{itemize}
	
	\item An integer $\owner[\sid] \in \{ 1, \dots, \numberParties \}$ that points to the party running the session.
  \item An integer $\peer[\sid] \in \{ 1, \dots, \numberParties \}$ that points to the session's intended communication partner.
	
  \item A string $\role[\sid] \in \{ \roleInit, \roleResp \}$
  indicating the party's role in the current session.
	
  \item A string $\sent[\sid]$, containing the message sent by the session.\footnote{\label{footnotesent}When a protocol has more messages, this is a list of strings.}

  \item A string $\received[\sid]$, containing the message received by the session.\textsuperscript{\labelcref{footnotesent}}

  \item A variable $\state[\sid]$, holding the state that will be revealed to the adversary.
    When a session accepts, this value is overwritten with the value \Accept, reflecting secure erasure. When a session aborts, this value is overwritten with the value \Reject.

  \item A variable $\sessionkeyArray[\sid]$, holding the session key (if the session is accepted), initialized at $\bot$.

\end{itemize}

\heading{Adversarial interaction}
To drive the protocol, the adversary interacts with the model via the following
queries.

\begin{itemize}
  \item $\orEst(i, j, \rho)$ establishes a new session $\sid$ with $\owner[\sid] = i$, $\peer[\sid] = j$, and $\role[\sid]=\rho$.
    The unique $\sid$ for this session is returned,
    which is generated by incrementing a global counter $\sidctr$.
  \item $\orSend(\sid, m)$ delivers a message $m$ to session $\sid$.
    The long-term IDs and keys of owner and peer are retrieved,
    as well as $\state[\sid]$. 
    The message $m$ is recorded in $\received[\sid]$,
    and then the protocol subroutine is called with the retrieved values,
    which returns an outgoing message $m'$ and an updated state $s'$.
    The message $m'$ is recorded in $\sent[\sid]$,
    and $\state[\sid]$ is overwritten with $s'$,
    then $m'$ is returned to the adversary.
    If the protocol accepts, the state is erased (overwritten with $\Accept$)
    and $\sessionkeyArray[\sid]$ is set with the computed \sessionkey,
    and the session is \emph{completed}.
    If the protocol \pcabort{}s, then the state is erased (overwritten with $\Reject$)
    and no message is returned.
    Messages and state can be empty ($\bot$) for the first or last step of the protocol.
  \item $\orTest(b, \sid)$ lets the adversary get a real-or-random key,
    depending on the game parameter $b$.
    The first call to \orTest{} defines the test session $\sid^*$,
    subsequent calls do nothing
    (the model enforces this by checking if $\sid^*$ was already set).
    If $b=0$ the adversary gets the real session key $\sessionkeyArray[\sid^*]$,
    otherwise (if $b=1$) the adversary gets a uniformly random bitstring.
    The goal for the adversary is to make a guess $b'$ for the value $b$:
    if $b' = b$ the adversary wins the game.
\end{itemize}

The above should be familiar to those readers acquainted with the $CK^{+}$ model.

\heading{Modelling QKD.}
We model QKD with an ideal functionality that can be accessed by honest users via (classical) oracle calls \enckey{} and \deckey{}.
In addition we provide the adversary with the \orQKDGet{} query to reveal QKD keys
and \orQKDSet{} to modify the keys.

The QKD oracle, described in detail in \cref{fig:qkd},
keeps track of some bookkeeping variables.

\begin{itemize}
  \item The global counter $\kidctr$ provides a unique key ID per key
    by incrementing the counter for every new key that is requested.
    Real QKD devices may instead combine a local counter with a unique device ID,
    or deliver randomly generated key IDs from a sufficiently large space:
    as long as the probability of a key ID collision is negligible, our model applies.
  \item the array $\key$ stores the secret bytes per key ID;
  \item the array $\flag$ stores per key ID if the key is unknown to the adversary ($\honest$),
    leaked to the adversary $(\leaked$), or even set by the adversary $(\corrupt)$.
  \item the array $\qsent$ stores per key ID which $\sid$ received the key via \enckey.
  \item the array $\qrecv$ stores per key ID a list of $\sid$s,
    to track which sessions received the key via \deckey.
    While \honest{} or \leaked{} keys will only be delivered once via this method,
    the oracle may deliver \corrupt{} keys more than once.
\end{itemize}
The oracle is initialized with \qkdInit,
which sets $\kidctr$ to zero, and initializes the arrays to the empty value.

An honest session $\sid$ gets access to the oracles $\enckey(\sid, \cdot)$ and
\deckey{}$(\sid, \cdot)$: 
\begin{itemize}
  \item $\enckey(\sid, \ell)$
    returns a pair of a unique key identifier $\kid$ and a uniformly random key $k$ of length $\ell$.
    The oracle generates a unique $\kid$ by incrementing $\kidctr$
    and records the calling $\sid$ in the $\qsent$ array.
    It then generates a new key and marks it \honest,
    but only if $\key[\kid]$ was not set before
    (which can only occur for a \corrupt{} key).
    Keys are always available via this query,
    therefore our model does not capture denial-of-service attacks on QKD.
  \item $\deckey(\sid, \kid)$
    returns $\key[\kid]$, if the QKD keys is indeed shared between the communicating parties,
    and then erases the stored value.
    To verify that the key was indeed shared between the relevant parties,
    the \owner and \peer of the calling session $\sid$
    are compared against the \peer and \owner of the session that was recorded in $\qsent[\kid]$.
    By binding the QKD keys to \owner/\peer arrays of the AKE model, we enforce end-to-end QKD.
    If a non-empty key is delivered,
    then the calling $\sid$ is appended to $\qrecv[\kid]$.
\end{itemize}

\nicoresetlinenr{}
\begin{figure}
  \fbox{\begin{varwidth}{\textwidth-2\fboxsep-2\fboxrule+\columnsep}\begin{multicols}{2}%
      \underline{\qkdInit()}
        \begin{nicodemus}
        \item $\kidctr := 0$
        \item $\key :=$ [\ ]
        \item $\flag :=$ [\ ]
        \item $\qsent :=$ [\ ]
        \item $\qrecv :=$ [\ ]
        \end{nicodemus}

      \vspace{1em}
        \underline{$\orQKDSids(\kid)$}
				\begin{nicodemus}
        \item $\pcreturn (\qsent[\kid], \qrecv[\kid])$
				\end{nicodemus}

      \vspace{1em}
        \underline{$\orQKDGet(\kid)$}
				\begin{nicodemus}
          \item \pcif $\flag[\kid] = \honest$
          \item \quad $\flag[\kid] := \leaked$
          \item \pcreturn $\key[\kid]$
				\end{nicodemus}

      \vspace{1em}
        \underline{$\orQKDSet(\kid, k)$}
				\begin{nicodemus}
          \item $\key[\kid] := k$
          \item $\flag[\kid] := \corrupt$
				\end{nicodemus}

      \columnbreak
      \underline{$\enckey(\sid, \ell)$}
				\begin{nicodemus}
        \item $\kidctr := \kidctr + 1$
        \item $\qsent[\kidctr] := \sid$
        \item \pcif $\key[\kidctr] = \bot$
        \item \quad $\key[\kidctr] \uni \{0,1\}^{\ell}$
        \item \quad $\flag[\kidctr] := \honest$
        \item $k := \key[\kidctr]$
        \item \pcreturn $(\kidctr, k)$
				\end{nicodemus}

      \vspace{1em}
        \underline{$\deckey(\sid, \kid)$}
				\begin{nicodemus}
          \item $\sidqsend := \qsent[\kid]$
          \item $\pcif (\owner[\sidqsend], \peer[\sidqsend]) \neq $

            $\hspace{\fill}(\peer[\sid], \owner[\sid])$
          \item \quad \pcreturn $\bot$
          \item $\pcif \key[\kid] = \bot$
          \item $\quad \pcreturn \bot$
          \item $\qrecv[\kidctr] \mathrel{+}= [\sid]$
          \item $k := \key[\kid]$
          \item $\key[\kid] := \bot$ \label[line]{line:eraseKey}
          \item \pcreturn $k$
				\end{nicodemus}

  \end{multicols}\end{varwidth}}
  \caption{QKD oracle model.
    Honest sessions have access to the $\enckey(\sid, \cdot)$ and $\deckey(\sid,\cdot)$ oracles (in {\color{qkdoraclecolor} purple}),
    while the adversary may query \orQKDSids, \orQKDGet, and \orQKDSet\ (in {\color{oraclecolor} red}).
    The party calling $\enckey$ specifies the key length $\ell$.
    $\keyID$ is a (predictable) key identifier.
    Besides the functionality, we add some {\color{bookkeepcolor} bookkeeping values (in green)} that are used in the proofs,
    but are not available to either the honest user or the adversary.
    Keys are removed in \cref{line:eraseKey} from the oracle when they are requested by the receiver (the party calling \deckey). 
  }\label{fig:qkd}
\end{figure}

We remark that \honest{} keys are uniformly random in the QKD oracle,
while real QKD keys have some negligibly small (but non-zero) statistical distance ($\varepsilon$)
to uniformly random keys.
We are already abstracting away many details of QKD,
and we expect that this additional simplification does not significantly impact the security of hybrid protocols.
For that reason we leave the proper modelling of $\varepsilon$-secure QKD as an open problem for future work.

Modelling QKD directly, including quantum and classical communication~\cite{PQCRYPTO:MosSteUst13},
provides stronger security guarantees than modelling QKD 
via an oracle as we do in this work.
Integration into an AKE protocol could then use the post-processing transcript
as an input to prevent attacks such as the \attackname attack.
However, that would also requires that the AKE protocol has access to (the transcript of) QKD post-processing.
Instead the typical ETSI 014 interface (see \cref{sssec:etsi}) provides only a key and corresponding ID.

\subsection{Modeling adversarial interaction} \label{subsec:AdvInteraction}

\heading{Secret-revealing oracles.} To account for the potential involvement of public-key algorithms, we assume---like in previous work on AKE---that public-key pairs, if such are used, have been distributed a priori. Given that we want our model to encompass hybrid QKD-PKC protocols, we want to be able to express security even upon failure of one of their components. We will model such failures by leakage of secret values like the involved parties' long-term secret keys, or secret state information, which the attacker obtains via access to additional oracles:
\begin{itemize}
  \item $\orReveal(\sid)$ leaks the (real) session key $\sessionkeyArray[\sid]$ to the adversary.
    The model sets the flag $\revealed[\sid]$ to record that this session key has been leaked.
\item $\orCorrupt(i)$ leaks the long-term secret key $\sk_i$ belonging to party $i$ to the adversary.
  The model sets the flag $\corrupted[\sid]$ to record that this party has been corrupted.
  The adversary can still interact with corrupted parties (for example via \orSend{} queries).
\item $\orRevState(\sid)$ leaks the state $\state[\sid]$ to the adversary.
  The model sets the flag $\revealedState[\sid]$ to record that the state of this session has been leaked.
\end{itemize} 

Malicious interaction with the QKD model is modelled via the following adversary oracle queries:
\begin{itemize}
  \item $\orQKDGet(\kid)$ leaks $\key[\kid]$ to the adversary.
    If the status was not already set (to \corrupt),
    the model sets $\flag[\kid] := \leaked$ to record that the QKD key has been leaked,
  \item $\orQKDSet(\kid, k)$ allows the adversary to overwrite $\key[\kid] := k$,
    The model sets $\flag[\kid] := \corrupt$ to record that the QKD key was set by the adversary.
\item $\orQKDSids(\kid)$ reveals to the adversary which sessions used the QKD key identified by $\kid$.
  This does not correspond any attack,
  but rather reflects that we consider this information to be public.
\end{itemize}

The $\orQKDGet$ query corresponds to passive attacks on QKD
where the adversary learns the honestly generated key.
For example, this could reflect side-channel leakage from an honest execution of QKD.
The $\orQKDSet$ query corresponds to active attacks on QKD,
where the adversary has some influence over the output that honest parties receive from the QKD device.
For example, this could reflect implementation errors, bad randomness, supply chain attacks,
or simply the fact that one of the executing QKD parties was corrupted.
Some of these attacks may only provide the adversary with some malleability of the output,
but letting the adversary choose these values freely only increases their power.

\subsection{Session cleanness and security definition}

If we would let the adversary call the above oracles without restrictions,
then no protocol can prevent the adversary from guessing $b$.
For example, revealing the test session key directly with $\orReveal(\sid^*)$
allows the adversary to compare directly against the output from the \orTest{} query,
thereby revealing the value of $b$.
An attack that would work against \emph{any} protocol is considered a trivial attack,
which we formalize using predicates, encoded as subroutines in \cref{fig:formal_game_trivial}.
The subroutine $\TrivialPQC$ defines what constitutes a trivial attack against the post-quantum cryptography in a protocol:
when the adversary executes an attack such that $\TrivialPQC$ does \emph{not} hold,
then the protocol should be secure as long as the post-quantum primitives that it uses are secure.
The subroutine $\TrivialQKD$ defines what are considered (trivial) attacks against QKD:
when $\TrivialQKD$ does not hold,
a protocol should be secure if the used QKD key and other primitives used in the protocol are secure.
The notions are independent, such that a protocol should get security from both PQC and QKD
against attacks that are neither $\TrivialPQC$ nor $\TrivialQKD$.
A session that is not trivially broken is also called \emph{clean}:
we only require security for a clean test session.
To define cleanliness, we first require a precise definition of matching sessions.

\heading{Matching sessions.}
Two sessions $\sid_1$ and $\sid_2$ are \emph{matching} if
\begin{itemize}
  \item $(\owner[\sid_1], \peer[\sid_1]) = (\peer[\sid_2], \owner[\sid_2])$: they intend to communicate with each other,
  \item $(\sent[\sid_1], \received[\sid_1]) = (\received[\sid_2], \sent[\sid_2])$: they have matching transcripts, and
  \item $\role[\sid_1] \neq \role[\sid_2]$: they have different roles.
\end{itemize}
We are only concerned with sessions that match with the test session.
The subroutine $\FindMatches(\sid^*)$ (defined in \cref{fig:formal_game_trivial}) creates
a set $\matchingSessions$ containing all sessions that match with $\sid^*$.

\heading{\TrivialPQC.}
The $\TrivialPQC$ notion is taken from~\cite{PKC:HKSU20}, with the exception that they inline $\FindMatches$.
$\TrivialPQC$ excludes certain combinations of queries because no PQC protocol can protect against them,
as indicated by the attacks below.
\begin{enumerate}
  \item If the test session did not accept, then
    the session key $\bot$ is trivially distinguished from random.
  \item If the test session was revealed, then
    the session key is known to the adversary and easily distinguished from random.
  \item If the long-term key and state of the test session were all leaked, then
    all secret values used to compute the test session key are known to the adversary.
  \item If there is no matching session and the peer has been corrupted,
    then the test session may have been executed directly with the adversary.
  \item If there is a matching session, that session may not be revealed.
  \item If there is a matching session, that session needs to have at least one secret value
    (long-term key or state) so the adversary cannot recompute the session key.
\end{enumerate}
In addition to the above list,
if the adversary manages to create more than one session that matches with $\sid^*$,
then the attack is considered non-trivial.
Note that this holds even if the adversary reveals (one of) those matching sessions,
so effectively an adversary that can create multiple matching sessions can always win the game.
This check formalizes a requirement of the CK$^{+}$ model~\cite{C:Krawczyk05},
which only considers protocols where the initial message is unique,
while allowing us to quantify the security loss induced by lack of randomness in the protocol.
Intuitively this check ensures that in a secure protocol the adversary cannot create multiple sessions that output the same key,
which might lead to confusion attacks, for example if the adversary swaps different messages encrypted under the same key.
The above is formally encoded by subroutine $\TrivialPQC$ in \cref{fig:formal_game_trivial}.

\heading{\TrivialQKD.}
We can only expect security from QKD if the test session used at least one $\honest$ QKD key.
Formally, we require that there exists a $\kid$ such that
\begin{itemize}
  \item $\flag[\kid] = \honest$, and
  \item $\sid^* = \qsent[\kid]$ or $\sid^* \in \qrecv[\kid]$.
\end{itemize}
We cannot guarantee security if the adversary revealed the test session or the matching session.
However, we also want to reward the adversary if they can create multiple \emph{completed} matching sessions,
but only if the test session used an honest QKD key
(otherwise a trivial adversary against QKD exists that can create multiple accepting sessions
by setting the QKD key with \orQKDSet and delivering the key to multiple recipients).

The above is encoded by the subroutine \TrivialQKD{} in \cref{fig:formal_game_trivial}.
Note that if $\flag[\kid] = \honest$,
then $\sid^* \in \qrecv[\kid]$ is equivalent to $[\sid^*] = \qrecv[\kid]$,
because $\qrecv[\kid]$ has length at most one,
since \honest{} keys are deleted after they are delivered once.
Also note that if the adversary manages to create multiple accepting matching sessions
(without breaking all QKD keys used),
then they are allowed to reveal those sessions and can therefore win the game.

In addition to the above, we also require that the test session
accepted,
it was not revealed
and if there is one matching session
then that can also not be revealed.

\heading{\Attack.}
Our protocol is vulnerable to an attack:
the adversary can impersonate the responder without compromising the responder long-term key,
instead they reveal the initiator state to reveal $\kbob$.
If the adversary also breaks QKD, they can complete the responder session themselves.
This is a known attack on the classical protocol (without QKD)~\cite{PKC:HKSU20},
and it is not considered a trivial attack
because there exist protocols that do protect against this scenario.
Our work aims to integrate QKD into an existing AKE protocol,
not to fix the classical protocol.
Note that QKD does help:
the adversary must break QKD to exploit this attack.
This attack is formally encoded by the subroutine $\Attack$ in \cref{fig:formal_game_trivial}.

\heading{Formal definition.}
\Cref{fig:formal_game} ties everything together and formalizes the above model.
The three games $\INDAAPQC$, $\INDStAAPQC$, and $\INDAAQKD$
only differ by which cleanliness/trivial predicate they use.
The games are defined relative to a challenge bit $b$,
indicating if the adversary will receive a real or a random key.
The challenger generates long-term public keys for each party
and initializes the QKD oracle,
then adversary receives the public keys and gets to interact with the parties
and their secrets via the earlier defined oracles.
The adversary then makes a guess $b'$ for the challenge bit,
which the game outputs if the adversary is considered non-trivial
(and otherwise the game outputs 0).
The adversary is said to \emph{win} the game if the game output equals $b$.

\nicoresetlinenr
\begin{figure}
  \fbox{\begin{minipage}{\dimexpr\textwidth-2\fboxsep-2\fboxrule\relax}\begin{multicols}{2}%
    {\color{gray}
      \underline{{\bf GAME} $\INDAAPQC_b(\Pi, A)$}
    \begin{nicodemus}
    \item $\varwrite{sID}_\varwrite{ctr} := 0$
    \item $\varwrite{sID}^* := 0$
      \item $\pcfor i \in \set{1, \dots, \numberParties}$
      \item $\quad (\pk_i, \sk_i) \gets \KeyGenStat()$
      \item $\algowrite{QkdInit}()$
      \item $b' \gets A^{\mathcal{O}}(\pk_1, \dots, \pk_\numberParties)$
      \item $\varwrite{matches} := \algowrite{FindMatches}(\varwrite{sID}^*)$
      \item $\pcif \algowrite{Trivial}(\varwrite{matches})$ \label[line]{line:calltrivial}
      \item $\quad \pcreturn 0$ \gcom{ adversary loses }
      \item $\pcreturn b'$
    \end{nicodemus}
  }

  \vspace{1em}
  \underline{{\bf GAME} $\INDStAAPQC_b(\Pi, A)$}
    \begin{nicodemus}
    \item $\sidctr := 0$
    \item $\sid^* := 0$
      \item $\pcfor i \in \set{1, \dots, \numberParties}$
      \item $\quad (\pk_i, \sk_i) \gets \KeyGenStat()$
      \item $\qkdInit()$
      \item $b' \gets A^\Oracle(\pk_1, \dots, \pk_\numberParties)$
      \item $\matchingSessions := \FindMatches(\sid^*)$
      \item $\pcif \Attack(\matchingSessions)$ \label[line]{line:callattack}
      \item $\quad \pcreturn 0$ \gcom{ adversary loses }
      \item $\pcreturn b'$
    \end{nicodemus}

  \vspace{1em}
  \underline{{\bf GAME} $\INDAAQKD_b(\Pi, A)$}
    \begin{nicodemus}
    \item $\sidctr := 0$
    \item $\sid^* := 0$
      \item $\pcfor i \in \set{1, \dots, \numberParties}$
      \item $\quad (\pk_i, \sk_i) \gets \KeyGenStat()$
      \item $\qkdInit()$
      \item $b' \gets A^\Oracle(\pk_1, \dots, \pk_\numberParties)$
      \item $\matchingSessions := \FindMatches(\sid^*)$
      \item $\pcif \TrivialQKD(\matchingSessions)$ \label[line]{line:calltrivialqkd}
      \item $\quad \pcreturn 0$ \gcom{ adversary loses }
      \item $\pcreturn b'$
    \end{nicodemus}

  \vspace{1em}
    \underline{$\orTest(b, \sessionID)$}
    \begin{nicodemus}
      \item $\pcif \sid = 0 \pcor \sid^* \neq 0:$ \gcom{ only one query}
      \item $\quad \pcreturn \bot$
      \item $\sessionID^* := \sessionID$
      \item $k_0^* := \sessionkeyArray[\sessionID^*]$
      \item $k_1^* \uni \{0,1\}^{\lensess}$
      \item $\pcreturn k_b^*$
    \end{nicodemus}

  \vspace{1em}
    \underline{$\orReveal(\sessionID)$}
    \begin{nicodemus}
      \item $\revealed[\sessionID] := \true$
      \item $\pcreturn \sessionkeyArray[\sessionID]$
    \end{nicodemus}

  \vspace{1em}
    \underline{$\orCorrupt(i)$}
    \begin{nicodemus}
      \item $\corrupted[i] := \true$
      \item $\pcreturn \sk_i$
    \end{nicodemus}

  \vspace{1em}
    \underline{$\orRevState(\sessionID)$}
    \begin{nicodemus}
      \item $\revealedState[\sessionID] := \true$
      \item $\pcreturn \state[\sessionID]$
    \end{nicodemus}

  \vspace{1em}
    \underline{$\orEst(i, j, \rho)$}
    \begin{nicodemus}
      \item $\sidctr := \sidctr + 1$
      \item $\owner[\sidctr] := i$
      \item $\peer[\sidctr] := j$
      \item $\role[\sidctr] := \rho$
      \item $\pcreturn \sidctr$
    \end{nicodemus}

  \vspace{1em}
    \underline{$\orSendInit(\sid)$}
    \begin{nicodemus}
      \item $(i, j) := (\owner[\sid], \peer[\sid])$
      \item $(\rho, s) := (\role[\sid], \state[\sid])$
      \item $\pcif i = \bot \pcor \rho \neq \roleInit \pcor s \neq \bot$
      \item $\quad \pcabort$
      \item $(m_1, s') \gets \Init(\pk_j)$
      \item $\sent[\sid] := m_1$
      \item $\state[\sid] := s'$
      \item $\pcreturn m_1$
    \end{nicodemus}

  \vspace{1em}
    \underline{$\orSendMOne(\sid, m_1)$}
    \begin{nicodemus}
      \item $(i, j) := (\owner[\sid], \peer[\sid])$
      \item $(\rho, s) := (\role[\sid], \state[\sid])$
      \item $\pcif i = \bot \pcor \rho \neq \roleResp \pcor s \neq \bot$
      \item $\quad \pcabort$
      \item $\received[\sid] := m_1$
      \item $(m_2, k) \gets \SendMOne^{\enckey(\sid, \cdot)}($

        $\hspace*{\fill}i, \sk_i, j, \pk_j, m_1)$
      \item $\sent[\sid] := m_2$
      \item $\sessionkeyArray[\sid] := k$
      \item $\state[\sid] := \Accept$
      \item $\pcreturn m_2$
    \end{nicodemus}

  \vspace{1em}
    \underline{$\orSendMTwo(\sid, m_2)$}
    \begin{nicodemus}
      \item $(i, j) := (\owner[\sid], \peer[\sid])$
      \item $(\rho, s) := (\role[\sid], \state[\sid])$
      \item $\pcif i = \bot \pcor \rho \neq \roleInit$
      \item $\quad \pcabort$
      \item $\received[\sid] := m_2$
      \item $k \gets \SendMTwo^{\deckey(\sid, \cdot)}($

        $\hspace{\fill}i, \sk_i, j, s, m_2)$
      \item $\sessionkeyArray[\sid] := k$
      \item $\state[\sid] := \Accept$
    \end{nicodemus}
\end{multicols}\end{minipage}}
\caption{
  Formal security game, adapted from~\cite{PKC:HKSU20}.
  Protocol $\Pi$ is defined by subroutines $(\Init, \SendMOne, \SendMTwo)$.
  Adversary $A$ gets access to oracle $\Oracle = (\orTest(b, \cdot)$, $\orReveal$, $\orCorrupt$, $\orRevState$,
  $\orEst$, $\orSendInit$, $\orSendMOne$, $\orSendMTwo$,
  $\orQKDSids$, $\orQKDGet$, $\orQKDSet)$.
}\label{fig:formal_game}
\end{figure}

\begin{figure}
  \fbox{\begin{minipage}{\dimexpr\textwidth-2\fboxsep-2\fboxrule\relax}
    \underline{$\Attack(\matchingSessions)$}
    \begin{nicodemus}
    \item $\pcreturn \TrivialPQC(\matchingSessions) \pcor (\matchingSessions = \emptyset \pcand \revealedState[\sessionID^*])$
    \end{nicodemus}

  \vspace{1em}
    \underline{$\TrivialPQC(\matchingSessions)$}
    \begin{nicodemus}
    \item $\pcif \sessionkeyArray[\sessionID^*] = \bot$
    \item $\quad \pcreturn \true$ \gcom{ test session incomplete}
    \item $\pcif \revealed[\sessionID^*]$
    \item $\quad \pcreturn \true$ \gcom{ test session key revealed directly}
    \item $\pcif \corrupted[\owner[\sid^*]] \pcand \revealedState[\sid^*]$\label{line:candrs}
    \item $\quad \pcreturn \true$ \gcom{ adversary can complete test session}
    \item $\pcif |\matchingSessions| > 1$\label[line]{line:multimatch}
    \item $\quad \pcreturn \false$ \gcom{ multiple matching sessions}
    \item $\pcif \matchingSessions = \emptyset$
    \item $\quad \pcreturn \corrupted[\peer[\sid^*]]$
      \gcom{ session completed directly with adversary}
    \item $\set{\sidmatch} := \matchingSessions$ \gcom{ only one match with session ID $\sidmatch$}
    \item $\pcreturn \revealed[\sidmatch] \pcor$ \gcom{ session key leaked directly via matching session}

      \hspace{4em} $(\corrupted[\owner[\sidmatch]] \pcand \revealedState[\sidmatch])$
      \gcom{ adversary can complete peer session}
    \end{nicodemus}

  \vspace{1em}
    \underline{$\TrivialQKD(\matchingSessions)$}
    \begin{nicodemus}
      \item $\pcif \sessionkeyArray[\sessionID^*] = \bot$
      \item $\quad \pcreturn \true$ \gcom{ test session incomplete}
      \item $\pcif \revealed[\sessionID^*]$
      \item $\quad \pcreturn \true$ \gcom{ test session key revealed directly}
      \item $\complete := \Completed(\matchingSessions)$
      \item $\pcif |\complete| = 1$
      \item $\quad \set{\sidmatch} := \complete$ \gcom{ only one completed match with session ID $\sidmatch$}
      \item $\quad \pcif \revealed[\sidmatch]$ \gcom{ session key leaked directly via matching session}
      \item $\qquad \pcreturn \true$
      \item $\pcfor 1 \leq \kid \leq \kidctr$
      \item $\quad \pcif \flag[\kid] = \honest \pcand (\sid^* = \qsent[\kid] \pcor [\sid^*] = \qrecv[\kid])$
      \item $\qquad \pcreturn \false$
        \gcom{ an honest QKD key was delivered to the test session}
      \item $\pcreturn \true$
        \gcom{ no honest QKD key was delivered to the test session}
    \end{nicodemus}

    \vspace{1em}
    \underline{$\Completed(\matchingSessions)$} \gcom{ filter out uncomplete sessions (from $\matchingSessions$)}
    \begin{nicodemus}
      \item $\complete := \emptyset$
      \item $\pcfor \sid \in \matchingSessions$
      \item $\quad \pcif \sessionkeyArray[\sid] \neq \bot$
        \gcom{ session completed}
      \item $\qquad \complete := \complete \cup \set{\sid}$
      \item $\pcreturn \complete$
    \end{nicodemus}

    \vspace{1em}
    \underline{$\FindMatches(\sid^*)$} \gcom{ create set of sessions matching with $\sid^*$}
    \begin{nicodemus}
      \item $\matchingSessions := \emptyset$
      \item $(i, j) := (\owner[\sid^*], \peer[\sid^*])$
      \item $\pcfor 1 \leq \sid \leq \sid_\ctr$
      \item $\quad \pcif \owner[\sid] = j \pcand \peer[\sid] = i \pcand$
          \gcom{ intended partners}

        \hspace{4em} $\received[\sid] = \sent[\sid^*] \pcand$

        \hspace{4em} $\sent[\sid] = \received[\sid^*] \pcand$
          \gcom{ matching transcripts}

        \hspace{4em} $\role[\sid] \neq \role[\sid^*]$
          \gcom{ different roles}
      \item $\quad \qquad \matchingSessions := \matchingSessions \cup \set{\sid}$
          \gcom{ add to set of matching sessions}
      \item $\pcreturn \matchingSessions$
    \end{nicodemus}
  \end{minipage}}
\caption{
  A protocol is considered trivially broken if too much information is revealed,
  such that no expectation of security remains for \emph{any} protocol.
\label{fig:formal_game_trivial} }
\end{figure}

\heading{Correctness definition.}
We require the protocol to be correct,
which intuitively means
that both participating parties will compute the same session key
in the presence of a passive adversary.
In the AKE literature~\cite{C:Krawczyk05,PKC:FSXY12} this requirement is usually formalized by requiring that
if two parties complete matching sessions, then they compute the same session key.
Since we allow an active adversary to change QKD keys via the \orQKDSet query,
we also need to exclude that capability to formally define what we mean by a passive adversary.
We enforce this by requiring that all QKD keys used by the matching sessions are not \corrupt.
\begin{definition}[AKE-correctness]\label{defn:correctness}
  An AKE protocol $\Pi$ is \emph{$\delta$-correct} if the following holds.
  If $\sid$ and $\sidmatch$ denote two sessions such that
  $\sessionkeyArray[\sid] \neq \bot$ and
  $\sessionkeyArray[\sidmatch] \neq \bot$ (they are complete),
  $\sidmatch \in \FindMatches(\sid)$ (they are matching), and
  for all $\kid$ such that $\kid = \qsent$ or $\kid \in \qrecv$
  it holds that $\flag[\kid] \neq \corrupt$ (QKD was not tampered with),
  then
  \[
    \Pr[\sessionkeyArray[\sid] \neq \sessionkeyArray[\sidmatch]] \leq \delta.
  \]
\end{definition}
We emphasize that both sessions need to be completed.
Without this requirement,
an adversary might duplicate a QKD key ID
such that multiple honest sessions will request the QKD key using \deckey{},
but only one will succeed.
We consider this to be an active attack and therefore exclude it from the correctness definition.

\heading{Security definition.}
For non-\TrivialPQC{} attacks, we use one of the two existing notions in our security definition:
the stronger notion of ``key \textbf{IND}istinguishability against \textbf{A}ctive \textbf{A}ttacks against \textbf{PQC}''
can be defined using the \INDAAPQC{} game,
and the weaker notion of ``key \textbf{IND}istinguishability against \textbf{A}ctive \textbf{A}ttacks
against \textbf{PQC} without \textbf{St}ate reveal in the test session''
is defined using the \INDStAAPQC{} game.
We include both games in \cref{fig:formal_game} for completeness,
but we only use the (weaker) \INDStAAPQC game in our analysis of the AKE protocol of \cref{fig:protocol}.
The difference from \INDStAAPQC{} is that it calls \TrivialPQC{} in~\Cref{line:calltrivial} instead of \Attack{} in~\Cref{line:callattack}.

\begin{definition}[StAA-PQC Advantage]
  Define the game {\INDStAAPQC}$_b$ for $b \in \{0,1\}$
  as in \cref{fig:formal_game,fig:formal_game_trivial}.
  The \emph{advantage} of an adversary $A$ against an AKE protocol $\Pi$ is
  \[
    \Adv_\Pi^{\INDStAAPQC}(A) =
    \big| \Pr[\INDStAAPQC_0(\Pi, A) \Rightarrow 1] - \Pr[\INDStAAPQC_1(\Pi, A) \Rightarrow 1] \big| \, .
  \]
\end{definition}

Against non-$\TrivialQKD$ attacks, we define a new security notion:
``key \textbf{IND}istinguishability against \textbf{A}ctive \textbf{A}ttacks against \textbf{QKD}'',
for which we use the $\INDAAQKD$ game.

\begin{definition}[AA-QKD Advantage]
  Define the game {\INDAAQKD}$_b$ for $b \in \{0,1\}$
  as in \cref{fig:formal_game,fig:formal_game_trivial}.
  The \emph{advantage} of an adversary $A$ against an AKE protocol $\Pi$ is
  \[
    \Adv_\Pi^{\INDAAQKD}(A) =
    \big| \Pr[\INDAAQKD_0(\Pi, A) \Rightarrow 1] - \Pr[\INDAAQKD_1(\Pi, A) \Rightarrow 1] \big| \, .
  \]
\end{definition}


\section{Security proof}\label{sec:security-proof}
	\cref{fig:formal_impl} formally describes our authenticated key exchange protocol with a QKD oracle from \cref{fig:protocol}.

\nicoresetlinenr
\begin{figure}
  \fbox{\begin{minipage}{\dimexpr\textwidth-2\fboxsep-2\fboxrule\relax}\begin{multicols}{2}%
    \underline{$\Init(\pk_j)$}
    \begin{nicodemus}
      \item $(\cbob, \kbob) \gets \EncapsStat(\pk_j)$
      \item $(\pk_e, \sk_e) \gets \KeyGenEph()$
      \item $m_1 := (\cbob, \pk_e)$
      \item $s' := (\kbob, \sk_e, m_1)$
      \item $\pcreturn(m_1, s')$
    \end{nicodemus}

    \vspace{1em}
    \underline{$\SendMOne^{\enckey(\sid, \cdot)}(i, \sk_i, j, \pk_j, m_1)$}
    \begin{nicodemus}
      \item $(\cbob, \pk_e) := m_1$ \gcom{ or abort}
      \item $\kbob' := \DecapsStat(\sk_i, \cbob)$
      \item $(\calice, \kalice) \gets \EncapsStat(\pk_j)$
      \item $(\ceph, \keph) \gets \EncapsEph(\pk_e)$
      \item $\kpqc := \KDF(\kbob', \kalice, \keph)$
      \item $(\kid, \kqkd) \gets \enckey(\sid, \lenqkd)$
      \item $t := (m_1, (\calice, \ceph, \kid))$
      \item $(\ksess, \tau_1, \tau_2) := \algowrite{Combine}(j, i, t, \kqkd, \kpqc)$
      \item $m_2 := (\calice, \ceph, \kid, \tau_1, \tau_2)$
      \item $\pcreturn (m_2, \ksess)$
    \end{nicodemus}

    \columnbreak
    \underline{$\SendMTwo^{\deckey(\sid, \cdot)}(i, \sk_i, j, s, m_2)$}
    \begin{nicodemus}
      \item $(\calice, \ceph, \kid, \tau_1', \tau_2') := m_2$ \gcom{ or abort}
      \item $(\kbob, \sk_e, m_1) := s$ \gcom{ or abort}
      \item $\kalice' := \DecapsStat(\sk_i, \calice)$
      \item $\keph' := \DecapsEph(\sk_e, \ceph)$
      \item $\kpqc := \KDF(\kbob, \kalice', \keph')$
      \item $\kqkd := $

        $\hspace{\fill}\deckey(\sid, \kid)$
      \item $\pcif \kqkd = \bot: \pcabort$
      \item $t := (m_1, (\calice, \ceph, \kid))$
      \item $(\ksess, \tau_1, \tau_2) := \algowrite{Combine}(i, j, t, \kqkd, \kpqc)$
      \item $\pcif \tau_1 \neq \tau_1' \pcor \tau_2 \neq \tau_2': \pcabort$
      \item $\pcreturn \ksess$
    \end{nicodemus}

    \vspace{1em}
    \underline{$\algowrite{Combine}(i_I, i_R, t, \kqkd, \kpqc)$}
    \begin{nicodemus}
      \item $(\kqkdm \concat \kqkds) := \kqkd$
      \item $(\kpqcm \concat \kpqcs) := \kpqc$
      \item $\ksess := \kqkds \oplus \kpqcs$
      \item $\tau_1 := \QKDMAC_{\kqkdm}((t, i_I, i_R))$
      \item $\tau_2 := \PQCMAC_{\kpqcm}((t, \tau_1, i_I, i_R))$
      \item $\pcreturn (\ksess, \tau_1, \tau_2)$
    \end{nicodemus}
\end{multicols}\end{minipage}}
\caption{
  A formal implementation of our AKE protocol $\Pi = (\Init, \SendMOne, \SendMTwo)$,
  using the QKD oracle $(\enckey, \deckey)$,
  in the case of end-to-end QKD.\label{fig:formal_impl}
}
\end{figure}

We use the notation $(x \concat y)$ to split keys into their required lengths: 
in all protocols the first part becomes a \MAC{} key,
so $x$ should have the length of a \MAC{} key and $y$ should be the remaining bits.
We use destructuring assignments to ``parse'' the incoming messages and state,
this should be interpreted as a shorthand notation for
checking if the parsed value has the correct format and aborting the session if it has not.

Our protocol uses two \MAC{}s,
the inner $\QKDMAC$ uses a key generated by QKD,
and we require statistical security:
strong unforgeability (see \cref{defn:adv-otsufcma}) must hold against computationally unbounded adversaries.
Indeed such \MAC{}s exist~\cite{WegCar81}
and are usually canonical,
but a non-canonical \MAC{} would suffice.
The outer $\PQCMAC$ uses a key generated by PQC primitives,
this \MAC must either have statistical security,
or it must be canonical (in which case computational security suffices).
This property protects against a (potentially unbounded) adversary
that changes a matching session into a non-matching one
by modifying the outer tag exchanged between two honest parties:
both statistically secure and canonical \MAC{}s will reject such modifications.
For simplicity, we assume both \MAC{}s are canonical.

Note that the order of nesting \MAC{}s is important.
Consider the alternative protocol that is identical to \cref{fig:formal_impl},
but computes \PQCMAC{} first and \QKDMAC{} second:
this protocol is vulnerable to the following attack.
An adversary can query \orQKDSet{} to change $\kqkd$ without changing $\kid$.
This modification of the key does not change the transcript between honest users,
so that \PQCMAC{} does not detect anything,
yet the honest parties will compute \QKDMAC{} with a different key
(so that the sent tag almost certainly differs from the one that is accepted).
The adversary (who knows $\kqkd$) can swap out the QKD tag,
so that both parties accept with the same session key,
but their sessions are non-matching,
and thus the adversary may reveal the peer session key. 

This attack does not apply to \cref{fig:formal_impl},
because the adversary cannot change $\kpqc$ without changing the transcript.
The reason is that \QKDMAC{} authenticates the full transcript,
and by correctness of the \KEM{}s,
both honest parties will (almost certainly) compute the same $\kpqc$
and thus the sent tag is the only one that is accepted.

First we prove that the protocol $\Pi$ defined in \cref{fig:formal_impl}
is correct when implemented with correct components.
\begin{theorem}[AKE-correctness]\label{thm:correctness}
  Let $\Pi$ be the AKE protocol defined in \cref{fig:formal_impl},
  where $\KEMeph$ is a $\deph$-correct KEM
  and $\KEMstat$ is a $\dstat$-correct KEM.
  Then $\Pi$ is a $(\deph + 2\dstat)$-correct AKE.
\end{theorem}

\begin{proof}
  Let $\sid$ and $\sidmatch$ be complete matching sessions with uncorrupted QKD keys,
  according to \cref{defn:correctness}.
  Since the sessions match, one of them is an initiator ($\sidinit$),
  the other is a responder ($\sidresp$)
  and the transcript contains a single QKD key ID ($\kid^*$).
  Because $\sidinit$ completed it received a QKD key: $\kqkd := \deckey(\sidresp, \kid^*) \neq \bot$,
  and since $\flag[\kid^*] \neq \corrupt$, both sessions received the same QKD key.
  
  For a matching session, the only other source of error is a decapsulation failure.
  The protocol contains three decapsulations,
  and so the probability of any failure is at most the sum of the probability
  that either one of them fails, so that
  \[
    \Pr[\sessionkeyArray[\sid] \neq \sessionkeyArray[\sidmatch]] \leq \deph + 2\dstat. 
  \] \qed
\end{proof}

We now give a bound on the advantage that any (computationally bounded) adversary has on
protocol $\Pi$ (defined in \cref{fig:formal_impl})
based on the advantage on its secure PQC components.

\begin{theorem}[PQC-based security] \label{thm:pqc-security}
  Let $\Pi$ be the protocol of \cref{fig:formal_impl},
  where $\KEMeph$ is $\deph$-correct and has key length $\leneph$,
  and $\KEMstat$ is $\dstat$-correct,
  has collision probabilities $\muEncStat$ and $\muSecStat$,
  and has key length $\lenstat$.
  Let $A$ be an $\INDStAAPQC$ adversary executing $\Pi$ with $\numberParties$ parties
  that establishes $\numberSessions$ sessions that makes at most $q_H$ calls
  to the key-derivation function $\KDF$ (modelled as a QROM).
  Then there exist
  \INDCPA{} adversary $B_1$ against $\KEMeph$,
  \INDCPA{} adversary $B_2$ against $\KEMstat$,
  \INDCCA adversary $B_3$ against $\KEMstat$, and
  \OTSUFCMA adversary $B_4$ against $\PQCMAC$ such that
  \begin{align*}
    & \Adv^{\INDStAAPQC}(\Pi,A) \le\\
    & \phantom{0} \mathbin{\hphantom{+}} 2\numberSessions\!^2 \left(
      \deph + \Adv_{\KEMeph}^{\INDCPA}(B_1) + 2q_H\frac{1}{\sqrt{2^{\leneph}}}\right)\\
    & \phantom{0} + 2\numberSessions\!^2\numberParties\left(\dstat + \Adv_{\KEMstat}^{\INDCCA}(B_2) + 2q_H\frac{1}{\sqrt{2^{\lenstat}}}\right)
    + 2\cdot \numberSessions \cdot \muEncStat\\
    & \phantom{0} + 8 \numberSessions \numberParties
    \left(\deltastat + \Adv_{\KEMstat}^{\INDCCA}(B_3)
    + \numberSessions\!^2 \muSecStat + \numberSessions\frac{2q_H}{\sqrt{2^{\lenstat}}}
    + \numberSessions\Adv_{\PQCMAC}^{\OTSUFCMA}(B_4)\right)
  \end{align*}
  where the runtime of $B_1,B_2,B_3,B_4$ is about that of $A$.
\end{theorem}

Readers familiar with the KEM-TLS protocol and its security proof~\cite{CCS:SchSteWig20},
might be surprised to see that \INDCPA{} security suffices for $\KEMeph$ in our protocol.
The reason our protocol does not require \INDOCCA{} (a one-time variant of \INDCCA{} security)
is that an initiator session \emph{immediately} aborts when an adversary modifies or replaces the ephemeral encapsulation,
because they cannot accompany that same message with the correct \MAC tags.

\begin{proof}[PQC-based security]
We split the proof into two cases.
The first case, \cref{case1}, concerns the case $\matchingSessions \ne \emptyset$.
For the second case (\cref{case2}), we look at $\matchingSessions = \emptyset$.
\begin{description}
  \item[\Cref{case1}: passive attack] is formally treated in \cref{sec:case1}.
    In this case, a matching session exists.

    We will prove that with overwhelming probability, this matching session must be unique.
    Since we are assuming that there are matching sessions,
    and since the responder session does not have a state,
    there is no benefit to the attacker in revealing the state of
    the responder. Therefore, assuming there is exactly one matching session,
    we can always assume that the
    state of the responder session is not revealed.
        
    We will split \cref{case1} into two sub-cases:
    \begin{description}
      \item[\ref{case1a}] The state of the initiator has not been revealed.
        Thus, the ephemeral secret key $\sk_e$ has not been revealed to the attacker,
        and we base security of the final session key on the security of the ephemeral KEM.
      \item[\ref{case1b}] The initiator has not been corrupted.
        Thus, the initiator's static key $\sk_i$ has not been revealed to the attacker.
        We base security of the final session key on the security of the initiator static
        KEM.
    \end{description}
    \begin{lemma}[Case 1] \label{lem:case1}
    There exist adversaries $B_1,B_2$ such that
    \begin{multline*}
    \big| \Pr[\INDStAAPQC_1(\Pi, A) \Rightarrow 1 \wedge \matchingSessions\ne \emptyset] -
    \Pr[\INDStAAPQC_0(\Pi, A) \Rightarrow 1 \wedge \matchingSessions\ne \emptyset]\big|\\
    \le
    2\numberSessions\!^2\cdot\left(
    \deph + \Adv_\KEM^{\INDCPA}(B_1) + 2q_H\frac{1}{\sqrt{2^{\leneph}}}+
    \numberParties\left(\dstat + \Adv_\KEM^{\INDCCA}(B_2) + 2q_H\frac{1}{\sqrt{2^{\lenstat}}}\right)
    \right)\\+2\cdot \numberSessions \cdot \muEncStat
    \end{multline*}
    and the runtime of $B_1,B_2$ is about that of $A$.
    \end{lemma}
  \item[\Cref{case2}: active attack] is formally treated in \cref{sec:case2}. This means that no matching session exists.
    Since we only consider non-trivial attacks, we in particular rule out that the test session's peer has been corrupted.
    It is thus safe to assume that their static key has not been revealed, and we base security of $\kpqc$ on the security of the peer's static
    KEM.
    It remains to consider the case that there is a non-matching session that computes the same $\kpqc$,
    for example if the adversary swapped out $\kid$ or changes one of the \MAC tags.
    However, then by the unforgeability of $\PQCMAC$,
    this will be detected by the initiator who will then abort.
    That means there is only one accepting (responder) session whose session key depends on $\kpqc$,
    and thus $\kpqc$ is independent from the adversary view.
    \begin{lemma}[Case 2] \label{lem:case2}
    There exist adversaries $B_3,B_4$ such that
    \begin{multline*}
      \big| \Pr[\INDStAAPQC_0(\Pi, A) \Rightarrow 1 \land \matchingSessions = \emptyset] \\
      - \Pr[\INDStAAPQC_1(\Pi, A) \Rightarrow 1 \land \matchingSessions = \emptyset] \big| \\
      \leq 8 \numberSessions \numberParties
      \Bigg(
        \deltastat + \Adv_{\KEMstat}^{\INDCCA}(B_3)
      + \numberSessions\!^2 \muSecStat + \numberSessions\frac{2q_H}{\sqrt{2^{\lenstat}}} + \numberSessions\Adv_{\PQCMAC}^{\OTSUFCMA}(B_4)\Bigg)
    \end{multline*}
    and the runtime of $B_3,B_4$ is about that of $A$.
    \end{lemma}
\end{description}

In order to bound
\[
  \left| \Pr[\INDStAAPQC_0(\Pi, A) \Rightarrow 1]
  - \Pr[\INDStAAPQC_1(\Pi, A) \Rightarrow 1] \right| \\
\]
we combine \cref{lem:case1} and \cref{lem:case2} by taking the sum of the bounds.
This gives
\begin{align*}
& | \Pr[\INDStAAPQC_0(\Pi, A) \Rightarrow 1]
  - \Pr[\INDStAAPQC_1(\Pi, A) \Rightarrow 1] | \\
& \phantom{0} \mathbin{\hphantom{+}} 2\numberSessions\!^2 \left(
  \deph + \Adv_{\KEMeph}^{\INDCPA}(B_1) + 2q_H\frac{1}{\sqrt{2^{\leneph}}}\right)\\
& \phantom{0} + 2\numberSessions\!^2\numberParties\left(\dstat + \Adv_{\KEMstat}^{\INDCCA}(B_2) + 2q_H\frac{1}{\sqrt{2^{\lenstat}}}\right)
+ 2\cdot \numberSessions \cdot \muEncStat\\
& \phantom{0} + 8 \numberSessions \numberParties
\left(\deltastat + \Adv_{\KEMstat}^{\INDCCA}(B_3)
+ \numberSessions\!^2 \muSecStat + \numberSessions\frac{2q_H}{\sqrt{2^{\lenstat}}}
+ \numberSessions\Adv_{\PQCMAC}^{\OTSUFCMA}(B_4)\right).
\end{align*}
\end{proof}

The proofs of the \cref{lem:case1,lem:case2} can be found in the following sections.

We now give a bound on the advantage that any (potentially unbounded) adversary has on
protocol $\Pi$ (defined in \cref{fig:formal_impl})
based on the advantage on its secure QKD components.

\begin{theorem}[QKD-based security] \label{thm:qkd-security}
  Let $\Pi$ be the protocol of \cref{fig:formal_impl},
  where $\KEMeph$ is $\deph$-correct
  and $\KEMstat$ is $\dstat$-correct.
  Let $A$ be an $\INDAAQKD$ adversary executing $\Pi$
  that establishes $\numberSessions$ sessions.
  Then there exist and \OTSUFCMA adversary $B_7$ against $\QKDMAC$ such that
  \begin{align*}
    & \Adv^{\INDAAQKD}(\Pi,A) \le
    2 \numberSessions \left( \numberSessions (2\deltastat + \deltaeph)+ \Adv_{\QKDMAC}^{\OTSUFCMA}(B_7) \right)
  \end{align*}
  where the runtime of $B_7$ is potentially unbounded.
\end{theorem}

\textit{Proof sketch (QKD-based security).}
The formal proof for \cref{thm:qkd-security} is given in \cref{sec:case3}.
The idea is that the QKD key has not leaked to (and has not been set by) the adversary.
Our QKD oracle guarantees that honest QKD keys can be handed out at most twice.
Thus, we focus on the test session and the session that uses the same QKD key.
In our protocol $\Pi$, this will always be an initiator/responder pair.
We base security of the final session key on the MAC,
this ensures that the responder session is a matching session, or the initiator will abort.
In both cases, the other session's session key cannot be revealed.
Thus, the QKD key (and thus the session key) looks uniformly random to the adversary.

\refstepcounter{pfcase}\label{case1}%
\subsection{Security proof for \texorpdfstring{\cref*{case1}}{Case 1}: passive attack, \texorpdfstring{\cref{lem:case1}}{Lemma 5.3}}\label{sec:case1}
Formally, we want to upper bound
\[
	\left|\Pr[\INDStAAPQC_1(\Pi, A) \Rightarrow 1 \wedge \matchingSessions\ne \emptyset] - 
	\Pr[\INDStAAPQC_0(\Pi, A) \Rightarrow 1 \wedge \matchingSessions\ne \emptyset]\right|
	\enspace ,
\]
which we will do via a sequence of games.

In the following section, we will first switch to game $\game{1}$ to ensure we are in the correct case,
after that the hop to game $\game{2}$ ensures that there is precisely one matching session.
Game $\game{3}$ makes a guess which two sessions will be the initiator and responder,
so that in any game hereafter we can talk about `the' initiator and responder sessions.

After that, we split the rest of the proof into two subcases:
\cref{case1a} where the ephemeral secret key of the initiator session has not been revealed and 
\cref{case1b} where the static secret key of initiator has not been revealed.

\nicoresetlinenr
\begin{figure}
  \centering
  \fbox{\begin{varwidth}{\dimexpr\textwidth-2\fboxsep-2\fboxrule\relax}
    \underline{{\bf GAME} $\game{0,b}$ - $\game{3,b}$}
    \begin{nicodemus}
      \item $\sidctr := 0$ \gcom{ session counter}
      \item $\sid^* := 0$ \gcom{ test session ID}
      \item $\sinit \uni \set{1, \dots, \numberSessions}$ \gcom{$\game{3,b}$: guess initiator}\label[line]{line:guessInit}
      \item $\sresp \uni \set{1, \dots, \numberSessions}$ \gcom{$\game{3,b}$: guess responder}\label[line]{line:guessResp}
      \item $\pcfor i \in \set{1, \dots, \numberParties}$
      \item $\quad (\pk_i, \sk_i) \gets \KeyGenStat()$
      \item $\qkdInit()$
      \item $b' \gets A^\Oracle(\pk_1, \dots, \pk_\numberParties)$
      \item $\matchingSessions := \FindMatches(\sid^*)$
      \item $\pcif \matchingSessions=\emptyset$: \pcreturn 0\gcom{$\game{1,b}$: ensure matching session exists}\label[line]{line:matchNonZero}
      \item $\pcif |\matchingSessions|>1$: \pcreturn 0\gcom{$\game{2,b}$: ensure at most one matching session}\label[line]{line:abort}
      \item $\pcif \Attack(\matchingSessions)$
      \item $\quad \pcreturn 0$

      \item Pick $\sidinit,\sidresp\in \{\sessionID^*,\overline{\sid}\}$

        \hspace{4em} s.t. $\role[\sidinit]=\roleInit$ and $\role[\sidresp]=\roleResp$\gcom{$\game{3,b}$}\label[line]{line:checkGuess0}
      \item $\pcif \sidinit \ne \sinit \pcor \sidresp \ne \sresp$
      \item $\quad \pcreturn 0$ \gcom{$\game{3,b}$}\label[line]{line:checkGuess1}
      \item $\pcreturn b'$
    \end{nicodemus}
  \end{varwidth}}
  \caption{$\game{0,b}$ - $\game{3,b}$. Any oracle unchanged is not included.\label{fig:Gnst}}
\end{figure}

\emph{Games $\game{0,b}$} For each challenge bit $b\in \{0,1\}$, game $\game{0,b}$ is the original game \INDStAAPQC{}$_b(\Pi, A)$, so
\begin{equation*}
\Pr[\INDStAAPQC_b(\Pi, A) \Rightarrow 1 \wedge \matchingSessions\ne \emptyset]
 =
\Pr[\gameb{0,b} \Rightarrow 1 \wedge \matchingSessions\ne \emptyset].
\end{equation*}

\emph{Games $\game{1,b}$}
In this game, we add the check that there is at least one matching session in \cref{line:matchNonZero},
to ensure we are in the right scenario.
For each challenge bit $b\in \{0,1\}$, the new game $\game{1,b}$ only differs from $\game{0,b}$ when $\matchingSessions = \emptyset$.
We're only considering the case where $\matchingSessions \ne \emptyset$ so we have
\begin{equation*}
	\Pr[\gameb{0,b} \Rightarrow 1 \wedge \matchingSessions\ne \emptyset]
=
\Pr[\gameb{1,b} \Rightarrow 1] \enspace .
\end{equation*}

\emph{Games $\game{2,b}$} For each challenge bit $b\in \{0,1\}$, we introduce a new game $\game{2,b}$ that returns 0 in \cref{line:abort} if there is more than one matching session.
Since $\gameb{1,b}$ and $\gameb{2,b}$ only differ upon this case, we have
\[
  \left| \Pr[\gameb{1,b} \Rightarrow 1]-\Pr[\gameb{2,b}\Rightarrow 1] \right| \le \Pr[\text{Return in \cref{line:abort}}] \enspace.
\]
We now bound $\Pr[\text{Return in \cref{line:abort}}]$ for each of the two possible roles the test session could have.
First, assume that $\role[\sid^*]=\roleInit$.
Having multiple matching sessions of $\sid^*$ means that there are two distinct \roleResp{} sessions $\sid_1$ and $\sid_2$ that sent the same first response.
Formally, this means $\sent[\sid_1]=[\calice,\ceph,\kid,\tau_1,\tau_2]=\sent[\sid_2]$.
Since all ciphertexts were generated honestly, the probability that
both the static and ephemeral ciphertexts are equal for both sessions is at most
$\muEncStat\cdot\muEncEph$ (see \cref{def:MuEnc}).
Since there are at most $(\numberSessions-1)$ possible sessions that could create this situation and since $\muEncEph\le 1$, we have
\begin{multline*}
  \Pr[\text{Return in \cref{line:abort}}\wedge \role[\sid^*]=\roleInit]\le (\numberSessions-1)\cdot \muEncEph\cdot\muEncStat\\
  \le \numberSessions \cdot \muEncStat.
\end{multline*}

Now consider the case that $\role[\sid^*]=\roleResp$.
Having multiple matching sessions of $\sid^*$ means there are two distinct \roleInit{} sessions
$\sid_1$ and $\sid_2$ which sent the same initial message.
Formally, this means $\sent[\sid_1]=[\cbob,\pk_e]=\sent[\sid_2]$.
This means that both sessions need to generate the same ciphertext to Bob's static key,
which happens with probability at most $\muEncStat$. Similarly, the probability that
both sessions generated the same public key $\pk_e$ is at most $\muKgEph$ (see \cref{def:MuKg}).
Since there are at most $(\numberSessions-1)$ sessions that could be matching, we have
\begin{multline*}
  \Pr[\text{Return in \cref{line:abort}}\wedge\role[\sid^*]=\roleResp]\le (\numberSessions-1)\cdot \muEncStat\cdot\muKgEph\\
  \le \numberSessions \cdot \muEncStat \enspace .
\end{multline*}
In total, this gives
\[
  \Pr[\text{Return in \cref{line:abort}}]\le \numberSessions \cdot \muEncStat \enspace .
\]

So far, we have
\[
	\left|\Pr[\INDStAAPQC_1(\Pi, A) \Rightarrow 1 \wedge \matchingSessions\ne \emptyset] - 
	\Pr[\INDStAAPQC_0(\Pi, A) \Rightarrow 1 \wedge \matchingSessions\ne \emptyset]\right|\]\[\le
	\left|\Pr[\gameb{2,1}\Rightarrow 1]-\Pr[\gameb{2,0}\Rightarrow 1]\right|+2\cdot \numberSessions \cdot \muEncStat
	\enspace .
\]

If games $\game{0,b}-\game{2,b}$ do not return early, we can safely assume that $|\matchingSessions| = 1$.
In other words, there is exactly one matching session of $\sid^*$, which we will denote by $\overline{\sid}$.

\emph{Games $\game{3,b}$}
At the start of both games $\game{3,b}$, we guess the session IDs of the init and responder sessions
in \cref{line:guessInit} and \cref{line:guessResp},
and we check the guesses are correct in \cref{line:checkGuess0} to \cref{line:checkGuess1}.
If the session IDs are guessed correctly, $\game{2,b}$ and $\game{3,b}$ are equal,
and if one of the sessions IDs has not been guessed correctly, $\game{3,b}$ outputs $0$.
Therefore
\[\Pr[\gameb{2,b}\Rightarrow 1]=\numberSessions\!^2\cdot \Pr[\gameb{3,b}\Rightarrow 1]\]
from which we get
\begin{multline}\label{eq:case1init}
	\left|\Pr[\INDStAAPQC_1(\Pi, A) \Rightarrow 1 \wedge \matchingSessions\ne \emptyset] - 
	\Pr[\INDStAAPQC_0(\Pi, A) \Rightarrow 1 \wedge \matchingSessions\ne \emptyset]\right| \\
	\le
	\numberSessions\!^2\cdot\left|\Pr[\gameb{3,1}\Rightarrow 1]-\Pr[\gameb{3,0}\Rightarrow 1]\right|+2\cdot \numberSessions \cdot \muEncStat
	\enspace .
\end{multline}

    If the attacker does any action that makes it so that
    $\Attack(\matchingSessions)$ returns \true, he will not be able
    to distinguish anything. Therefore it is safe to assume that
    the attacker's actions do not cause $\Attack(\matchingSessions)$ to
    return \true.
    In the case where $|\matchingSessions| = 1$, and we name the matching session $\sidmatch$,
    this comes down to:
    \begin{enumerate}
       \item $\sessionkeyArray[\sid^*]\ne \bot$ (the session actually finished)
       \item $\neg\revealed[\sid^*]$ (the test session's key was not revealed)
       \item $\neg\corrupted[\owner[\sid^*]]$ or $\neg\revealedState[\sid^*]$ (either the test session's owner was not corrupted or the test session was not revealed)\label{item:c1test}
       \item $\neg\revealed[\overline{\sid}]$ (the matching session's key is not revealed)
       \item $\neg\corrupted[\owner[\overline{\sid}]]$ or $\neg\revealedState[\overline{\sid}]$ (either the matching session's owner was not corrupted or the matching session was not revealed)\label{item:c1match}
    \end{enumerate}
    Since the responder session has no state,
    we may assume $\neg\revealedState[\sidresp]$.
    And since $\{\sidinit,\allowbreak \sidresp\} = \set{\sid^*, \sidmatch}$,
    the conjuction of properties \ref{item:c1test} and \ref{item:c1match}
    is equivalent to $\neg \corrupted[\owner[\sidinit]] \lor \neg\revealedState[\sidinit]$.
    Knowing this, we have to prove the security
    of the protocol in the following case:
    $\sessionkeyArray[\sid^*]\ne \bot$, $\neg\revealed[\sid^*]$,
    $\neg\revealed[\overline{\sid}]$, and either $\neg\corrupted[\sidinit]$
    or $\neg\revealedState[\sidinit]$ where $\sidinit\in \{\sid^*,\overline{\sid}\}$
    such that $\role[\sidinit]=\roleInit$.

We now look at two possible cases:
\begin{itemize}
\item\cref{case1a}: where we assume that the state of the initiator has not been revealed.
\item\cref{case1b}: where we assume that the initiator has not been corrupted.
\end{itemize}

In \cref{case1a} we assume that $\neg\revealedState[\sidinit]$,
where $\sidinit$ is the session id of the initiator.
We denote this case with $\neg\text{st}$.
Since the state of the initiator is not revealed, we can assume that with overwhelming probability,
the adversary does not know the secret key $\sk_e$ and therefore also not $\keph$.
Since $\keph$ is always part of the input of the $\KDF$ and since the $\KDF$ is modelled as a random
oracle,
the resulting session key is secure.

In \cref{case1b} we assume that $\neg\corrupted[\sidinit]$,
where again $\sidinit$ is the session id of the initiator.
We denote this case with $\neg sk(i)$.
Since the long-term secret key of the initiator is not known to the attacker,
we can assume that with overwhelming probability they do not know $\kalice$.
And again, since $\kalice$ is part of the input of the $\KDF$,
the resulting session key is secure.
\subsubsection{\texorpdfstring{\Cref*{case1a}}{Case 1a}: State not revealed}
\refstepcounter{pfsubcase}\label{case1a}
\begin{figure}
  \centering
\nicoresetlinenr
\fbox{\begin{varwidth}{\dimexpr\textwidth-2\fboxsep-2\fboxrule\relax}
\underline{{\bf GAME} $\gamenst{3,b}$ - $\gamenst{7,b}$}
\begin{nicodemus}
  \item $\sidctr := 0$ \gcom{ session counter}
  \item $\sid^* := 0$ \gcom{ test session ID}
  \item $\sinit \uni \set{1, \dots, \numberSessions}$ 
  \item $\sresp \uni \set{1, \dots, \numberSessions}$ 
  \item $\pcfor i \in \set{1, \dots, \numberParties}$
  \item $\quad (\pk_i, \sk_i) \gets \KeyGenStat()$
  \item $\qkdInit()$
  \item $(\widetilde{\pk_e}, \widetilde{\sk_e}) \gets \KeyGenEph()$ \gcom{$\gamenst{5,b}$} \label[line]{line:nst:keygen}
  \item $(\widetilde{c_e}, \widetilde{\keph}) \gets \EncapsEph(\widetilde{\pk_e})$ \gcom{$\gamenst{5,b}$} \label[line]{line:nst:enc}
  \item $\widetilde{\keph} \uni \{0,1\}^{\leneph}$ \gcom{$\gamenst{6,b}$} \label[line]{line:nst:ephRand}
  \item $\widetilde{\kpqc} \uni \{0,1\}^{\lenpqc}$ \gcom{$\gamenst{7,b}$} \label[line]{line:nst:pqcRand}
  \item $b' \gets A^\Oracle(\pk_1, \dots, \pk_N)$
  \item $\matchingSessions := \FindMatches(\sid^*)$
  \item $\pcif |\matchingSessions| \ne 1: \pcreturn 0$
  \item $\pcif \Attack(\matchingSessions)$
  \item $\quad \pcreturn 0$
  \item $\pcif \revealedState[\sessionID^*]: \pcreturn 0$ \gcom{$\gamenst{4,b}$} \label[line]{line:nst:stateNotRev}
  \item Pick $\sidinit,\sidresp\in \{\sessionID^*,\overline{\sid}\}$
s.t. $\role[\sidinit]=\roleInit$ and $\role[\sidresp]=\roleResp$
  \item $\pcif \sidinit \ne \sinit \pcor \sidresp \ne \sresp$
  \item $\quad \pcreturn 0$ 
  \item $\pcreturn b'$
\end{nicodemus}
\end{varwidth}}
\caption{$\gamenst{3,b}$ - $\gamenst{7,b}$. Any oracle not included is unchanged.}\label{fig:Gnst2}
\end{figure}
\begin{figure}
\ContinuedFloat
  \fbox{\begin{minipage}{\dimexpr\textwidth-2\fboxsep-2\fboxrule\relax}
    \underline{$\orSendInit(\sid)$}
    \begin{nicodemus}
      \item $(i, j, \rho, s) := (\owner[\sid], \peer[\sid], \role[\sid], \state[\sid])$
      \item $\pcif i = \bot \pcor \rho \neq \roleInit \pcor s \neq \bot: \pcabort$
      \item $(\cbob, \kbob) \gets \EncapsStat(\pk_j)$
      \item $(\pk_e, \sk_e) \gets \KeyGenEph()$
      \item $\pcif \sid = \sinit$
      \item $\quad (\pk_e, \sk_e) := (\widetilde{\pk_e},\widetilde{\sk_e})$ \gcom{$\gamenst{5,b}$}\label[line]{line:nst:keygen1}
      \item $m_1 := (\cbob, \pk_e)$
      \item $\sent[\sid] := m_1$ \gcom{ record outgoing message}
      \item $\state[\sid] := (\kbob, \sk_e, m_1)$ \gcom{ update internal state}
      \item $\pcreturn m_1$
    \end{nicodemus}
  \end{minipage}}
  \fbox{\begin{minipage}{\dimexpr\textwidth-2\fboxsep-2\fboxrule\relax}
    \underline{$\orSendMOne(\sid,m_1)$}
    \begin{nicodemus}
      \item $(i, j, \rho, s) := (\owner[\sid], \peer[\sid], \role[\sid], \state[\sid])$
      \item $\pcif i = \bot \pcor \rho \neq \roleResp \pcor s \neq \bot: \pcabort$
      \item $\received[\sid] := m_1$ \gcom{ record incoming message}
      \item $(\cbob, \pk_e) := m_1$ \gcom{ or abort}
      \item $\kbob' := \DecapsStat(\sk_i, \cbob)$
      \item $(\calice, \kalice) \gets \EncapsStat(\pk_j)$
      \item $(c_e, \keph) \gets \EncapsEph(\pk_e)$
      \item $\pcif \sid = \sresp$
      \item $\quad (c_e,k_e) := (\widetilde{c_e},\widetilde{k_e})$ \gcom{$\gamenst{5,b}$}\label[line]{line:nst:enc1}
      \item $\kpqc := \KDF(\kbob', \kalice, \keph)$
      \item $\pcif \sessionID = \sresp$ 
      \item $\quad \kpqc := \widetilde{\kpqc}$ \gcom{$\gamenst{7,b}$}\label[line]{line:nst:pqcRand1}
      \item $(\kid, \kqkd) := \enckey(\sid, \lenqkd)$ 
      \item $t := (m_1, (\calice, c_e, \kid))$
      \item $\kqkdm \concat \kqkds := \kqkd$
      \item $\kpqcm \concat \kpqcs := \kpqc$
      \item $\ksess := \kqkds \oplus \kpqcs$
      \item $\tau_1 := \QKDMAC_{\kqkdm}((t, j, i))$
      \item $\tau_2 := \PQCMAC_{\kpqcm}((t, \tau_1, j, i))$
      \item $m_2 = (\calice, \ceph, \kid, \tau_1, \tau_2)$
      \item $\sent[\sid] := m_2$ \gcom{ record outgoing message}
      \item $\sessionkeyArray[\sid] := \ksess$ \gcom{ set session key}
      \item $\state[\sid] := \Accept$ \gcom{ accept session}
      \item $\pcreturn m_2$
    \end{nicodemus}
  \end{minipage}}
\caption{(cont.) $\gamenst{3,b}$ - $\gamenst{7,b}$. Any oracle not included is unchanged.}
\end{figure}
\begin{figure}
  \centering
\ContinuedFloat
  \fbox{\begin{minipage}{\dimexpr\textwidth-2\fboxsep-2\fboxrule\relax}
    \underline{$\orSendMTwo(\sid, m_2)$}
    \begin{nicodemus}
      \item $(i, j, \rho, s) := (\owner[\sid], \peer[\sid], \role[\sid], \state[\sid])$
      \item $\pcif i = \bot \pcor \rho \neq \roleInit: \pcabort$ \gcom{ session not established}
      \item $\received[\sid] := m_2$ \gcom{ record incoming message}
      \item $(\calice, \ceph, \kid, \tau_1', \tau_2') := m_2$ \gcom{ or abort}
      \item $(\kbob, \sk_e, m_1) := s$ \gcom{ or abort}
      \item $\kalice' := \DecapsStat(\sk_i, \calice)$
      \item $\keph := \DecapsEph(\sk_e, c_e)$
      \item $\pcif \sid = \sinit$
      \item $\quad \keph := \widetilde{\keph}$ \gcom{$\gamenst{5,b}$}\label[line]{line:nst:enc2}
      \item $\kpqc := \KDF(\kbob, \kalice', \keph')$
      \item $\pcif \sessionID = \sinit$ 
      \item $\quad \kpqc := \widetilde{\kpqc}$ \gcom{$\gamenst{7,b}$}\label[line]{line:nst:pqcRand2}
      \item $\kqkd := \deckey(\sid, \kid)$
      \item $\pcif \kqkd = \bot: \pcabort$
      \item $t := (m_1, (\calice, \ceph, \kid))$
      \item $\kqkdm \concat \kqkds := \kqkd$
      \item $\kpqcm \concat \kpqcs := \kpqc$
      \item $\ksess := \kqkds \oplus \kpqcs$
      \item $\tau_1 := \QKDMAC_{\kqkdm}((t, i, j))$
      \item $\tau_2 := \PQCMAC_{\kpqcm}((t, \tau_1, i, j))$
      \item $\pcif \tau_1 \neq \tau_1' \pcor \tau_2 \neq \tau_2': \pcabort$
      \item $\sessionkeyArray[\sid] := \ksess$
      \item $\state[\sid] := \Accept$ \gcom{ clear revealable state}
    \end{nicodemus}
  \end{minipage}}
  \caption{(cont.) $\gamenst{3,b}$ - $\gamenst{7,b}$. Any oracle not included is unchanged.}
\end{figure}

\emph{Games $\gamenst{3,b}$ - $\gamenst{7,b}$ are defined in \cref{fig:Gnst2}.}
Games $\game{3,b}$ and $\gamenst{3,b}$ are the same for both $b$, so
\[\left| \Pr[\gameb{3,1} \Rightarrow 1 \wedge \neg\text{st}] - \Pr[\gameb{3,0}\Rightarrow 1 \wedge \neg\text{st}]\right|
=\left| \Pr[\gamenstb{3,1}\Rightarrow 1 \wedge \neg\text{st}] -
\Pr[\gamenstb{3,0}\Rightarrow 1 \wedge \neg\text{st}]\right|.\]

\emph{Games $\gamenst{4,b}$}
We add the condition $\neg\text{st}$ in game $\gamenst{4,b}$ at \cref{line:nst:stateNotRev}, to ensure we are in the correct case. Since
\[\Pr[\gamenstb{3,b}\Rightarrow 1 \wedge \neg\text{st}]=
\Pr[\gamenstb{4,b}\Rightarrow 1]\]
for both $b$, which gives
\[\left| \Pr[\gamenstb{3,1}\Rightarrow 1 \wedge \neg\text{st}] -
\Pr[\gamenstb{3,0}\Rightarrow 1 \wedge \neg\text{st}]\right|=
\left| \Pr[\gamenstb{4,1}\Rightarrow 1] - \Pr[\gamenstb{4,0}\Rightarrow 1]\right|.\]

\emph{Games $\gamenst{5,b}$}
For $\gamenst{5,b}$ we move the calls to $\KeyGenEph$ and $\EncapsEph$ of the
ephemeral $\KEM$ to the game itself, \cref{line:nst:keygen,line:nst:enc}.
These values then replace the ephemeral public key, secret key, key and ciphertext in \cref{line:nst:keygen1,line:nst:enc1,line:nst:enc2}.
This change is only noticed by the distinguisher
if $\DecapsEph$ returns the wrong key. 
Since the probability that this happens is at most $\deph$ (see \cref{def:KemCorrectness}), 
we have
\begin{equation} \label{eq:game45}
\left|\Pr[\gamenstb{4,b}\Rightarrow 1]-\Pr[\gamenstb{5,b}\Rightarrow 1]\right|\le \deph.
\end{equation}

\emph{Games $\gamenst{6,b}$}
In this game, we replace the ephemeral key of the test sessions generated by $\EncapsEph$
with a uniformly random key, in \cref{line:nst:ephRand}.
To justify replacing the ephemeral key $\keph$ with a uniformly random key,
we now relate the potential additional advantage to \INDCPA security of the ephemeral \KEM.
We do this by showing that we can use any adversary $C$ that distinguishes between $\gamenst{5,b}$
and $\gamenst{6,b}$ to create an adversary $B_1$ of the ephemeral \KEM that has 
a runtime about that of $C$.
The adversary $C$ gets access to either $\gamenstb{5,b}$ or $\gamenst{6,b}$ and then has
to output either $0$ or $1$.
In the $\INDCPA$ game, the input to $B_1$ consists of $\pk$, $c$ and $k_i$,
where $\pk$ and $c$ are generated by the $\KeyGenEph$ and $\EncapsEph$ respectively,
and $k_i$ is generated by $\EncapsEph$ if $i=0$, and uniformly random if $i=1$.
This attacker $B_1$ can use the adversary $C$ as follows: 
replace the $\widetilde{\pk_e}$, $\widetilde{c_e}$ and $\widetilde{k_e}$
after \cref{line:nst:ephRand} with the inputs of $B_1$. Since these values were generated with
the same algorithms, this does not change their distribution.
Now $C$, with access to these algorithms, outputs $0$ or $1$.
Let $B_1$ then give the same output as $C$.

The difference between $\gamenst{5,b}$ and $\gamenst{6,b}$ is whether $\widetilde{k_e}$ is
generated by $\EncapsEph$ or uniformly random.
Since $C$ distinguishes between the honestly generated key and a uniformly random key,
replacing this key with the key $k_i$ for which we have the same question
allows distinguishing an honestly generated key from a uniformly random key for
the $\INDCPA$ game.
Since this is true for all adversaries $C$, we have
\begin{equation} \label{eq:game56}
	\left|\Pr[\gamenstb{5,b} \Rightarrow 1] - \Pr[\gamenstb{6,b}
  \Rightarrow 1]\right| \le \Adv_{\KEMeph}^{\INDCPA}(B_1) \enspace.
\end{equation}

\emph{Games $\gamenst{7,b}$}
Here, we replace the output of the $\KDF$ of the test session with a uniformly random
value. The value is generated at \cref{line:nst:pqcRand} and used at \cref{line:nst:pqcRand1,line:nst:pqcRand2}.

Since we model $\KDF:\{0,1\}^{2\lenstat + \leneph}\to \{0,1\}^{\lenpqc}$ as a random oracle, we can use the one-way to hiding \cref{lem:ow2h}
to bound the distance between $\gamenst{6,b}$ and $\gamenst{7,b}$.
Define $H(\cdot) = \KDF(\kalice,\kbob,\cdot)$ where $\kalice$ and $\kbob$ are the
keys used in session $\sinit$.
Then for any fixed $\kalice$, $\kbob$ we can replace sampling
$\KDF\uni (\{0,1\}^{2\lenstat + \leneph}\to \{0,1\}^{\lenpqc})$ with sampling
$H\uni (\{0,1\}^{\leneph} \to \{0,1\}^{\lenpqc})$.
Let $B^{H}(x,y)$ be the execution of $\gamenstb{6,b}$
where we replace $\keph$ of the test session with $x$,
and $\kpqc$ of the test session with $y$.
Then
\[
	\Pr[\gamenstb{6,b} \Rightarrow 1] = 
  \Pr[b'=1:b'\gets B^{H}(x,H(x))]
\]
and
\[
	\Pr[\gamenstb{7,b} \Rightarrow 1] =
  \Pr[b'=1:b'\gets B^{H}(x,y), y\uni \{0,1\}^{\lenpqc}].
\]
where we also take the average over 
$H\uni (\{0,1\}^{\leneph} \to \{0,1\}^{\lenpqc})$
and $x\uni \{0,1\}^{\leneph}$.
As in \cref{lem:ow2h}, let $C$ be an oracle algorithm that on input $x$ does the following: pick
$i\uni \{1,...,q_h\}$ and $y\uni \{0,1\}^m$, run $B^{H}(x,y)$ until just before
the $i$-th query, measure the argument of the query in the computational basis,
and output the measurement outcome.
We define
\[P_C=\Pr[x=x':x'\gets C^{H}(x)]\]
where the probability is also over
$H\uni (\{0,1\}^{\leneph}\to \{0,1\}^{\lenpqc})$ and $x\uni\{0,1\}^{\leneph}$.
But since $x$ is not available to $C$ at any point, the best the attacker can do is guess,
which means that $P_C=\frac{1}{2^{\leneph}}$.

Then by \cref{lem:ow2h} we have
\begin{equation} \label{eq:game67}
	\left|\Pr[\gamenstb{6,b} \Rightarrow 1] - 
	\Pr[\gamenstb{7,b}\Rightarrow 1]\right|\le 2q_H\sqrt{P_C}=2q_H\frac{1}{\sqrt{2^{\leneph}}}.
\end{equation}

Next, we want to bound
\[\left|\Pr[\gamenstb{7,0}\Rightarrow 1] -
\Pr[\gamenstb{7,1}\Rightarrow 1]\right|.\]
Key $k_0^*$ is generated as a part $\kpqc \oplus \kqkd$, where $\kpqc$ is uniformly random.
Furthermore, this key is only used in the matching session of the test session, which 
the adversary is not allowed to reveal.
Key $k_1^*$ is generated uniformly random. Therefore the adversary cannot distinguish between
$\gamenst{7,0}$ and $\gamenst{7,1}$, and we have
\begin{equation} \label{eq:game7equal}
  \Pr[\gamenstb{7,0}\Rightarrow 1]=
\Pr[\gamenstb{7,1}\Rightarrow 1].
\end{equation}
Putting equations \eqref{eq:game45}, \eqref{eq:game56} and \eqref{eq:game67} together, we get that
\[\left|\Pr[\gamenstb{4,b} \Rightarrow 1] - 
\Pr[\gamenstb{7,b}\Rightarrow 1]\right|\le \deph + \Adv_{\KEMeph}^{\INDCPA}(B_1) + 2q_H\frac{1}{\sqrt{2^{\leneph}}}.\]
Combining this with \eqref{eq:game7equal}, we get that
\begin{equation}\label{eq:case1a}
  \left|\Pr[\gamenstb{3,0} \wedge \neg st\Rightarrow 1] - 
\Pr[\gamenstb{3,1}\wedge \neg st\Rightarrow 1]\right|\le 2\deph + 2\Adv_{\KEMeph}^{\INDCPA}(B_1) + 4q_H\frac{1}{\sqrt{2^{\leneph}}}.\end{equation}

\subsubsection{\texorpdfstring{\Cref*{case1b}}{Case 1b}: Initiator not corrupted}%
\refstepcounter{pfsubcase}\label{case1b}
\begin{figure}
\nicoresetlinenr
\fbox{\begin{minipage}{\dimexpr\textwidth-2\fboxsep-2\fboxrule\relax}
\underline{{\bf GAME} $\gamenski{3,b}$ - $\gamenski{7,b}$}
\begin{nicodemus}
  \item $\sidctr := 0$ \gcom{ session counter}
  \item $\sid^* := 0$ \gcom{ test session ID}
  \item $\sinit \uni \set{1, \dots, \numberSessions}$ 
  \item $\sresp \uni \set{1, \dots, \numberSessions}$ 
  \item $j'\uni \set{1, \dots, \numberParties}$ \gcom{$\gamenski{5,b}$}\label[line]{line:nski:guessParty}
  \item $\pcfor i \in \set{1, \dots, \numberParties}$
  \item $\quad (\pk_i, \sk_i) \gets \KeyGenStat()$
  \item $(\widetilde{\pk_{j'}}, \widetilde{\sk_{j'}}) \gets \Pi.\KeyGenStat()$ \gcom{$\gamenski{6,b}$} \label[line]{line:nski:enc}
  \item $(\widetilde{\calice}, \widetilde{\kalice}) \gets \EncapsStat(\widetilde{\pk}_{j'})$ \gcom{$\gamenski{6,b}$} \label[line]{line:nski:enc0}
  \item $\widetilde{\kalice} \uni \{0,1\}^{\lenstat}$ \gcom{$\gamenski{7,b}$} \label[line]{line:nski:statRand}
  \item $\widetilde{\kpqc} \uni \{0,1\}^{\lenpqc}$ \gcom{$\gamenski{8,b}$} \label[line]{line:nski:pqcRand}
  \item $\qkdInit()$
  \item $b' \gets A^\Oracle(\pk_1, \dots, \pk_{\numberParties})$
  \item $\matchingSessions := \FindMatches(\sid^*)$
  \item $\pcif |\matchingSessions| \ne 1: \pcreturn 0$
  \item $\pcif \Attack(\matchingSessions)$
  \item $\quad \pcreturn 0$
  \item Pick $\sidinit,\sidresp\in \{\sessionID^*,\overline{\sid}\}$
s.t. $\role[\sidinit]=\roleInit$ and $\role[\sidresp]=\roleResp$
  \item $\pcif \sidinit \ne \sinit \pcor \sidresp \ne \sresp$
  \item $\quad \pcreturn 0$
  \item $\pcif j'\ne \owner[\sidinit]$ \label[line]{line:nski:guessParty1}
  \item $\quad \pcreturn 0$ \gcom{$\gamenski{5,b}$}
  \item $\pcif \corrupted[\sidinit]: \pcreturn 0$ \gcom{$\gamenski{4,b}$} \label[line]{line:nski:initNotCor}
  \item $\pcreturn b'$
\end{nicodemus}
\end{minipage}}
\caption{$\gamenski{3,b}$ - $\gamenski{8,b}$. Any oracle not included is unchanged.}\label{fig:Gnski}
\end{figure}
\begin{figure}
\ContinuedFloat
  \fbox{\begin{minipage}{\dimexpr\textwidth-2\fboxsep-2\fboxrule\relax}
    \underline{$\orSendInit(\sid)$}
    \begin{nicodemus}
      \item $(i, j, \rho, s) := (\owner[\sid], \peer[\sid], \role[\sid], \state[\sid])$
      \item $\pcif i = \bot \pcor \rho \neq \roleInit \pcor s \neq \bot: \pcabort$
      \item $(\cbob, \kbob) \gets \EncapsStat(\pk_j)$
      \item $(\pk_e, \sk_e) \gets \KeyGenEph()$
      \item $m_1 := (\cbob, \pk_e)$
      \item $\sent[\sid] := m_1$ \gcom{ record outgoing message}
      \item $\state[\sid] := (\kbob, \sk_e, m_1)$ \gcom{ set state}
      \item $\pcreturn m_1$
    \end{nicodemus}
  \end{minipage}}
  \fbox{\begin{minipage}{\dimexpr\textwidth-2\fboxsep-2\fboxrule\relax}
    \underline{$\orSendMOne(\sid,m_1)$}
    \begin{nicodemus}
      \item $(i, j, \rho, s) := (\owner[\sid], \peer[\sid], \role[\sid], \state[\sid])$
      \item $\pcif i = \bot \pcor \rho \neq \roleResp \pcor s \neq \bot: \pcabort$
      \item $\received[\sid] := m_1$ \gcom{ record incoming message}
      \item $(\cbob, \pk_e) := m_1$ \gcom{ or abort}
      \item $\kbob' := \DecapsStat(\sk_i, \cbob)$
      \item $(\calice, \kalice) \gets \EncapsStat(\pk_j)$
      \item $\pcif \sid = \sresp$
      \item $\quad (\calice, \kalice) := (\widetilde{\calice},\widetilde{\kalice})$ \gcom{$\gamenski{6,b}$}\label[line]{line:nski:enc1}
      \item $(c_e, \keph) \gets \EncapsEph(\pk_e)$
      \item $\kpqc := \KDF(\kbob', \kalice, \keph)$
      \item $\pcif \sessionID = \sresp$ 
      \item $\quad \kpqc := \widetilde{\kpqc}$ \gcom{$\gamenski{8,b}$}\label[line]{line:nski:pqcRand1}
      \item $(\kid, \kqkd) := \enckey(\sid,\lenqkd)$
      \item $t := (m_1, (\calice, \ceph, \kid))$
      \item $\kqkdm \concat \kqkds := \kqkd$
      \item $\kpqcm \concat \kpqcs := \kpqc$
      \item $\ksess := \kqkds \oplus \kpqcs$
      \item $\tau_1 := \QKDMAC_{\kqkdm}((t, j, i))$
      \item $\tau_2 := \PQCMAC_{\kpqcm}((t, \tau_1, j, i))$
      \item $m_2 = (\calice, \ceph, \kid, \tau_1, \tau_2)$
      \item $\sent[\sid] := m_2$ \gcom{ record outgoing message}
      \item $\sessionkeyArray[\sid] := \ksess$
      \item $\state[\sid] := \Accept$
      \item $\pcreturn m_2$
    \end{nicodemus}
  \end{minipage}}
  \caption{(cont.) $\gamenski{3,b}$ - $\gamenski{7,b}$. Any oracle not included is unchanged.}
\end{figure}
\begin{figure}
\ContinuedFloat
  \fbox{\begin{minipage}{\dimexpr\textwidth-2\fboxsep-2\fboxrule\relax}
    \underline{$\orSendMTwo(\sid, m_2)$}
    \begin{nicodemus}
      \item $(i, j, \rho, s) := (\owner[\sid], \peer[\sid], \role[\sid], \state[\sid])$
      \item $\pcif i = \bot \pcor \rho \neq \roleInit: \pcabort$ \gcom{ session not established}
      \item $\received[\sid] := m_2$ \gcom{ record incoming message}
      \item $(\calice, \ceph, \kid, \tau_1', \tau_2') := m_2$ \gcom{ or abort}
      \item $(\kbob, \sk_e, m_1) := s$ \gcom{ or abort}
      \item $\kalice' := \DecapsStat(\sk_i, \calice)$
      \item $\pcif \sid = \sinit \pcand \calice = \widetilde{\calice}$
      \item $\quad \kalice' := \widetilde{\kalice}$ \gcom{$\gamenski{6,b}$}\label[line]{line:nski:enc2}
      \item $\pcif \sid = \sinit \pcand \calice \ne \widetilde{\calice}$
      \item $\quad \kalice' := \DecapsStat(\widetilde{\sk_i}, \calice)$ \gcom{$\gamenski{6,b}$}\label[line]{line:nski:enc3}
      \item $\keph := \DecapsEph(\sk_e, \ceph)$
      \item $\kpqc := \KDF(\kbob, \kalice', \keph')$
      \item $\pcif \sessionID = \sinit$ 
      \item $\quad \kpqc := \widetilde{\kpqc}$ \gcom{$\gamenski{8,b}$}\label[line]{line:nski:pqcRand2}
      \item $\kqkd := \deckey(\sid, \kid)$
      \item $\pcif \kqkd = \bot: \pcabort$
      \item $t := (m_1, (\calice, \ceph, \kid))$
      \item $\kqkdm \concat \kqkds := \kqkd$
      \item $\kpqcm \concat \kpqcs := \kpqc$
      \item $\ksess := \kqkds \oplus \kpqcs$
      \item $\tau_1 := \QKDMAC_{\kqkdm}((t, i, j))$
      \item $\tau_2 := \PQCMAC_{\kpqcm}((t, \tau_1, i, j))$
      \item $\pcif \tau_1 \neq \tau_1' \pcor \tau_2 \neq \tau_2': \pcabort$
      \item $\sessionkeyArray[\sid] := \ksess$
      \item $\state[\sid] := \Accept$ \gcom{ clear revealable state}
    \end{nicodemus}
  \end{minipage}}
  \caption{(cont.) $\gamenski{3,b}$ - $\gamenski{8,b}$. Any oracle not included is unchanged.}
\end{figure}

\emph{Games $\gamenski{3,b}$ - $\gamenski{8,b}$ are defined in \cref{fig:Gnski}.}
Games $\game{3,b}$ and $\gamenski{3,b}$ are the same for both $b$, so
\begin{multline*}\left| \Pr[\gameb{3,1} \Rightarrow 1 \wedge \neg\text{sk}(i)] - \Pr[\gameb{3,0}\Rightarrow 1 \wedge \neg\text{sk}(i)]\right|
\\=\left| \Pr[\gamenski{3,1}\Rightarrow 1 \wedge \neg\text{sk}(i)] -
\Pr[\gamenski{3,0}\Rightarrow 1 \wedge \neg\text{sk}(i)]\right|.\end{multline*}

\emph{Games $\gamenski{4,b}$}
We add the condition $\neg\text{sk}(i)$ in game $\gamenski{4,b}$ at \cref{line:nski:initNotCor}, to ensure we are in the correct case. Since 
\[\Pr[\gamenski{3,b}\Rightarrow 1 \wedge \neg\text{sk}(i)]=
\Pr[\gamenski{4,b}\Rightarrow 1]\]
for both $b$, which gives
\begin{equation}\label{eq:gamenski34}
\left| \Pr[\gamenski{3,1}\Rightarrow 1 \wedge \neg\text{sk}(i)] -
\Pr[\gamenski{3,0}\Rightarrow 1 \wedge \neg\text{sk}(i)]\right|=
\left| \Pr[\gamenski{4,1}\Rightarrow 1] - \Pr[\gamenski{4,0}\Rightarrow 1]\right|.
\end{equation}

\emph{Games $\gamenski{5,b}$}
For $\gamenski{5,b}$ we guess which party is the owner of $\sidinit$. 
In line~\ref{line:nski:guessParty} we add the guess for the owner of $\sidinit$,
and in line~\ref{line:nski:guessParty1} we check if it is correct.
If the party is guessed incorrectly, the game returns $0$.
If the party is guessed correctly, the game returns $1$ if and only if
$\gamenski{4,b}$ would have returned $1$.
Since the party is guessed uniformly at random, we get
\begin{equation}\label{eq:gamenski45}
\Pr[\gamenski{4,b}\Rightarrow 1]=\numberParties\cdot \Pr[\gamenski{5,b}\Rightarrow 1]\enspace.
\end{equation}
\emph{Games $\gamenski{6,b}$}
For $\gamenski{6,b}$ we move the call $\EncapsStat$ of the
static $\KEMstat$ to the $\orEst$ oracle, \cref{line:nski:enc}.
These values then replace the static key and ciphertext in \cref{line:nski:enc1,line:nski:enc2}.
This change is only noticed by the distinguisher
if $\DecapsStat$ returns the wrong key. 
Since the probability of returning the wrong key is at most $\dstat$ (see \cref{def:KemCorrectness}), 
we have
\begin{equation}\label{eq:gamenski56}
\left|\Pr[\gamenski{5,b}\Rightarrow 1]-\Pr[\gamenski{6,b}\Rightarrow 1]\right|\le \dstat.
\end{equation}

\emph{Games $\gamenski{7,b}$}
In this game, we replace Bob's static key of the test sessions generated by $\EncapsStat$
with a uniformly random key, in \cref{line:nski:statRand}.
To justify replacing the static key $\kbob$ with a uniformly random key,
we now relate the potential additional advantage to \INDCCA security of the static \KEM.
We do this by showing that we can use any adversary $C$ that distinguishes between $\gamenski{5,b}$
and $\gamenski{6,b}$ to create an adversary $B_2$ of the static \KEM that has 
a runtime about that of $C$.
The adversary $C$ gets access to either $\gamenski{5,b}$ or $\gamenski{6,b}$ and then has
to output either $0$ or $1$.
In the $\INDCCA$ game, the input to $B_2$ consists of $\pk$, $c$ and $k_i$,
where $\pk$ and $c$ are generated by the $\KeyGenStat$ and $\EncapsStat$ respectively,
and $k_i$ is generated by $\EncapsStat$ if $i=0$, and uniformly random if $i=1$.
This attacker $B_2$ can use the distinguisher $C$ as follows: 
replace the $\widetilde{\pk_b}$, $\widetilde{\cbob}$ and $\widetilde{\kbob}$
after \cref{line:nski:statRand} with the inputs of $B_2$. Since these values were generated with
the same algorithms, this does not change their distribution.
Furthermore, any call to $\DecapsStat$ using key $\widetilde k_b$ gets replaced by
a call to the $\Decaps$ oracle of the $\IND-\CCA$ game.
Now $C$, with access to these algorithms, outputs $0$ or $1$.
Let $B_2$ then give the same output as $C$. 

The difference between $\gamenski{6,b}$ and $\gamenski{7,b}$ is whether $\widetilde{\kbob}$ is
generated by $\EncapsStat$ or uniformly random.
Since $C$ distinguishes between the honestly generated key and a uniformly random key,
replacing this key with the key $k_i$ for which we have the same question
allows distinguishing an honestly generated key from a uniformly random key for
the $\INDCCA$ game.
Since this is true for all adversaries $C$, we have
\begin{equation}\label{eq:gamenski67}
	\left|\Pr[\gamenski{6,b} \Rightarrow 1] - \Pr[\gamenski{7,b}
  \Rightarrow 1]\right| \le \Adv_{\KEMstat}^{\INDCCA}(B_2) \enspace.
\end{equation}

\emph{Games $\gamenski{8,b}$}
Here, we replace the output of the $\KDF$ of the test session with a uniformly random
value. The value is generated at \cref{line:nski:pqcRand} and used at \cref{line:nski:pqcRand1,line:nski:pqcRand2}.

Since we model $\KDF:\{0,1\}^{2\lenstat + \leneph}\to \{0,1\}^{\lenpqc}$ as a random oracle, we can use one-way to hiding (\cref{lem:ow2h})
to bound the distance between $\gamenski{7,b}$ and $\gamenski{8,b}$.
Define $H(\cdot) = \KDF(\kalice,\cdot,\keph)$ where $\kalice$ and $\keph$ are the
keys used in session $\sinit$.
Then for any fixed $\kalice$, $\keph$ we can replace sampling
$\KDF\uni (\{0,1\}^{2\lenstat + \leneph}\to \{0,1\}^{\lenpqc})$ with sampling
$H\uni (\{0,1\}^{\lenstat} \to \{0,1\}^{\lenpqc})$.
Let $B^{H}(x,y)$ be the execution of $\gamenski{7,b}$
where we replace $\kbob$ of the test session with $x$,
and $\kpqc$ of the test session with $y$.
Then
\[
	\Pr[\gamenski{7,b} \Rightarrow 1] = 
  \Pr[b' = 1 : b'\gets B^{H}(x,H(x))]
\]
and
\[
	\Pr[\gamenski{8,b} \Rightarrow 1] =
  \Pr[b' = 1 : y\uni \{0,1\}^{\lenpqc},b' \gets B^{H}(x,y)]
\]
where we also take the probability over $H\uni (\{0,1\}^{\lenstat} \to \{0,1\}^{\lenpqc})$
and $x\uni \{0,1\}^{\lenstat}$.
As in \cref{lem:ow2h}, let $C$ be an oracle algorithm that on input $x$ does the following: pick
$i\uni \{1,...,q_h\}$ and $y\uni \{0,1\}^m$, run $B^{H}(x,y)$ until just before
the $i$-th query, measure the argument of the query in the computational basis,
and output the measurement outcome.
We define
\[P_C=\Pr[x=x':x'\gets C^{H}(x)]\]
where we also take the probability over $H\uni (\{0,1\}^{\lenstat}\to \{0,1\}^{\lenpqc})$ and $x\uni\{0,1\}^{\lenstat}$.
But since $x$ is not available to $C$ at any point, the best the attacker can do is guess,
which means that $P_C=\frac{1}{2^{\lenstat}}$.

Then by \cref{lem:ow2h} we have
\begin{equation}\label{eq:gamenski78}
	\left|\Pr[\gamenski{7,b} \Rightarrow 1] - 
	\Pr[\gamenski{8,b}\Rightarrow 1]\right|\le 2q_H\sqrt{P_C}=2q_H\frac{1}{\sqrt{2^{\lenstat}}}.
\end{equation}

Next, we want to bound
\[\left|\Pr[\gamenski{8,0}\Rightarrow 1] -
\Pr[\gamenski{8,1}\Rightarrow 1]\right|.\]
Key $k_0^*$ is generated as a part $\kpqc \oplus \kqkd$, where $\kpqc$ is uniformly random.
Furthermore, this key is only used in the matching session of the test session, which 
the adversary is not allowed to reveal.
Key $k_1^*$ is generated uniformly random. Therefore the adversary cannot distinguish between
$\gamenski{8,0}$ and $\gamenski{8,1}$, and we have
\begin{equation} \label{eq:game8equal}
  \Pr[\gamenski{8,0}\Rightarrow 1]=
\Pr[\gamenski{8,1}\Rightarrow 1].
\end{equation}
Combining equations \eqref{eq:gamenski56}, \eqref{eq:gamenski67} and \eqref{eq:gamenski78}, we get that
\[\left|\Pr[\gamenski{5,b} \Rightarrow 1] - 
\Pr[\gamenski{8,b}\Rightarrow 1]\right|\le \dstat + \Adv_{\KEMstat}^{\INDCCA}(B_2) + 2q_H\frac{1}{\sqrt{2^{\lenstat}}}.\]
Combining this with \eqref{eq:game8equal}, we get that
\[\left|\Pr[\gamenski{5,0} \Rightarrow 1] - 
\Pr[\gamenski{5,1}\Rightarrow 1]\right|\le 2\dstat + 2\Adv_{\KEMstat}^{\INDCCA}(B_2) + 4q_H\frac{1}{\sqrt{2^{\lenstat}}}\]
which, when combined with equations \eqref{eq:gamenski34} and \eqref{eq:gamenski45}, gives
\begin{multline}\label{eq:case1b}
  \left|\Pr[\gamenski{3,0} \wedge \neg sk(i)\Rightarrow 1] - 
\Pr[\gamenski{3,1} \wedge \neg sk(i)\Rightarrow 1]\right|
\\\le 2\numberParties\left(\dstat + \Adv_{\KEMstat}^{\INDCCA}(B_2) + 2q_H\frac{1}{\sqrt{2^{\lenstat}}}\right)
\end{multline}

\subsubsection{Combining \texorpdfstring{\cref{case1a}}{Case 1a} and \texorpdfstring{\cref{case1b}}{Case 1b}}%
Combining \cref{eq:case1a,eq:case1b} gives
\begin{multline*}
\left|\Pr[\gamenski{3,0}\Rightarrow 1] - 
\Pr[\gamenski{3,1}\Rightarrow 1]\right| \\
\le 
2\deph + 2\Adv_{\KEMeph}^{\INDCPA}(B_1) + 4q_H\frac{1}{\sqrt{2^{\leneph}}}+
2\numberParties\left(\dstat + \Adv_{\KEMstat}^{\INDCCA}(B_2) + 2q_H\frac{1}{\sqrt{2^{\lenstat}}}\right)
\enspace .
\end{multline*}
If we fill this in \cref{eq:case1init}, we have proven the claim of \cref{lem:case1}
\[
\left|\Pr[\INDStAAPQC_1(\Pi, A) \Rightarrow 1 \wedge \matchingSessions\ne \emptyset] - 
\Pr[\INDStAAPQC_0(\Pi, A) \Rightarrow 1 \wedge \matchingSessions\ne \emptyset]\right|\]\[\le
\numberSessions\!^2\cdot\left|\Pr[\gameb{3,1}\Rightarrow 1]-\Pr[\gameb{3,0}\Rightarrow 1]\right|+2\cdot \numberSessions \cdot \muEncStat
\]\[
\le
2\numberSessions\!^2\cdot\left(
  \deph + \Adv_{\KEMeph}^{\INDCPA}(B_1) + 2q_H\frac{1}{\sqrt{2^{\leneph}}}+
  \numberParties\left(\dstat + \Adv_{\KEMstat}^{\INDCCA}(B_2) + 2q_H\frac{1}{\sqrt{2^{\lenstat}}}\right)
\right)\]\[+2\cdot \numberSessions \cdot \muEncStat
\]

\refstepcounter{pfcase}\label{case2}%
\subsection{Security proof for \texorpdfstring{\cref*{case2}}{Case 2}: active attack, \texorpdfstring{\cref{lem:case2}}{Lemma 5.4}}\label{sec:case2}
We consider the case that $\matchingSessions = \emptyset$,
so we are proving an upper bound for:
\begin{multline*}
  | \Pr[\INDStAAPQC_0(\Pi, A) \Rightarrow 1 \land \matchingSessions = \emptyset] \\
  - \Pr[\INDStAAPQC_1(\Pi, A) \Rightarrow 1 \land \matchingSessions = \emptyset] |.
\end{multline*}
\begin{proof}
We consider the initiator and responder session separately,
but due to the symmetry in the protocol the proofs are very similar.
We consider only adversaries for which $\Attack(\matchingSessions)$ does not hold.
We know $\matchingSessions = \emptyset$ by assumption of the case,
so that by definition of \Attack, we have
\begin{multline*}
  \neg \Attack(\emptyset)
  \Leftrightarrow \\ 
  \neg \revealedState[\sid^*]
  \land \sessionkeyArray[\sid^*] \neq \bot
  \land \neg \revealed[\sid^*]
  \land \neg \corrupted[\peer[\sid^*]].
\end{multline*}
In other words, the encapsulation to the peer's static key
leaks neither directly via a state reveal,
nor indirectly via corruption of the peer.
We prove this by reduction to the \INDCCA security of $\KEMstat$.
There may be a partially matching session that computes the same $\kpqc$
(we consider that in \cref{case2a1}).
Since we are considering active attacks,
the full transcript must differ somewhere between these sessions,
so (unless a forgery against $\PQCMAC$ was made),
the initiator will detect this difference and abort.
If the test session is a responder,
then this means that its value $\kpqcs$ is independent of the adversary view
and therefore the session key is indistinguishable from uniformly random.
If there does not exist a partially matching session
(we consider this in \cref{case2a2}),
then $\kpqcs$ looks random for the adversary
and therefore so does the session key.

\nicoresetlinenr
\begin{figure}
  \fbox{\begin{minipage}{\dimexpr\textwidth-2\fboxsep-2\fboxrule\relax}\begin{multicols}{2}
    \underline{{\bf GAME} $\gamenm{1,b}$ - $\gamenm{12,b}$}
    \begin{nicodemus}
    \item $\sidctr := 0$
    \item $\sid^* := 0$
    \item $s' \uni \set{1, \dots, \numberSessions}$ \gcom{ \gamenm{2,b}}
    \item $s'' \uni \set{1, \dots, \numberSessions} \setminus \set{s'}$ \gcom{ \gamenmik{9,b}}
    \item $j' \uni \set{1, \dots, \numberParties}$ \gcom{ \gamenm{3,b}}
    \item $\pcfor i \in \set{1, \dots, \numberParties}$
    \item $\quad (\pk_i, \sk_i) \gets \KeyGenStat()$
    \item $\qkdInit()$
    \item $(\widetilde{\cbob}, \widetilde{\kbob}) := \EncapsStat(\pk_{j'})$ \gcom{ \gamenmi{5,b}}\label[line]{line:nmi-encaps-sidstar}
    \item $\widetilde{\kbob} \uni \{0,1\}^{\lenstat}$ \gcom{ \gamenmi{6,b}}
    \item $\widetilde{\kpqc} \uni \{0,1\}^{\lenpqc}$ \gcom{ \gamenmik{11,b}}
    \item $b' \gets A^\Oracle(\pk_1, \dots, \pk_{\numberParties})$\label[line]{line:nmi-calla}
    \item $\matchingSessions := \FindMatches(\sid^*)$
    \item $\pcif \matchingSessions \neq \emptyset: \pcreturn 0$ \gcom{ $\gamenm{1,b}$}
    \item $\pcif \Attack(\matchingSessions)$
    \item $\quad \pcreturn 0$
    \item $\pcif s' \neq \sid^*: \pcreturn 0$ \gcom{ $\gamenm{2,b}$}
    \item $\pcif j' \neq \peer[\sid^*]: \pcreturn 0$ \gcom{ $\gamenm{3,b}$}
    \item $\pcif \role[\sid^*] \neq \roleInit$
    \item $\quad \pcreturn 0$ \gcom{ $\gamenmi{4,b}$}

    \item $\kdfMatch = \false$
    \item $\pcfor \sid \in [S] \setminus \set{\sid^*}$
    \item $\quad \pcif \kdfIn[\sid] = \kdfIn[\sid^*]$
    \item $\qquad \kdfMatch = \true$
    \item $\pcif \neg\kdfMatch: \pcreturn 0$ \gcom{ \gamenmik{8,b}} \label[line]{line:nmi-not-kdfMatch}

    \item $\pcif \kdfIn[s''] \neq \kdfIn[\sid^*]$
    \item $\quad \pcreturn 0$ \gcom{ \gamenmik{9,b}}
    \item $\pcfor \sid \in \set{1, \dots, \numberSessions} \setminus \set{s'', \sid^*}$
    \item $\quad \pcif \kdfIn[\sid] = \kdfIn[s'']$
    \item $\qquad \pcreturn 0$ \gcom{ \gamenmik{10,b}}\label[line]{line:kdfUnique}
    \item $\pcreturn b'$
    \end{nicodemus}

    \vspace{1em}
    \underline{$\orSendInit(\sid)$}
    \begin{nicodemus}
      \item $(i, j) := (\owner[\sid], \peer[\sid])$
      \item $(\rho, s) := (\role[\sid], \state[\sid])$
      \item $\pcif i = \bot \pcor \rho \neq \roleInit \pcor s \neq \bot$
      \item $\quad \pcabort$

      \item $(\cbob, \kbob) := \EncapsStat(\pk_j)$\label[line]{line:nmi-encaps}
      \item $\pcif \sid = s'$
      \item $\quad (\cbob, \kbob) := (\widetilde{\cbob}, \widetilde{\kbob})$ \gcom{ \gamenmi{5,b}}
      \item $(\pk_e, \sk_e) := \KeyGenEph()$
      \item $m_1 := (\cbob, \pk_e)$
      \item $\sent[\sid] := m_1$
      \item $\state[\sid] := (\kbob, \sk_e, m_1)$
      \item $\pcreturn m_1$
    \end{nicodemus}

    \vspace{1em}
    \underline{$\orSendMOne(\sid, m_1)$}
    \begin{nicodemus}
      \item $(i, j) := (\owner[\sid], \peer[\sid])$
      \item $(\rho, s) := (\role[\sid], \state[\sid])$
      \item $\pcif i = \bot \pcor \rho \neq \roleResp \pcor s \neq \bot$
      \item $\quad \pcabort$
      \item $\received[\sid] := m_1$

      \item $(\cbob, \pk_e) := m_1$
      \item $\kbob' := \DecapsStat(\sk_i, \cbob)$\label[line]{line:nmi-decaps}

      \item $\pcif i = j' \pcand \cbob = \widetilde{\cbob}$
      \item $\quad \kbob' := \widetilde{\kbob}$ \gcom{ \gamenmi{5,b}}

      \item $(\calice, \kalice) \gets \EncapsStat(\pk_j)$
      \item $(\ceph, \keph) \gets \EncapsEph(\pk_e)$

      \item $\kdfIn[\sid] = (\kbob', \kalice, \keph)$ \gcom{ \gamenmi{7,b}}

      \item $\kpqc := \KDF(\kbob', \kalice, \keph)$
      \item $\pcif \sid = s''$
      \item $\quad \kpqc := \widetilde{\kpqc}$ \gcom{ \gamenmik{11,b}}

      \item $(\kid, \kqkd) := \enckey(\sid, \lenqkd)$
      \item $t := (m_1, (\calice, \ceph, \kid))$

      \item $(\kpqcm \concat \kpqcs) := \kpqc$
      \item $(\kqkdm \concat \kqkds) := \kqkd$
      \item $\ksess := \kpqcs \oplus \kqkds$

      \item $\tau_1 = \QKDMAC_{\kqkdm}((t, j, i))$
      \item $\tau_2 = \PQCMAC_{\kpqcm}((t, \tau_1, j, i))$\label[line]{line:mac2responder}

      \item $m_2 := (\calice, \ceph, \kid, \tau_1, \tau_2)$

      \item $\sent[\sid] := m_2$
      \item $\sessionkeyArray[\sid] := \ksess$
      \item $\state[\sid] := \Accept$
      \item $\pcreturn m_2$
    \end{nicodemus}

    \vspace{1em}
    \underline{$\orSendMTwo(\sid, m_2)$}
    \begin{nicodemus}
      \item $(i, j) := (\owner[\sid], \peer[\sid])$
      \item $(\rho, s) := (\role[\sid], \state[\sid])$
      \item $\pcif i = \bot \pcor \rho \neq \roleInit$
      \item $\quad \pcabort$
      \item $\received[\sid] := m_2$

      \item $(\calice, \ceph, \kid, \tau_1', \tau_2') := m_2$
      \item $(\kbob, \sk_e, m_1) := s$
      \item $\kalice' := \DecapsStat(\sk_i, \calice)$
      \item $\keph' := \DecapsEph(\sk_e, \ceph)$

      \item $\pcif \sid = s'$
      \item $\quad \kdfIn[\sid] = (\kbob, \kalice', \keph')$ \gcom{ \gamenmi{7,b}}

      \item $\kpqc := \KDF(\kbob, \kalice', \keph')$
      \item $\pcif \sid = s'$
      \item $\quad \kpqc := \widetilde{\kpqc}$ \gcom{ \gamenmik{11,b}}

      \item $\kqkd := $

        $\hspace{\fill}\deckey(\sid, \kid)$
      \item $\pcif \kqkd = \bot: \pcabort$
      \item $t := (m_1, (\calice, \ceph, \kid))$

      \item $(\kpqcm \concat \kpqcs) := \kpqc$
      \item $(\kqkdm \concat \kqkds) := \kqkd$
      \item $\ksess := \kpqcs \oplus \kqkds$

      \item $\tau_1 = \QKDMAC_{\kqkdm}((t, i, j))$
      \item $\tau_2 = \PQCMAC_{\kpqcm}((t, \tau_1, i, j))$\label[line]{line:mac2initiator}

      \item $\pcif \sid = s': \pcabort$ \gcom{ \gamenmik{12,b}}
      \item $\pcif \tau_1 \neq \tau_1' \pcor \tau_2 \neq \tau_2': \pcabort$\label[line]{line:verifytags}

      \item $\sessionkeyArray[\sid] := \ksess$
      \item $\state[\sid] := \Accept$
    \end{nicodemus}
\end{multicols}\end{minipage}}
\caption{Game $\gamenm{\cdot, b}$: non-matching sessions, initiator session, matching $\kdfIn$ exists}%
\label{fig:proof_nomatch}
\end{figure}

The game hops are given in \cref{fig:proof_nomatch}.
First we formalize the assumption of \cref{case2}:
we encode $\matchingSessions = \emptyset$
in \gamenm{1,b}, so that
\begin{equation}\label{eq:c2h1}
  \Pr[\INDStAAPQC_b(\Pi, A) \Rightarrow 1 \land \matchingSessions = \emptyset]
  = \Pr[\gamenm{1,b} \Rightarrow 1].
\end{equation}

Next, \gamenm{2,b} makes a random guess $s'$ for the test session $\sid^*$,
and since $\Pr[s' = \sid^*] = 1/\numberSessions$ and we return 0 if the guess was wrong:
\begin{equation}\label{eq:c2h2}
  \Pr[\gamenm{1,b} \Rightarrow 1] = \numberSessions \Pr[\gamenm{2,b} \Rightarrow 1]. 
\end{equation}
\gamenm{3,b} makes a random guess $j'$ for the test session $\peer[\sid^*]$,
and since $\Pr[j' = \peer[\sid^*]] = 1/\numberParties$:
\begin{equation}\label{eq:c2h3}
  \Pr[\gamenm{2,b} \Rightarrow 1] = \numberParties \Pr[\gamenm{3,b} \Rightarrow 1].
\end{equation}

\refstepcounter{pfsubcase}\label{case2a}%
\heading{\Cref*{case2a}: Test session is an initiator.}%
We first consider the case that $\role[\sid^*] = \roleInit$,
encoded in \gamenmi{4,b}:
\begin{equation}\label{eq:c2ah4}
  \Pr[\gamenm{3,b} \Rightarrow 1 \land \role[\sid^*] = \roleInit]
  = \Pr[\gamenmi{4,b} \Rightarrow 1].
\end{equation}

In game $\gamenmi{5,b}$, we sample the test-session encapsulation in advance,
and we also replace the decapsulated value with the encapsulated one.
The games are identical unless a decapsulation failure occurs in $\gamenmi{4,b}$,
which by \cref{def:KemCorrectness} happens with probability at most $\deltastat$, so we have
\begin{equation}\label{eq:c2ah5}
  | \Pr[\gamenmi{4,b} \Rightarrow 1] - \Pr[\gamenmi{5,b} \Rightarrow 1] | \leq \deltastat.
\end{equation}

Game $\gamenmi{6,b}$ samples a uniformly random $\widetilde{\kbob}$
and overwrites the encapsulated value.
Given an adversary $D$ that distinguishes
$\gamenmi{5,b}$ from $\gamenmi{6,b}$
with advantage $\Adv(D)$,
we define an adversary $B_3$ that uses $D$
and wins the \INDCCA game of $\KEMstat$ with $\Adv_{\KEMstat}^{\INDCCA}(B_3) = \Adv(D)$.

Adversary $B_3$ gets input $(\hat{\pk}, \hat{c}, \hat{k})$
and can query the decapsulation oracle $\oracleDecaps(\hat{\sk}, \cdot)$
on all inputs except $\hat{c}$.
$B_3$ wins if they can correctly say whether $\hat{c}$ is an encapsulation of $\hat{k}$
or if $\hat{k}$ is a uniformly random value.
$B_3$ runs \gamenmi{5,b}
except that they replace the public key of the guessed session $\pk_{j'}$ with $\hat{\pk}$
in \cref{line:nmi-calla,line:nmi-encaps},
they replace $(\widetilde{\cbob}, \widetilde{\kbob})$ with $(\hat{c}, \hat{k})$ on \cref{line:nmi-encaps-sidstar},
and instead of decapsulating with $\sk_{j'}$ they query the oracle in \cref{line:nmi-decaps}
(making sure to skip the query on input $\hat{c}$ and directly output $\hat{k}$ instead).
Finally, $B_3$ outputs whatever $D$ outputs.

If $B_3$ receives a real encapsulation they are running \gamenmi{5,b},
otherwise $\hat{k}$ was random and they are running \gamenmi{6,b}.
We also see that $B_3$ is an \INDCCA adversary,
so that
\begin{equation}\label{eq:c2ah6}
  | \Pr[\gamenmi{5,b} \Rightarrow 1]
  - \Pr[\gamenmi{6,b} \Rightarrow 1] | \leq \Adv_{\KEMstat}^{\INDCCA}(B_3).
\end{equation}

To prepare for the next case distinction,
in \gamenmi{7,b} we record the input to the \KDF in the $\kdfIn$ associative array.
Since this bookkeeping does not change the adversary view, we have
\begin{equation}\label{eq:c2ah7}
  \Pr[\gamenmi{6,b} \Rightarrow 1] = \Pr[\gamenmi{7,b} \Rightarrow 1].
\end{equation}

\heading{\Cref*{case2a1}. A KDF input match exists.}
\refstepcounter{pfsubsubcase}\label{case2a1}
In this case, we assume that there exists an honest session $\sid'$ such that
$\sid' \neq \sid^*$ and $\kdfIn[\sid^*] = \kdfIn[\sid']$.
We formalize this condition in \gamenmi{8,b} on \cref{line:nmi-not-kdfMatch}.
\begin{equation}\label{eq:c2hop8}
  \Pr[\gamenmi{7,b} \Rightarrow 1 \land \exists \sid' \neq \sid^*: \kdfIn[\sid^*] = \kdfIn[\sid']]
  = \Pr[\gamenmik{8,b} \Rightarrow 1].
\end{equation}

Game \gamenmik{9,b} makes a guess $s''$ for the session such that
$\kdfIn[s''] = \kdfIn[\sid^*]$,
and we return 0 if our guess turned out to be wrong.
Since $\gamenmik{8,b}$ established that such a session must exist
and our guess is uniformly random over all sessions other than $\sid^*$, we have that
\begin{equation}\label{eq:c2hop9}
  \Pr[\gamenmik{8,b} \Rightarrow 1 ]
  = (\numberSessions-1) \Pr[\gamenmik{9,b} \Rightarrow 1]
\end{equation}

Game \gamenmik{10,b} establishes that $s''$ is unique.
Game \gamenmik{9,b} and \gamenmik{10,b} are identical
unless \gamenmik{10,b} returns 0 on \cref{line:kdfUnique}.
A collision in $\kdfIn$ requires at least a collision in $\kalice$,
but by \cref{def:MuSec} we know that such a collision occurs with
probability at most $\muSecStat$.
There are at most $\numberSessions-2$ sessions that can collide,
so (also by the difference lemma) we get
\begin{multline}\label{eq:c2hop10}
  | \Pr[\gamenmik{9,b} \Rightarrow 1]
  - \Pr[\gamenmik{10,b} \Rightarrow 1]| \\
  \leq \Pr[\gamenmik{10,b} \text{ returns on \cref{line:kdfUnique}}]
  \leq (\numberSessions-2)\muSecStat
  \leq \numberSessions\muSecStat.
\end{multline}

Game \gamenmik{11,b} replaces $\kpqc$ with the same uniformly random sampled value
in both $\sid^*$ and $s''$.
Since we model $\KDF:\{0,1\}^{2\lenstat + \leneph}\to \{0,1\}^{\lenpqc}$ as a random oracle, we can use the one-way to hiding \cref{lem:ow2h}.
Similarly to the hop from $\gamenst{4,b}$ to $\gamenst{5,b}$, we
define $H(\cdot)=\KDF(\kalice,\cdot,\keph)$ where $\kalice$ and
$\keph$ are the keys used in $\kdfIn[s'']$.

Then for any fixed $\kalice$, $\keph$ we can replace sampling
$\KDF\uni (\{0,1\}^{2\lenstat + \leneph}\to \{0,1\}^{\lenpqc})$ with sampling
$H\uni (\{0,1\}^{\lenstat}\to \{0,1\}^{\lenpqc})$.
Let $B^{H}(x,y)$ be the execution of $\gamenmik{10,b}$
where we replace $\widetilde{\kbob}$ of the test session with $x$,
and $\widetilde{\kpqc}$ of the test session with $y$.
Then
\[
	\Pr[\gamenmik{10,b} \Rightarrow 1] = 
  \Pr[b'=1:b' \gets B^{H}(x,H(x)) ]
\]
and
\[\Pr[\gamenmik{11,b} \Rightarrow 1] =
\Pr[b'=1:b'\gets B^{H'}(x,y),y\uni \{0,1\}^{\lenpqc}]\]
where we also take the average over $H\uni (\{0,1\}^{\lenstat} \to \{0,1\}^{\lenpqc})$ and $x\uni \{0,1\}^{\lenstat}$.
As in \cref{lem:ow2h}, let $C$ be an oracle algorithm that on input $x$ does the following: pick
$i\uni \{1,...,q_h\}$ and $y\uni \{0,1\}^{\lenpqc}$, run $B^{H}(x,y)$ until just before
the $i$-th query, measure the argument of the query in the computational basis,
and output the measurement outcome.

We define
\[P_C=\Pr[x=x':x'\gets C^{H}(x)]\]
where the probability is also over $H\uni (\{0,1\}^{\lenstat}\to \{0,1\}^{\lenpqc})$ and $x\uni \{0,1\}^{\lenstat}$.
But since $x$ is not available to $C$ at any point, the best the attacker can do is guess,
which means that $P_C=\frac{1}{2^{\lenstat}}$.

Then by \cref{lem:ow2h} we have
\begin{equation}\label{eq:c2hop11}
  \left|\Pr[\gamenmik{10,b} \Rightarrow 1] - 
    \Pr[\gamenmik{11,b}\Rightarrow 1]\right|
  \le 2q_H\sqrt{P_C}=\frac{2q_H}{\sqrt{2^{\lenstat}}}.
\end{equation}

Game \gamenmik{12,b} always aborts the test session.
To prove this is indistinguishable from \gamenmik{11,b} we consider two cases,
based on whether the transcript matches on all values before the second MAC tag.
Specifically,
let $\sent[\sid^*] = m_1^*$ and $\received[\sid^*] = m_2^* = (\calice^*, \ceph^*, \kid^*, \tau_1^*, \tau_2^*)$
be the transcript of the test session,
then we denote $t^* = (m_1^*, (\calice^*, \ceph^*, \kid^*))$.
Similarly we write $(t'', \tau_1'', \tau_2'')$ for the values recorded in 
$\sent[s'']$ and $\received[s'']$.

\heading{\Cref*{case2a1a}.}%
\refstepcounter{pfsubsubsubcase}\label{case2a1a}
In this case the session $s''$ \emph{almost} matches the test session $\sid^*$:
the transcript matches up to and including the first tag,
and both sessions have indicated each other as their peers.
Specifically in this case $(t^*, \tau_1^*, \owner[\sid^*], \peer[\sid^*]) = (t'', \tau_1'', \peer[s''], \owner[s''])$.
The tag $\tau_2'' =\allowbreak \PQCMAC_{\kpqcm}((t'', \tau_1'',\allowbreak \peer[s''],\allowbreak \owner[s'']))$ was generated in session $s''$.
Since $s''$ is not a matching session of $\sid^*$,
it must hold that $\sid^*$ received a different tag: $\tau_2^* \neq \tau_2''$.
But since $\kpqc$ (and therefore $\kpqcm$) has the same value in both $\sid^*$ and $s''$,
both the key and message of the $\PQCMAC$ are identical
and $\sid^*$ will recompute $\tau_2''$ during verification, and therefore reject $\tau_2^*$.
In this case, $\sid^*$ will therefore abort with certainty:
\begin{multline}\label{eq:2a1a}
  \Pr[\gamenmik{11,b} \Rightarrow 1 \land (t^*, \tau_1^*, \owner[\sid^*], \peer[\sid^*]) = (t'', \tau_1'', \peer[s''], \owner[s''])]
  \\ = \Pr[\gamenmik{12,b} \Rightarrow 1].
\end{multline}

\heading{\Cref*{case2a1b}.}%
\refstepcounter{pfsubsubsubcase}\label{case2a1b}
In this case the session $s''$ differs from $\sid^*$ somewhere in the partial transcript,
or the sessions do not indicate each other as peers.
This means the input to $\PQCMAC_{\kpqcm}$ will be different,
from which we can give a reduction to the \OTSUFCMA security of $\PQCMAC$.
Specifically in this case $(t^*, \tau_1^*, \owner[\sid^*], \peer[\sid^*]) \neq (t'', \tau_1'', \peer[s''], \owner[s''])$.
Note that \gamenmik{11,b} and \gamenmik{12,b} are identical,
unless the failure event occurs
where \gamenmik{11,b} does \emph{not} abort session $\sid^*$ on \cref{line:verifytags},
so we bound the probability of this failure event.
Let $D$ be the adversary that triggers the failure event.
We construct the following adversary against the $\OTSUFCMA^{\PQCMAC, B_4}$ game.
$B_4 = (B_{4,0}, B_{4,1})$ runs \gamenmik{11,b} using $D$ as a subroutine, but with the following modifications:
in session $s''$ on \cref{line:mac2responder}, instead of computing \PQCMAC,
$B_{4,0}$ outputs $m = (t'', \tau_1'', \peer[s''], \owner[s''])$ and gets back a tag $\tau$ from the \OTSUFCMA game:
$B_4$ sets $\tau_2'' = \tau$ in \gamenmik{11,b}.
Then on $\orSendMTwo(s', m_2)$,
$B_{4,1}$ outputs $(m', \tau') = ((t^*, \tau_1^*, \owner[\sid^*], \peer[\sid^*]), \tau_2^*)$.
Note that both $\kpqcm$ in \gamenmik{11,b} and $k$ in the \OTSUFCMA game are uniformly random keys,
that $m \neq m'$ by the assumption of \cref{case2a1b},
and that $B_4$ created a valid forgery exactly when \gamenmik{11,b} would not abort, so that
\begin{multline*}
  \Pr\big[\gamenmik{11,b} \text{ does not abort on \cref{line:verifytags}} \land \\
  (t^*, \tau_1^*, \owner[\sid^*], \peer[\sid^*]) \neq (t'', \tau_1'', \peer[s''], \owner[s''])\big]
  = \Adv_{\PQCMAC}^{\OTSUFCMA}(B_4)
\end{multline*}
and thus by the difference lemma
\begin{multline}\label{eq:2a1b}
  \Big| \Pr\left[\gamenmik{11,b} \Rightarrow 1 \land
    (t^*, \tau_1^*, \owner[\sid^*], \peer[\sid^*]) \neq (t'', \tau_1'', \peer[s''], \owner[s''])\right]\\
  - \Pr[\gamenmik{12,b} \Rightarrow 1] \Big|
  \leq \Adv_{\PQCMAC}^{\OTSUFCMA}(B_4).
\end{multline}

We combine disjoint \cref{case2a1a,case2a1b} by taking the sum over \cref{eq:2a1a,eq:2a1b},
and we get
\begin{equation}\label{eq:c2hop12}
  | \Pr[\gamenmik{11,b} \Rightarrow 1]
  - \Pr[\gamenmik{12,b} \Rightarrow 1] |
  \leq \Adv_{\PQCMAC}^{\OTSUFCMA}(B_4).
\end{equation}

Since both \gamenmik{12,0} and \gamenmik{12,1} always abort the test session,
they are indistinguishable:
\begin{equation}\label{eq:c2-12}
  \Pr[\gamenmik{12,0} \Rightarrow 1] = \Pr[\gamenmik{12,1} \Rightarrow 1] = 0.
\end{equation}

Combining \cref{eq:c2hop8,eq:c2hop9,eq:c2hop10,eq:c2hop11,eq:c2hop12,eq:c2-12} concludes \cref{case2a1}, we get
\begin{multline}\label{eq:2a1}
  \big| \Pr[\gamenmi{7,0} \Rightarrow 1 \land \exists \sid' \neq \sid^*: \kdfIn[\sid^*] = \kdfIn[\sid']] \\
  - \Pr[\gamenmi{7,1} \Rightarrow 1 \land \exists \sid' \neq \sid^*: \kdfIn[\sid^*] = \kdfIn[\sid']] \big| \\
  \leq 2 (\numberSessions-1) \left( \numberSessions \muSecStat + \frac{2q_H}{\sqrt{2^{\lenstat}}} + \Adv_{\PQCMAC}^{\OTSUFCMA}(B_4) \right)
\end{multline}

\heading{\Cref*{case2a2}. No KDF input match exists.}%
\refstepcounter{pfsubsubcase}\label{case2a2}

\nicoresetlinenr
\begin{figure}
  \fbox{\begin{minipage}{\dimexpr\textwidth-2\fboxsep-2\fboxrule\relax}\begin{multicols}{2}
    \underline{{\bf GAME} $\gamenmi{7,b}$ - $\gamenmink{12,b}$}
    \begin{nicodemus}
    \item $\sidctr := 0$
    \item $\sid^* := 0$
    \item $s' \uni \set{1, \dots, \numberSessions}$ {\color{gray}\gcom{ \gamenm{2,b}}}
    \item $j' \uni [\numberParties]$ {\color{gray}\gcom{ \gamenm{3,b}}}
    \item $\pcfor i \in \set{1, \dots, \numberParties}$
    \item $\quad (\pk_i, \sk_i) \gets \KeyGenStat()$
    \item $\qkdInit()$
    \item $(\widetilde{\cbob}, \widetilde{\kbob}) := \EncapsStat(\pk_{j'})$ {\color{gray}\gcom{ \gamenmi{5,b}}}
    \item $\widetilde{\kbob} \uni \{0,1\}^{\lenstat}$ {\color{gray}\gcom{ \gamenmi{6,b}}}
    \item $\widetilde{\kpqc} \uni \{0,1\}^{\lenpqc}$ \gcom{ \gamenmink{9,b}}
    \item $\widetilde{\ksess} \uni \{0,1\}^{\lensess}$ \gcom{ \gamenmink{10,b}}
    \item $b' \gets A^\Oracle(\pk_1, \dots, \pk_{\numberParties})$
    \item $\matchingSessions := \FindMatches(\sid^*)$
    \item $\pcif \matchingSessions \neq \emptyset: \pcreturn 0$ {\color{gray}\gcom{ $\gamenm{1,b}$}}
    \item $\pcif \Attack(\matchingSessions)$
    \item $\quad \pcreturn 0$
    \item $\pcif s' \neq \sid^*: \pcreturn 0$ {\color{gray}\gcom{ $\gamenm{2,b}$}}
    \item $\pcif j' \neq \peer[\sid^*]: \pcreturn 0$ {\color{gray}\gcom{ $\gamenm{3,b}$}}
    \item $\pcif \role[\sid^*] \neq \roleInit: \pcreturn 0$ {\color{gray}\gcom{ $\gamenmi{4,b}$}}
    \item $\pcfor \sid \in \set{1, \dots, \numberSessions} \setminus \set{\sid^*}$
    \item $\quad \pcif \kdfIn[\sid] = \kdfIn[\sid^*]$
    \item $\qquad \pcreturn 0$ \gcom{ \gamenmink{8,b}}
    \item $\pcreturn b'$
    \end{nicodemus}

    \vspace{1em}
    \underline{$\orSendInit(\sid)$}
    \begin{nicodemus}
      \item $(i, j) := (\owner[\sid], \peer[\sid])$
      \item $(\rho, s) := (\role[\sid], \state[\sid])$
      \item $\pcif i = \bot \pcor \rho \neq \roleInit \pcor s \neq \bot$
      \item $\quad \pcabort$

      \item $(\cbob, \kbob) := \EncapsStat(\pk_j)$
      \item $\pcif \sid = s'$
      \item $\quad (\cbob, \kbob) := (\widetilde{\cbob}, \widetilde{\kbob})$ {\color{gray}\gcom{ \gamenmi{5,b}}}
      \item $(\pk_e, \sk_e) := \KeyGenEph()$
      \item $m_1 := (\cbob, \pk_e)$
      \item $\sent[\sid] := m_1$
      \item $\state[\sid] := (\kbob, \sk_e, m_1)$
      \item $\pcreturn m_1$
    \end{nicodemus}

    \vspace{1em}
    \underline{$\orSendMOne(\sid, m_1)$}
    \begin{nicodemus}
      \item $(i, j) := (\owner[\sid], \peer[\sid])$
      \item $(\rho, s) := (\role[\sid], \state[\sid])$
      \item $\pcif i = \bot \pcor \rho \neq \roleResp \pcor s \neq \bot$
      \item $\quad \pcabort$
      \item $\received[\sid] := m_1$

      \item $(\cbob, \pk_e) := m_1$
      \item $\kbob' := \DecapsStat(\sk_i, \cbob)$

      \item $\pcif i = j' \pcand \cbob = \widetilde{\cbob}$
      \item $\quad \kbob' := \widetilde{\kbob}$ {\color{gray}\gcom{ \gamenmi{5,b}}}

      \item $(\calice, \kalice) \gets \EncapsStat(\pk_j)$
      \item $(\ceph, \keph) \gets \EncapsEph(\pk_e)$

      \item $\kdfIn[\sid] = (\kbob', \kalice, \keph)$ {\color{gray}\gcom{ \gamenmi{7,b}}}

      \item $\kpqc := \KDF(\kbob', \kalice, \keph)$

      \item $(\kid, \kqkd) := \enckey(\sid, \lenqkd)$
      \item $t := (m_1, (\calice, \ceph, \kid))$

      \item $(\kpqcm \concat \kpqcs) := \kpqc$
      \item $(\kqkdm \concat \kqkds) := \kqkd$
      \item $\ksess := \kpqcs \oplus \kqkds$

      \item $\tau_1 = \QKDMAC_{\kqkdm}((t, j, i))$
      \item $\tau_2 = \PQCMAC_{\kpqcm}((t, \tau_1, j, i))$

      \item $m_2 := (\calice, \ceph, \kid, \tau_1, \tau_2)$

      \item $\sent[\sid] := m_2$
      \item $\sessionkeyArray[\sid] := \ksess$
      \item $\state[\sid] := \Accept$
      \item $\pcreturn m_2$
    \end{nicodemus}

    \vspace{1em}
    \underline{$\orSendMTwo(\sid, m_2)$}
    \begin{nicodemus}
      \item $(i, j) := (\owner[\sid], \peer[\sid])$
      \item $(\rho, s) := (\role[\sid], \state[\sid])$
      \item $\pcif i = \bot \pcor \rho \neq \roleInit$
      \item $\quad \pcabort$
      \item $\received[\sid] := m_2$

      \item $(\calice, \ceph, \kid, \tau_1', \tau_2') := m_2$
      \item $(\kbob, \sk_e, m_1) := s$
      \item $\kalice' := \DecapsStat(\sk_i, \calice)$
      \item $\keph' := \DecapsEph(\sk_e, \ceph)$

      \item $\pcif \sid = s'$
      \item $\quad \kdfIn[\sid] = (\kbob, \kalice', \keph')$ {\color{gray}\gcom{ \gamenmi{7,b}}}

      \item $\kpqc := \KDF(\kbob, \kalice', \keph')$
      \item $\pcif \sid = s'$
      \item $\quad \kpqc := \widetilde{\kpqc}$ \gcom{ \gamenmink{9,b}}

      \item $\kqkd :=$

        $\hspace{\fill}\deckey(\sid, \kid)$
      \item $\pcif \kqkd = \bot: \pcabort$
      \item $t := (m_1, (\calice, \ceph, \kid))$

      \item $(\kpqcm \concat \kpqcs) := \kpqc$
      \item $(\kqkdm \concat \kqkds) := \kqkd$
      \item $\ksess := \kpqcs \oplus \kqkds$

      \item $\pcif \sid = s'$
      \item $\quad \ksess := \widetilde{\ksess}$ \gcom{ \gamenmink{10,b}}

      \item $\tau_1 = \QKDMAC_{\kqkdm}((t, i, j))$
      \item $\tau_2 = \PQCMAC_{\kpqcm}((t, \tau_1, i, j))$

      \item $\pcif \tau_1 \neq \tau_1' \pcor \tau_2 \neq \tau_2': \pcabort$

      \item $\sessionkeyArray[\sid] := \ksess$
      \item $\state[\sid] := \Accept$
    \end{nicodemus}
\end{multicols}\end{minipage}}
\caption{Game $\gamenmik{\cdot, b}$: non-matching sessions, initiator session, matching $\kdfIn$ does not exist}%
\label{fig:proof_nmik}
\end{figure}
If no KDF input match exists,
then $\kdfIn[\sid] \neq \kdfIn[\sid^*]$
for all $\sid \neq \sid^*$.
We encode this in \gamenmink{8,b} in \cref{fig:proof_nmik}, so that
\begin{equation}\label{eq:c2a2h8}
  \Pr[\gamenmi{7,b} \Rightarrow 1 \land \forall \sid \neq \sid^*: \kdfIn[\sid] \neq \kdfIn[\sid^*]]
  = \Pr[\gamenmink{8,b} \Rightarrow 1].
\end{equation}

Game \gamenmink{9,b} replaces $\kpqc$ with a random value $\widetilde{\kpqc}$.
Since this step replaces a single QROM output,
identical to the gamehop from $\gamenmik{12,b}$ to $\gamenmik{13,b}$,
we have
\begin{multline}\label{eq:c2a2h9}
  | \Pr[\gamenmink{8,b} \Rightarrow 1] - \Pr[\gamenmink{9,b} \Rightarrow 1] | 
  = | \Pr[\gamenmik{10,b} \Rightarrow 1] - \Pr[\gamenmik{11,b} \Rightarrow 1] | \\
  \leq 2q_H\sqrt{P_C}=\frac{2q_H}{\sqrt{2^{\lenstat}}}.
\end{multline}

From here we could prove that the initiator will abort,
unless the adversary creates a forgery against $\PQCMAC$.
However, even a forgery will not help the adversary,
because the key that would be accepted is uniformly random,
which we prove next.

Game \gamenmink{10,b} replaces $\ksess$ with a random value.
Since $\kpqcs$ acts as a one-time pad on $\kqkds$,
this replacement does not change the view of the adversary, so that
\begin{equation}\label{eq:c2a2h10}
  \Pr[\gamenmink{9,b} \Rightarrow 1]
  = \Pr[\gamenmink{10,b} \Rightarrow 1].
\end{equation}

Both $k^*_0$ and $k^*_1$ are uniformly random keys in $\gamenmink{10,b}$, 
so that
\begin{equation}\label{eq:c2a2-10}
  \Pr[\gamenmink{10,0} \Rightarrow 1]
  = \Pr[\gamenmink{10,1} \Rightarrow 1].
\end{equation}

So that for \cref{case2a2}, we combine \cref{eq:c2a2h8,eq:c2a2h9,eq:c2a2h10,eq:c2a2-10} to get
\begin{multline}\label{eq:2a2}
  | \Pr[\gamenmi{7,0} \Rightarrow 1 \land \forall \sid \neq \sid^*: \kdfIn[\sid] \neq \kdfIn[\sid^*]] \\ 
  - \Pr[\gamenmi{7,1} \Rightarrow 1 \land \forall \sid \neq \sid^*: \kdfIn[\sid] \neq \kdfIn[\sid^*]] |
  \leq \frac{4q_H}{\sqrt{2^{\lenstat}}}.
\end{multline}

To conclude mutually exclusive \cref{case2a1,case2a2}, we take the sum over \cref{eq:2a1,eq:2a2} and get
\begin{multline}\label{eq:2a-7}
  | \Pr[\gamenmi{7,0} \Rightarrow 1]
  - \Pr[\gamenmi{7,1} \Rightarrow 1] | \\
  \leq 2 (\numberSessions-1) \left( \numberSessions\muSecStat + \frac{2q_H}{\sqrt{2^{\lenstat}}} + \Adv_{\PQCMAC}^{\OTSUFCMA}(B_4) \right)
  + \frac{4q_H}{\sqrt{2^{\lenstat}}} \\ 
  \leq 2 \numberSessions \left( \numberSessions\muSecStat + \frac{2q_H}{\sqrt{2^{\lenstat}}} + \Adv_{\PQCMAC}^{\OTSUFCMA}(B_4) \right).
\end{multline}
so that for \cref{case2a} we combine \cref{eq:c2ah4,eq:c2ah5,eq:c2ah6,eq:c2ah7,eq:2a-7} to conclude 
\begin{multline}\label{eq:2a}
  | \Pr[\gamenm{3,0} \Rightarrow 1 \land \role[\sid^*] = \roleInit]
  - \Pr[\gamenm{3,1} \Rightarrow 1 \land \role[\sid^*] = \roleInit] | \\
  \leq 2 \deltastat + 2 \Adv_{\KEMstat}^{\INDCCA}(B_3) +
    2 \numberSessions \left( \numberSessions \muSecStat + \frac{q_H}{\sqrt{2^{\lenstat}}} + \Adv_{\PQCMAC}^{\OTSUFCMA}(B_4) \right) \, .
\end{multline}

\refstepcounter{pfsubcase}\label{case2b}%
\heading{\Cref*{case2b}: test session is a responder.}
The proof is mostly analogous to \cref{case2a}, except that the responder itself will never abort.
Instead, we will require that any non-matching initiator session that computes the same \kpqc{} will abort.
Also, \cref{case2b1} requires one extra gamehop: $\gamenmrk{13,b}$, encoded in \cref{line:c2bg13a,line:c2bg13b}.

\nicoresetlinenr
\begin{figure}
  \fbox{\begin{minipage}{\dimexpr\textwidth-2\fboxsep-2\fboxrule\relax}\begin{multicols}{2}
    \underline{{\bf GAME} $\gamenmr{\cdot,b}$}
    \begin{nicodemus}
    \item $\sidctr := 0$
    \item $\sid^* := 0$
    \item $s' \uni \set{1, \dots, \numberSessions}$ {\color{gray}\gcom{ \gamenm{2,b}}}
    \item $s'' \uni \set{1, \dots, \numberSessions} \setminus \set{s'}$ \gcom{ \gamenmrk{9,b}}
    \item $j' \uni \set{1, \dots, \numberParties}$ {\color{gray}\gcom{ \gamenm{3,b}}}
    \item $\pcfor i \in \set{1, \dots, \numberParties}$
    \item $\quad (\pk_i, \sk_i) \gets \KeyGenStat()$
    \item $\qkdInit()$
    \item $(\widetilde{\calice}, \widetilde{\kalice}) := \EncapsStat(\pk_{j'})$ \gcom{ \gamenmr{5,b}}\label[line]{line:nmr-encaps-sidstar}
    \item $\widetilde{\kalice} \uni \{0,1\}^{\lenstat}$ \gcom{ \gamenmr{6,b}}
    \item $\widetilde{\kpqc} \uni \{0,1\}^{\lenpqc}$ \gcom{ \gamenmrk{11,b}}
    \item $\widetilde{\ksess} \uni \{0,1\}^{\lensess}$\label[line]{line:c2bg13a}\gcom{ \gamenmrk{13,b}}
    \item $b' \gets A^\Oracle(\pk_1, \dots, \pk_{\numberParties})$\label[line]{line:nmr-calla}
    \item $\matchingSessions := \FindMatches(\sid^*)$
    \item $\pcif \matchingSessions \neq \emptyset: \pcreturn 0$ {\color{gray}\gcom{ $\gamenm{1,b}$}}
    \item $\pcif \Attack(\matchingSessions)$
    \item $\quad \pcreturn 0$
    \item $\pcif s' \neq \sid^*: \pcreturn 0$ {\color{gray}\gcom{ $\gamenm{2,b}$}}
    \item $\pcif j' \neq \peer[\sid^*]: \pcreturn 0$ {\color{gray}\gcom{ $\gamenm{3,b}$}}
    \item $\pcif \role[\sid^*] \neq \roleResp$
    \item $\quad \pcreturn 0$ \gcom{ $\gamenmr{4,b}$}

    \item $\kdfMatch := \false$
    \item $\pcfor \sid \in [S] \setminus \set{\sid^*}$
    \item $\quad \pcif \kdfIn[\sid] = \kdfIn[\sid^*]$
    \item $\qquad \kdfMatch := \true$
    \item $\pcif \neg\kdfMatch: \pcreturn 0$ \gcom{ \gamenmrk{8,b}}

    \item $\pcif \kdfIn[s''] \neq \kdfIn[\sid^*]$
    \item $\quad \pcreturn 0$ \gcom{ \gamenmrk{9,b}}
    \item $\pcfor \sid \in \set{1, \dots, \numberSessions} \setminus \set{s'', \sid^*}$
    \item $\quad \pcif \kdfIn[\sid] = \kdfIn[s'']$
    \item $\qquad \pcreturn 0$ \gcom{ \gamenmrk{10,b}}\label[line]{line:nmrkKdfUnique}
    \item $\pcreturn b'$
    \end{nicodemus}

    \vspace{1em}
    \underline{$\orSendMOne(\sid, m_1)$}
    \begin{nicodemus}
      \item $(i, j) := (\owner[\sid], \peer[\sid])$
      \item $(\rho, s) := (\role[\sid], \state[\sid])$
      \item $\pcif i = \bot \pcor \rho \neq \roleResp \pcor s \neq \bot$
      \item $\quad \pcabort$
      \item $\received[\sid] := m_1$

      \item $(\cbob, \pk_e) := m_1$
      \item $\kbob' := \DecapsStat(\sk_i, \cbob)$

      \item $(\calice, \kalice) \gets \EncapsStat(\pk_j)$\label[line]{line:nmr-encaps}
      \item $\pcif \sid = s'$
      \item $\quad (\calice, \kalice) := (\widetilde{\calice}, \widetilde{\kalice})$ \gcom{ \gamenmr{5,b}}
      \item $(\ceph, \keph) \gets \EncapsEph(\pk_e)$

      \item $\pcif \sid = s'$
      \item $\quad \kdfIn[\sid] = (\kbob', \kalice, \keph)$ \gcom{ \gamenmr{7,b}}

      \item $\kpqc := \KDF(\kbob', \kalice, \keph)$
      \item $\pcif \sid = s'$
      \item $\quad \kpqc := \widetilde{\kpqc}$ \gcom{ \gamenmrk{11,b}}

      \item $(\kid, \kqkd) := \enckey(\sid, \lenqkd)$
      \item $t := (m_1, (\calice, \ceph, \kid))$

      \item $(\kpqcm \concat \kpqcs) := \kpqc$
      \item $(\kqkdm \concat \kqkds) := \kqkd$
      \item $\ksess := \kpqcs \oplus \kqkds$

      \item $\pcif \sid = s'$
      \item $\quad \ksess = \widetilde{\ksess}$\label[line]{line:c2bg13b}\gcom{ \gamenmrk{13,b}}

      \item $\tau_1 = \QKDMAC_{\kqkdm}((t, j, i))$
      \item $\tau_2 = \PQCMAC_{\kpqcm}((t, \tau_1, j, i))$\label[line]{line:2bmac2responder}

      \item $m_2 := (\calice, \ceph, \kid, \tau_1, \tau_2)$

      \item $\sent[\sid] := m_2$
      \item $\sessionkeyArray[\sid] := \ksess$
      \item $\state[\sid] := \Accept$
      \item $\pcreturn m_2$
    \end{nicodemus}

    \vspace{1em}
    \underline{$\orSendMTwo(\sid, m_2)$}
    \begin{nicodemus}
      \item $(i, j) := (\owner[\sid], \peer[\sid])$
      \item $(\rho, s) := (\role[\sid], \state[\sid])$
      \item $\pcif i = \bot \pcor \rho \neq \roleInit: \pcabort$
      \item $\received[\sid] := m_2$

      \item $(\calice, \ceph, \kid, \tau_1', \tau_2') := m_2$
      \item $(\kbob, \sk_e, m_1) := s$
      \item $\kalice' := \DecapsStat(\sk_i, \calice)$\label[line]{line:nmr-decaps}
      \item $\pcif i = j' \pcand \calice = \widetilde{\calice}$
      \item $\quad \kalice' := \widetilde{\kalice}$ \gcom{ \gamenmr{5,b}}
      \item $\keph' := \DecapsEph(\sk_e, \ceph)$

      \item $\kdfIn[\sid] = (\kbob, \kalice', \keph')$ \gcom{ \gamenmr{7,b}}
      
      \item $\kpqc := \KDF(\kbob, \kalice', \keph')$
      \item $\pcif \sid = s''$
      \item $\quad \kpqc := \widetilde{\kpqc}$ \gcom{ \gamenmrk{11,b}}

      \item $\kqkd := $

        $\hspace{\fill}\deckey(\sid, \kid)$
      \item $\pcif \kqkd = \bot: \pcabort$
      \item $t := (m_1, (\calice, \ceph, \kid))$

      \item $(\kpqcm \concat \kpqcs) := \kpqc$
      \item $(\kqkdm \concat \kqkds) := \kqkd$
      \item $\ksess := \kpqcs \oplus \kqkds$

      \item $\tau_1 = \QKDMAC_{\kqkdm}((t, i, j))$
      \item $\tau_2 = \PQCMAC_{\kpqcm}((t, \tau_1, i, j))$

      \item $\pcif \sid = s'': \pcabort$ \gcom{ \gamenmrk{12,b}}
      \item $\pcif \tau_1 \neq \tau_1' \pcor \tau_2 \neq \tau_2': \pcabort$\label[line]{line:2bverifytags}

      \item $\sessionkeyArray[\sid] := \ksess$
      \item $\state[\sid] := \Accept$
    \end{nicodemus}
\end{multicols}\end{minipage}}
\caption{Game $\gamenmr{\cdot, b}$: non-matching sessions, responder test session}%
\label{fig:proof_nomatch_resp}
\end{figure}

Formally, \cref{case2b} occurs when $\role[\sid^*] = \roleResp$,
this is encoded in \gamenmr{4,b} (see \cref{fig:proof_nomatch_resp}):
\begin{equation}\label{eq:2bh4}
  \Pr[\gamenm{3,b} \Rightarrow 1 \land \role[\sid^*] = \roleResp]
  = \Pr[\gamenmr{4,b} \Rightarrow 1].
\end{equation}

In game $\gamenmr{5,b}$, we sample the test-session static encapsulation in advance,
and in initiator sessions we replace the decapsulated value with the encapsulated one.
The games are identical unless a decapsulation failure occurs in $\gamenmr{4,b}$,
which by \cref{def:KemCorrectness} is at most $\deltastat$, so we have
\begin{equation}\label{eq:2bh5}
  | \Pr[\gamenmr{4,b} \Rightarrow 1] - \Pr[\gamenmr{5,b} \Rightarrow 1] | \leq \deltastat.
\end{equation}

Game $\gamenmr{6,b}$ samples a uniformly random $\widetilde{\kalice}$
and overwrites the encapsulated value.
Given an adversary $D$ that distinguishes
$\gamenmr{5,b}$ from $\gamenmr{6,b}$
with advantage $\Adv(D)$,
we define an adversary $B_5$ that uses $D$
and wins the \INDCCA game of $\KEMstat$ with $\Adv_{\KEMstat}^{\INDCCA}(B_5) = \Adv(D)$.

Adversary $B$ gets input $(\hat{\pk}, \hat{c}, \hat{k})$
and can query the decapsulation oracle $\oracleDecaps(\hat{\sk}, \cdot)$
on all inputs except $\hat{c}$.
$B_5$ wins if they can correctly say whether $\hat{c}$ is an encapsulation of $\hat{k}$
or if $\hat{k}$ is a uniformly random value.
$B_5$ runs \gamenmr{5,b}
except that they replace the public key of the guessed session $\pk_{j'}$ with $\hat{\pk}$
in \cref{line:nmr-calla,line:nmr-encaps},
they replace $(\widetilde{\calice}, \widetilde{\kalice})$ with $(\hat{c}, \hat{k})$ on \cref{line:nmr-encaps-sidstar},
and instead of decapsulating with $\sk_{j'}$ they query the oracle in \cref{line:nmr-decaps}
(making sure to skip the query on input $\hat{c}$ and directly output $\hat{k}$ instead).
Finally, $B_5$ outputs whatever $D$ outputs.

If $B_5$ receives a real encapsulation they are running \gamenmi{5,b},
otherwise $\hat{k}$ was random and they are running \gamenmi{6,b}.
We also see that $B_5$ is an \INDCCA adversary,
so that
\begin{equation}\label{eq:2bh6}
  | \Pr[\gamenmr{5,b} \Rightarrow 1]
  - \Pr[\gamenmr{6,b} \Rightarrow 1] | \leq \Adv_{\KEMstat}^{\INDCCA}(B_5).
\end{equation}

To prepare for the next case distinction,
\gamenmr{7,b} records the input to the \KDF in the $\kdfIn$ associative array.
Since this bookkeeping does not change the adversary view, we have
\begin{equation}\label{eq:2bh7}
  \Pr[\gamenmr{6,b} \Rightarrow 1] = \Pr[\gamenmr{7,b} \Rightarrow 1].
\end{equation}

\heading{\Cref*{case2b1}. A KDF input match exists.}%
\refstepcounter{pfsubsubcase}\label{case2b1}
In this case, we assume that there exists an honest session $\sid'$ such that
$\sid' \neq \sid^*$ and $\kdfIn[\sid^*] = \kdfIn[\sid']$.
We formalize this condition in \gamenmr{8,b}:
\begin{equation}\label{eq:2b1h8}
  \Pr[\gamenmr{7,b} \Rightarrow 1 \land \exists \sid' \neq \sid^*: \kdfIn[\sid^*] = \kdfIn[\sid']]
  = \Pr[\gamenmrk{8,b} \Rightarrow 1].
\end{equation}

Game \gamenmrk{9,b} makes a guess $s''$ for the session such that
$\kdfIn[s''] = \kdfIn[\sid^*]$,
and we return 0 if our guess turned out to be wrong.
Since $\gamenmrk{8,b}$ established that such a session must exist
and our guess is uniformly random over all sessions other than $\sid^*$, we have that
\begin{equation}\label{eq:2b1h9}
  \Pr[\gamenmrk{8,b} \Rightarrow 1 ]
  = (\numberSessions-1) \Pr[\gamenmrk{9,b} \Rightarrow 1].
\end{equation}

Game \gamenmrk{10,b} establishes that $s''$ is unique.
Game \gamenmrk{9,b} and \gamenmrk{10,b} are identical
unless \gamenmrk{10,b} returns on \cref{line:nmrkKdfUnique}.
A collision in $\kdfIn$ requires at least a collision in $\kbob$,
but by \cref{def:MuSec}, we know that such a collision occurs with
probability at most $\muSecStat$.
There are at most $\numberSessions-2$ sessions that can collide,
so (also by the difference lemma) we get
\begin{multline}\label{eq:2b1h10}
  | \Pr[\gamenmrk{9,b} \Rightarrow 1]
  - \Pr[\gamenmrk{10,b} \Rightarrow 1]|
  \leq \Pr[\gamenmrk{10,b} \text{ returns on \cref{line:nmrkKdfUnique}}] \\
  \leq (\numberSessions-2)\muSecStat
  \leq \numberSessions\muSecStat.
\end{multline}

Game \gamenmrk{11,b} replaces $\kpqc$ with the same uniformly random sampled value
in both $\sid^*$ and $s''$.
We are replacing an output of the QROM,
which is analogous to the gamehop from \gamenmik{11,b} to \gamenmik{12,b},
so we have
\begin{multline}\label{eq:2b1h11}
  \left| \Pr[\gamenmrk{10,b} \Rightarrow 1]
  - \Pr[\gamenmrk{11,b} \Rightarrow 1] \right|
  = \left| \Pr[\gamenmik{10,b} \Rightarrow 1]
  - \Pr[\gamenmik{11,b} \Rightarrow 1] \right| \\
  \leq 2q_H\sqrt{P_C}=\frac{2q_H}{\sqrt{2^{\lenstat}}}.
\end{multline}

Game \gamenmrk{12,b} always aborts initiator session $s''$.
To prove this is indistinguishable from \gamenmrk{11,b} we consider two cases,
based on whether the transcript matches on all values before the second MAC tag.
Specifically,
let $\received[\sid^*] = m_1^*$ and $\sent[\sid^*] = m_2^* = (\calice^*, \ceph^*, \kid^*, \tau_1^*, \tau_2^*)$
be the transcript of the test session,
then we denote $t^* = (m_1^*, (\calice^*, \ceph^*, \kid^*))$.
Similarly we write $(t'', \tau_1'', \tau_2'')$ for the values recorded in 
$\sent[s'']$ and $\received[s'']$.

\heading{\Cref*{case2b1a}.}%
\refstepcounter{pfsubsubsubcase}\label{case2b1a}
In this case the session $s''$ \emph{almost} matches the test session $\sid^*$:
the transcript matches up to and including the first tag,
and both sessions have indicated each other as their peers.
Specifically in this case $(t^*, \tau_1^*, \peer[\sid^*], \owner[\sid^*]) = (t'', \tau_1'', \owner[s''], \peer[s''])$.
The tag $\tau_2^* =\allowbreak \PQCMAC_{\kpqcm}((t^*, \tau_1^*, \peer[\sid^*], \owner[\sid^*]))$
was generated in session $\sid^*$.
Since $s''$ is not a matching session of $\sid^*$,
it must hold that $s''$ received a different tag: $\tau_2'' \neq \tau_2^*$.
But since $\kpqc$ (and therefore $\kpqcm$) has the same value in both $\sid^*$ and $s''$,
both the key and message of the $\PQCMAC$ are identical
and $\sid''$ will recompute $\tau_2^*$ during verification, and therefore reject $\tau_2''$.
In this case, $s''$ will therefore abort with certainty:
\begin{multline}\label{eq:2b1a}
  \Pr[\gamenmrk{11,b} \Rightarrow 1 \land
(t^*, \tau_1^*, \peer[\sid^*], \owner[\sid^*]) = (t'', \tau_1'', \owner[s''], \peer[s''])] \\
  = \Pr[\gamenmrk{12,b} \Rightarrow 1].
\end{multline}

\heading{\Cref*{case2b1b}.}%
\refstepcounter{pfsubsubsubcase}\label{case2b1b}
In this case the session $s''$ differs from $\sid^*$ somewhere in the partial transcript,
or the sessions do not indicate each other as peers.
This means the input to $\PQCMAC_{\kpqcm}$ will be different,
from which we can give a reduction to the \OTSUFCMA security of $\PQCMAC$.
Specifically in this case $(t^*, \tau_1^*, \peer[\sid^*], \owner[\sid^*]) \neq (t'', \tau_1'', \owner[s''], \peer[s''])$.
Note that \gamenmrk{11,b} and \gamenmrk{12,b} are identical,
unless the failure event occurs
where \gamenmrk{11,b} does \emph{not} abort session $s''$ on \cref{line:2bverifytags},
so we bound the probability of this failure event.
Let $D$ be the adversary that triggers the failure event.
We construct the following adversary against the $\OTSUFCMA^{\PQCMAC, B_6}$ game.
$B_6 = (B_{6,0}, B_{6,1})$ runs \gamenmrk{11,b} using $D$ as a subroutine, but with the following modifications:
in session $s'$ (which equals $\sid^*$) on \cref{line:2bmac2responder}, instead of computing \PQCMAC,
$B_{6,0}$ outputs $m = (t^*, \tau_1^*, \peer[\sid^*], \owner[\sid^*])$ and gets back a tag $\tau$ from the \OTSUFCMA game:
$B_6$ sets $\tau_2^* = \tau$ in \gamenmrk{11,b}.
Then on \orSendMTwo{}$(s'', m_2)$,
$B_{6,1}$ outputs $(m', \tau') = ((t'', \tau_1'', \owner[s''], \peer[s'']), \tau_2'')$.
Note that both $\kpqcm$ in \gamenmrk{11,b} and $k$ in the \OTSUFCMA game are uniformly random keys,
that $m \neq m'$ by the assumption of \cref{case2b1b},
and that $B_6$ created a valid forgery exactly when \gamenmrk{11,b} would not abort, so that
\begin{multline*}
  \Pr[\gamenmrk{11,b} \text{ does not abort on \cref{line:2bverifytags}} \,\land \\
  (t^*, \tau_1^*, \peer[\sid^*], \owner[\sid^*]) \neq (t'', \tau_1'', \owner[s''], \peer[s''])]
  = \Adv_{\PQCMAC}^{\OTSUFCMA}(B_6)
\end{multline*}
and thus by the difference lemma
\begin{multline}\label{eq:2b1b}
  \Big| \Pr[\gamenmrk{11,b} \Rightarrow 1 \land
  (t^*, \tau_1^*, \peer[\sid^*], \owner[\sid^*]) \neq (t'', \tau_1'', \owner[s''], \peer[s''])]\\
  - \Pr[\gamenmrk{12,b} \Rightarrow 1] \Big|
  \leq \Adv_{\PQCMAC}^{\OTSUFCMA}(B_6).
\end{multline}

We combine disjoint \cref{case2b1a,case2b1b} by taking the sum over \cref{eq:2b1a,eq:2b1b},
we get
\begin{equation}\label{eq:2b1h12}
  | \Pr[\gamenmrk{11,b} \Rightarrow 1]
  - \Pr[\gamenmrk{12,b} \Rightarrow 1] |
  \leq \Adv_{\PQCMAC}^{\OTSUFCMA}(B_6).
\end{equation}

We established that the initiator session $s''$ always aborts in \gamenmrk{12,b}.
While this sufficed in \cref{case2a1}, here we require one additional gamehop.
In \gamenmrk{13,b} we replace $\ksess$ with a uniformly random value.
Since \gamenmrk{12,b} has established that $\widetilde{\kpqc}$ does not occur
in the view of the adversary via session $s''$,
$\kpqcs$ of the test session acts as a one-time pad in \gamenmrk{13,b}, so that
\begin{equation}\label{eq:2b1h13}
  \Pr[\gamenmrk{12,b} \Rightarrow 1]
  = \Pr[\gamenmrk{13,b} \Rightarrow 1]
\end{equation}
which means that $k^*_0$ and $k^*_1$ are uniformly random keys in $\gamenmrk{13,b}$, 
and
\begin{equation}\label{eq:2b1-13}
  \Pr[\gamenmrk{13,0} \Rightarrow 1]
  = \Pr[\gamenmrk{13,1} \Rightarrow 1].
\end{equation}

Combining \cref{eq:2b1h8,eq:2b1h9,eq:2b1h10,eq:2b1h11,eq:2b1h12,eq:2b1h13,eq:2b1-13} for \cref{case2b1}, we get
\begin{multline}\label{eq:2b1}
  \big| \Pr[\gamenmr{7,0} \Rightarrow 1 \land \exists \sid' \neq \sid^*: \kdfIn[\sid^*] = \kdfIn[\sid']] \\
  - \Pr[\gamenmr{7,1} \Rightarrow 1 \land \exists \sid' \neq \sid^*: \kdfIn[\sid^*] = \kdfIn[\sid']] \big| \\
  \leq 2 (\numberSessions-1) \left( \numberSessions \muSecStat + \frac{2q_H}{\sqrt{2^{\lenstat}}} + \Adv_{\PQCMAC}^{\OTSUFCMA}(B_6) \right)
\end{multline}

\heading{\Cref*{case2b2}. No KDF input match exists.}%
\refstepcounter{pfsubsubcase}\label{case2b2}
\nicoresetlinenr
\begin{figure}
  \fbox{\begin{minipage}{\dimexpr\textwidth-2\fboxsep-2\fboxrule\relax}\begin{multicols}{2}
    \underline{{\bf GAME} $\gamenmrnk{\cdot,b}$}
    \begin{nicodemus}
    \item $\sidctr := 0$
    \item $\sid^* := 0$
    \item $s' \uni [S]$ {\color{gray}\gcom{ \gamenm{2,b}}}
    \item $j' \uni [N]$ {\color{gray}\gcom{ \gamenm{3,b}}}
    \item $\pcfor i \in \set{1, \dots, N}$
    \item $\quad (\pk_i, \sk_i) \gets \KeyGenStat()$
    \item $\qkdInit()$
    \item $(\widetilde{\calice}, \widetilde{\kalice}) := \EncapsStat(\pk_{j'})$ {\color{gray}\gcom{ \gamenmr{5,b}}}
    \item $\widetilde{\kalice} \uni \{0,1\}^{\lenstat}$ {\color{gray}\gcom{ \gamenmr{6,b}}}
    \item $\widetilde{\kpqc} \uni \{0,1\}^{\lenpqc}$ \gcom{ \gamenmrnk{9,b}}
    \item $\widetilde{\ksess} \uni \{0,1\}^{\lensess}$ \gcom{ \gamenmrnk{10,b}}
    \item $b' \gets A^\Oracle(\pk_1, \dots, \pk_N)$
    \item $\matchingSessions := \FindMatches(\sid^*)$
    \item $\pcif \matchingSessions \neq \emptyset: \pcreturn 0$ {\color{gray}\gcom{ $\gamenm{1,b}$}}
    \item $\pcif \Attack(\matchingSessions)$
    \item $\quad \pcreturn 0$
    \item $\pcif s' \neq \sid^*: \pcreturn 0$ {\color{gray}\gcom{ $\gamenm{2,b}$}}
    \item $\pcif j' \neq \peer[\sid^*]: \pcreturn 0$ {\color{gray}\gcom{ $\gamenm{3,b}$}}
    \item $\pcif \role[\sid^*] \neq \roleResp$
    \item $\quad \pcreturn 0$ {\color{gray}\gcom{ $\gamenmr{4,b}$}}

    \item $\pcfor \sid \in [S] \setminus \set{\sid^*}$
    \item $\quad \pcif \kdfIn[\sid] = \kdfIn[\sid^*]$
    \item $\qquad \pcreturn 0$ \gcom{ \gamenmrnk{8,b}}

    \item $\pcreturn b'$
    \end{nicodemus}

    \vspace{1em}
    \underline{$\orSendMOne(\sid, m_1)$}
    \begin{nicodemus}
      \item $(i, j) := (\owner[\sid], \peer[\sid])$
      \item $(\rho, s) := (\role[\sid], \state[\sid])$
      \item $\pcif i = \bot \pcor \rho \neq \roleResp \pcor s \neq \bot$
      \item $\quad \pcabort$
      \item $\received[\sid] := m_1$

      \item $(\cbob, \pk_e) := m_1$
      \item $\kbob' := \DecapsStat(\sk_i, \cbob)$

      \item $(\calice, \kalice) \gets \EncapsStat(\pk_j)$
      \item $\pcif \sid = s'$
      \item $\quad (\calice, \kalice) := (\widetilde{\calice}, \widetilde{\kalice})$ {\color{gray}\gcom{ \gamenmr{5,b}}}
      \item $(\ceph, \keph) \gets \EncapsEph(\pk_e)$

      \item $\pcif \sid = s'$
      \item $\quad \kdfIn[\sid] = (\kbob', \kalice, \keph)$ {\color{gray}\gcom{ \gamenmr{7,b}}}

      \item $\kpqc := \KDF(\kbob', \kalice, \keph)$
      \item $\pcif \sid = s'$
      \item $\quad \kpqc := \widetilde{\kpqc}$ \gcom{ \gamenmrnk{9,b}}

      \item $(\kid, \kqkd) := \enckey(\sid, \lenqkd)$
      \item $t := (m_1, (\calice, \ceph, \kid))$

      \item $(\kpqcm \concat \kpqcs) := \kpqc$
      \item $(\kqkdm \concat \kqkds) := \kqkd$
      \item $\ksess := \kpqcs \oplus \kqkds$

      \item $\pcif \sid = s'$
      \item $\quad \ksess = \widetilde{\ksess}$\gcom{ \gamenmrnk{10,b}}

      \item $\tau_1 = \QKDMAC_{\kqkdm}((t, j, i))$
      \item $\tau_2 = \PQCMAC_{\kpqcm}((t, \tau_1, j, i))$

      \item $m_2 := (\calice, \ceph, \kid, \tau_1, \tau_2)$

      \item $\sent[\sid] := m_2$
      \item $\sessionkeyArray[\sid] := \ksess$
      \item $\state[\sid] := \Accept$
      \item $\pcreturn m_2$
    \end{nicodemus}

    \vspace{1em}
    \underline{$\orSendMTwo(\sid, m_2)$}
    \begin{nicodemus}
      \item $(i, j) := (\owner[\sid], \peer[\sid])$
      \item $(\rho, s) := (\role[\sid], \state[\sid])$
      \item $\pcif i = \bot \pcor \rho \neq \roleInit: \pcabort$
      \item $\received[\sid] := m_2$

      \item $(\calice, \ceph, \kid, \tau_1', \tau_2') := m_2$
      \item $(\kbob, \sk_e, m_1) := s$
      \item $\kalice' := \DecapsStat(\sk_i, \calice)$
      \item $\pcif i = j' \pcand \calice = \widetilde{\calice}$
      \item $\quad \kalice' := \widetilde{\kalice}$ {\color{gray}\gcom{ \gamenmr{5,b}}}
      \item $\keph' := \DecapsEph(\sk_e, \ceph)$

      \item $\kdfIn[\sid] = (\kbob, \kalice', \keph')$ {\color{gray}\gcom{ \gamenmr{7,b}}}
      
      \item $\kpqc := \KDF(\kbob, \kalice', \keph')$

      \item $\kqkd := $

        $\hspace{\fill}\deckey(\sid, \kid)$
      \item $\pcif \kqkd = \bot: \pcabort$
      \item $t := (m_1, (\calice, \ceph, \kid))$

      \item $(\kpqcm \concat \kpqcs) := \kpqc$
      \item $(\kqkdm \concat \kqkds) := \kqkd$
      \item $\ksess := \kpqcs \oplus \kqkds$

      \item $\tau_1 = \QKDMAC_{\kqkdm}((t, i, j))$
      \item $\tau_2 = \PQCMAC_{\kpqcm}((t, \tau_1, i, j))$

      \item $\pcif \tau_1 \neq \tau_1' \pcor \tau_2 \neq \tau_2': \pcabort$

      \item $\sessionkeyArray[\sid] := \ksess$
      \item $\state[\sid] := \Accept$
    \end{nicodemus}
\end{multicols}\end{minipage}}
\caption{Game $\gamenmrnk{\cdot, b}$: non-matching sessions, responder test session, no matching \kdfIn}%
\label{fig:proof_nomatch_resp_nokdf}
\end{figure}

If no KDF input match exists,
then $\kdfIn[\sid] \neq \kdfIn[\sid^*]$
for all $\sid \neq \sid^*$.
We encode this in \gamenmrnk{8,b} in \cref{fig:proof_nomatch_resp_nokdf}, so that
\begin{equation}\label{eq:2b2h8}
  \Pr[\gamenmr{7,b} \Rightarrow 1 \land \forall \sid \neq \sid^*: \kdfIn[\sid] \neq \kdfIn[\sid^*]]
  = \Pr[\gamenmrnk{8,b} \Rightarrow 1].
\end{equation}

Game \gamenmrnk{9,b} replaces $\kpqc$ with a random value $\widetilde{\kpqc}$.
Since this step replaces a single QROM output,
identical to the gamehop from $\gamenmrk{10,b}$ to $\gamenmrk{11,b}$,
we have
\begin{multline}\label{eq:2b2h9}
  | \Pr[\gamenmrnk{8,b} \Rightarrow 1] - \Pr[\gamenmrnk{9,b} \Rightarrow 1] | 
  = | \Pr[\gamenmrk{10,b} \Rightarrow 1] - \Pr[\gamenmrk{11,b} \Rightarrow 1] | \\
  \leq 2q_H\sqrt{P_C}=\frac{2q_H}{\sqrt{2^{\lenstat}}}.
\end{multline}

Game \gamenmrnk{10,b} replaces $\ksess$ with a random value.
Since $\kpqcs$ acts as a one-time pad on $\kqkds$,
this does not change the view of the adversary, so that
\begin{equation}\label{eq:2b2h10}
  \Pr[\gamenmrnk{9,b} \Rightarrow 1]
  = \Pr[\gamenmrnk{10,b} \Rightarrow 1].
\end{equation}

Both $k^*_0$ and $k^*_1$ are uniformly random keys in $\gamenmrnk{10,b}$, 
so that
\begin{equation}\label{eq:2b2-10}
  \Pr[\gamenmrnk{10,0} \Rightarrow 1]
  = \Pr[\gamenmrnk{10,1} \Rightarrow 1].
\end{equation}

So that combining \cref{eq:2b2h8,eq:2b2h9,eq:2b2h10,eq:2b2-10} for \cref{case2b2}, we get
\begin{multline}\label{eq:2b2}
  | \Pr[\gamenmi{7,0} \Rightarrow 1 \land \forall \sid \neq \sid^*: \kdfIn[\sid] \neq \kdfIn[\sid^*]] \\ 
  - \Pr[\gamenmi{7,1} \Rightarrow 1 \land \forall \sid \neq \sid^*: \kdfIn[\sid] \neq \kdfIn[\sid^*]] |
  \leq \frac{4q_H}{\sqrt{2^{\lenstat}}}.
\end{multline}

To conclude mutually exclusive \cref{case2b1,case2b2}, we take the sum over \cref{eq:2b1,eq:2b2} and get
\begin{multline}\label{eq:2b7}
  | \Pr[\gamenmr{7,0} \Rightarrow 1]
  - \Pr[\gamenmr{7,1} \Rightarrow 1] | \\
  \leq 
  2 (\numberSessions-1) \left( \numberSessions \muSecStat + \frac{2q_H}{\sqrt{2^{\lenstat}}} + \Adv_{\PQCMAC}^{\OTSUFCMA}(B_6) \right)
  + \frac{4q_H}{\sqrt{2^{\lenstat}}} \\
  \leq 2 \numberSessions \left( \numberSessions \muSecStat + \frac{2q_H}{\sqrt{2^{\lenstat}}} + \Adv_{\PQCMAC}^{\OTSUFCMA}(B_6) \right)
\end{multline}
so that for \cref{case2b} we combine \cref{eq:2bh4,eq:2bh5,eq:2bh6,eq:2bh7,eq:2b7} to conclude 
\begin{multline}\label{eq:2b}
  | \Pr[\gamenm{3,0} \Rightarrow 1 \land \role[\sid^*] = \roleResp]
  - \Pr[\gamenm{3,1} \Rightarrow 1 \land \role[\sid^*] = \roleResp] | \\
  \leq 2\deltastat + 2\Adv_{\KEMstat}^{\INDCCA}(B_5)
  + 2 \numberSessions \left( \numberSessions \muSecStat + \frac{2q_H}{\sqrt{2^{\lenstat}}} + \Adv_{\PQCMAC}^{\OTSUFCMA}(B_6) \right).
\end{multline}

For the mutually exclusive events encoded in \cref{case2a,case2b},
we can take the sum over \cref{eq:2a,eq:2b} 
and see that
\begin{multline}\label{eq:2-3}
  | \Pr[\gamenm{3,0} \Rightarrow 1]
  - \Pr[\gamenm{3,1} \Rightarrow 1| \\
  \leq 
    4\deltastat + 4\Adv_{\KEMstat}^{\INDCCA}(B_3)
  + 4 \numberSessions \left( \numberSessions \muSecStat + \frac{2q_H}{\sqrt{2^{\lenstat}}} + \Adv_{\PQCMAC}^{\OTSUFCMA}(B_4) \right)
\end{multline}
Combining \cref{eq:c2h1,eq:c2h2,eq:c2h3,eq:2-3} lets us conclude \cref{case2}, proving \cref{lem:case2}
\begin{multline*}
  | \Pr[\INDStAAPQC_0(\Pi, A) \Rightarrow 1 \land \matchingSessions = \emptyset]
  - \Pr[\INDStAAPQC_1(\Pi, A) \Rightarrow 1 \land \matchingSessions = \emptyset] | \\
  \leq 8 \numberSessions \numberParties
  \Bigg(
    \deltastat + \Adv_{\KEMstat}^{\INDCCA}(B_3)
  + \numberSessions \left( \numberSessions \muSecStat + \frac{2q_H}{\sqrt{2^{\lenstat}}} + \Adv_{\PQCMAC}^{\OTSUFCMA}(B_4) \right)\Bigg).
\end{multline*}
\end{proof}

\refstepcounter{pfcase}\label{case3}%
\subsection{Security proof for QKD-based security, \texorpdfstring{\cref{thm:qkd-security}}{Theorem 5.2}}\label{sec:case3}
\begin{proof}[QKD-based security]
We consider security for the $\INDAAQKD$ game, as defined in \cref{fig:formal_game_trivial},
so that we are proving an upper bound for
\[
  \left| \Pr[\INDAAQKD_0(\Pi, A) \Rightarrow 1]
  - \Pr[\INDAAQKD_1(\Pi, A) \Rightarrow 1] \right|.
\]

By the assumption that QKD is not (trivially) broken,
the QKD key used by the test session is \honest,
so it is uniformly random and has not leaked to the adversary directly via an \orQKDGet{} or an \orQKDSet{} query.
The QKD key may have been delivered to one other session,
so the main hurdle for proving \cref{thm:qkd-security} is to prove that the other session did not leak the QKD key.
Since the QKD key is not part of the state and $\kqkds$ is independent from other values,
this could only occur through a \orReveal{} query.
If the other session is a completed matching session,
then $\orReveal$ would constitute a trivial attack.
Otherwise the transcript of the other session differs from that of the test session,
in that case the \roleInit{} will reject (unless a \QKDMAC{} was forged).

\nicoresetlinenr
\begin{figure}
  \fbox{\begin{minipage}{\dimexpr\textwidth-2\fboxsep-2\fboxrule\relax}\begin{multicols}{2}
    \underline{{\bf GAME} $\gamenqz{\cdot,b}$}
    \begin{nicodemus}
    \item $\sidctr := 0$
    \item $\sid^* := 0$
    \item $\sqsend' \uni \set{1, \dots, \numberSessions}$ \gcom{ $\gamenqz{3,b}$}
    \item $\widetilde{\kqkd} \uni \{0,1\}^{\lensess}$ \gcom{ \gamenqz{4,b}}
    \item $\sqrec' \uni \{1, \dots, \numberSessions\} \setminus \{\sqsend'\}$ \gcom{$\gamenqt{7,b}$}
    \item $\widetilde{\ksess} \uni \{0,1\}^{\lensess}$ \gcom{ \gamenqz{9,b}}
    \item $\pcfor i \in \set{1, \dots, \numberParties}$
    \item $\quad (\pk_i, \sk_i) \gets \KeyGenStat()$
    \item $\qkdInit()$
    \item $b' \gets A^\Oracle(\pk_1, \dots, \pk_\numberParties)$
    \item $\matchingSessions := \FindMatches(\sid^*)$
    \item $\pcif \TrivialQKD(\matchingSessions)$
    \item $\quad \pcreturn 0$
    \item $\pcif |\kidUsed[\sid^*]| \neq 1$\label[line]{line:kidstarunique}
    \item $\quad \pcreturn 0$ \gcom{ \gamenqz{2,b}}
    \item $\kidstar := \kidUsed[\sid^*]$ \gcom{ \gamenqz{2,b}}
    \item $\pcif \sqsend' \neq \qsent[\kidstar]$\label[line]{line:sqsend-check} 
    \item $\quad \pcreturn 0$ \gcom{ \gamenqz{3,b}}\label[line]{line:sqsend-abort}
    \item $\pcif \qsent[\kidstar] = [\ ]$
    \item $\quad \pcreturn 0$ \gcom{ \gamenqr{5,b}}
    \item $\pcif \sqrec' \neq \qrecv[\kidstar]$
    \item $\quad \pcreturn 0$ \gcom{ \gamenqt{7,b}}\label[line]{line:sqrecv-abort}
    \item $\pcreturn b'$
    \end{nicodemus}

    \vspace{1em}
    \underline{$\orSendInit(\sid)$}
    \begin{nicodemus}
      \item $(i, j) := (\owner[\sid], \peer[\sid])$
      \item $(\rho, s) := (\role[\sid], \state[\sid])$
      \item $\pcif i = \bot \pcor \rho \neq \roleInit \pcor s \neq \bot: \pcabort$

      \item $(\cbob, \kbob) := \EncapsStat(\pk_j)$
      \item $\pcif \sid = \sqrec'$
      \item $\quad \arwrite{kb} := \kbob$ \gcom{ \gamenqt{8,b}}
      \item $(\pk_e, \sk_e) := \KeyGenEph()$
      \item $m_1 := (\cbob, \pk_e)$
      \item $\sent[\sid] := m_1$
      \item $\state[\sid] := (\kbob, \sk_e, m_1)$
      \item $\pcreturn m_1$
    \end{nicodemus}

    \vspace{1em}
    \underline{$\orSendMOne(\sid, m_1)$}
    \begin{nicodemus}
      \item $(i, j) := (\owner[\sid], \peer[\sid])$
      \item $(\rho, s) := (\role[\sid], \state[\sid])$
      \item $\pcif i = \bot \pcor \rho \neq \roleResp \pcor s \neq \bot$
      \item $\quad \pcabort$
      \item $\received[\sid] := m_1$

      \item $(\cbob, \pk_e) := m_1$
      \item $\kbob' := \DecapsStat(\sk_i, \kbob)$
      \item $\pcif \sid = \sqsend'$
      \item $\quad \kbob' := \arwrite{kb}$ \gcom{ \gamenqt{8,b}}
      \item $(\calice, \kalice) \gets \EncapsStat(\pk_j)$
      \item $(\ceph, \keph) \gets \EncapsEph(\pk_e)$
      \item $\pcif \sid = \sqsend'$
      \item $\quad (\arwrite{ka}, \arwrite{ke}) := (\kalice, \keph)$ \gcom{ \gamenqt{8,b}}
      \item $\kpqc := \KDF(\kbob', \kalice, \keph)$
      \item $(\kid, \kqkd) := \enckey(\sid, \lenqkd)$\label[line]{line:c3-qkdresp}
      \item $\kidUsed[\sid] \mathrel{+}= [\kid]$\gcom{ \gamenqz{1,b}}\label[line]{line:kidstar-resp}
      \item $\pcif \sid = \sqsend'$\label[line]{line:send-m1-kidused-check}
      \item $\quad \kqkd := \widetilde{\kqkd}$ \gcom{ \gamenqz{4,b}}
      \item $t := (m_1, (\calice, \ceph, \kid))$
      \item $\pcif \sid = \sqsend'$
      \item $\quad \arwrite{tresp} = t$ \gcom{ \gamenqnt{5,b}}

      \item $(\kqkdm \concat \kqkds) := \kqkd$
      \item $(\kpqcm \concat \kpqcs) := \kpqc$
      \item $\ksess := \kqkds \oplus \kpqcs$
      \item $\pcif \sid = \sqsend'$
      \item $\quad \ksess := \widetilde{\ksess}$ \gcom{ \gamenqz{9,b}}
      \item $\tau_1 = \QKDMAC_{\kqkdm}((t, j, i))$\label[line]{line:m1-tau}
      \item $\tau_2 = \PQCMAC_{\kpqcm}((t, \tau_1, j, i))$

      \item $m_2 := (\calice, \ceph, \tau_1, \tau_2)$

      \item $\sent[\sid] := m_2$
      \item $\sessionkeyArray[\sid] := \ksess$
      \item $\state[\sid] := \Accept$
      \item $\pcreturn m_2$

    \end{nicodemus}

    \vspace{1em}
    \underline{$\orSendMTwo(\sid, m_2)$}
    \begin{nicodemus}
      \item $(i, j) := (\owner[\sid], \peer[\sid])$
      \item $(\rho, s) := (\role[\sid], \state[\sid])$
      \item $\pcif i = \bot \pcor \rho \neq \roleInit: \pcabort$
      \item $\received[\sid] := m_2$

      \item $(\calice, \ceph, \kid, \tau_1', \tau_2') := m_2$
      \item $(\kbob, \sk_e, m_1) := s$ \gcom{ or abort}\label[line]{line:sendm2-state}
      \item $\kalice' := \DecapsStat(\sk_i, \calice)$
      \item $\keph' := \DecapsEph(\sk_e, \ceph)$
      \item $\pcif \sid = \sqrec'$
      \item $\quad (\kalice', \keph') := (\arwrite{ka}, \arwrite{ke})$ \gcom{ \gamenqt{8,b}}
      \item $\kpqc := \KDF(\kbob, \kalice', \keph')$

      \item $\kqkd := $

        $\hspace{\fill}\deckey(\sid, \kid)$\label[line]{line:c3-qkdinit}
      \item $\pcif \kqkd = \bot: \pcabort$\label[line]{line:c3-qkdabort}
      \item $\kidUsed[\sid] \mathrel{+}= [\kid]$\gcom{ \gamenqz{1,b}}\label[line]{line:kidstar-init}
      \item $\pcif \sqsend' = \qsent[\kid]$\label[line]{line:c3-checkinit}
      \item $\quad \kqkd := \widetilde{\kqkd}$ \gcom{ \gamenqz{4,b}}
      \item $t := (m_1, (\calice, \ceph, \kid))$
      \item $\pcif \sqsend' = \qsent[\kid] \pcand \arwrite{tresp} = t$
      \item $\quad \pcabort$ \gcom{ \gamenqnt{6,b}}
      \item $\pcif \sqsend' = \qsent[\kid] \pcand \arwrite{tresp} \neq t$
      \item $\quad \pcabort$ \gcom{ \gamenqt{6,b}}
      \item $(\kqkdm \concat \kqkds) := \kqkd$
      \item $(\kpqcm \concat \kpqcs) := \kpqc$
      \item $\ksess := \kqkds \oplus \kpqcs$
      \item $\pcif \sqsend' = \qsent[\kid]$
      \item $\quad \ksess := \widetilde{\ksess}$ \gcom{ \gamenqz{9,b}}
      \item $\tau_1 = \QKDMAC_{\kpqcm}((t, i, j))$
      \item $\tau_2 = \PQCMAC_{\kpqcm}((t, \tau_1, i, j))$

      \item $\pcif \sqsend' = \qsent[\kid]$
      \item $\quad \pcabort$ \gcom{ \gamenqnt{7,b}}
      \item $\pcif \tau_1 \neq \tau_1' \pcor \tau_2 \neq \tau_2': \pcabort$\label[line]{line:c3-tauabort}

      \item $\sessionkeyArray[\sid] := \ksess$
      \item $\state[\sid] := \Accept$
    \end{nicodemus}
\end{multicols}\end{minipage}}
\caption{
  Game $\gamenqz{}$: QKD is secure.
}%
\label{fig:proof_qkd_match}
\end{figure}

Game $\gamenqz{0}$ equals $\INDAAQKD_b$:
\begin{equation}\label{eq:3h0}
  \Pr[\INDAAQKD_b(\Pi, A) \Rightarrow 1]
  = \Pr[\gamenqz{0,b} \Rightarrow 1].
\end{equation}

Game \gamenqz{1,b} adds \cref{line:kidstar-resp,line:kidstar-init},
tracking per session which QKD key IDs they use in the associative array \kidUsed,
which are initialized at $\kidUsed[\sid] = [\ ]$ for all $\sid$.
Since this is only doing some bookkeeping, we have
\begin{equation}\label{eq:3h1a}
  \Pr[\gamenqz{0,b} \Rightarrow 1]
  = \Pr[\gamenqz{1,b} \Rightarrow 1].
\end{equation}

Game \gamenqz{2,b} asserts that the test session used just a single \kid,
and labels it \kidstar.
The Send queries of accepting sessions
must be queried exactly once and in order:
\orSendInit{} and \orSendMTwo{} for an initiator
and \orSendMOne{} for a responder.
This is enforced by the \state{} variable,
where we note that for an initiator this enforcement is implicit
since \cref{line:sendm2-state} of $\SendMTwo$
aborts if the \state{} does not hold a triple.
In the protocol a session queries the QKD oracle only once
(a responder in \cref{line:c3-qkdresp} and an initiator in \cref{line:c3-qkdinit}),
so that each session uses \emph{at most} one QKD key,
and each accepting session uses \emph{exactly} one QKD key.
Thus the check on \cref{line:kidstarunique} will always be false, so that
\begin{equation}\label{eq:3h2a}
  \Pr[\gamenqz{0,b} \Rightarrow 1]
  = \Pr[\gamenqz{1,b} \Rightarrow 1].
\end{equation}

Next we consider the other session that shares the QKD key with the test session,
if it exists.
First, define $\sqsend = \qsent[\kidstar]$.
Note that this session always exists ($\sqsend \neq \bot$):
only an honest query to \enckey{} can set $\flag[\kidstar]$ to \honest,
and this also sets $\qsent[\kidstar]$.
Second, if $\qrecv[\kidstar] \neq [\ ]$, then define \sqrec{} such that $[\sqrec] = \qrecv[\kidstar]$.
Note that $|\qrecv[\kidstar]| \leq 1$ even if $\sqrec \neq \sid^*$,
because $\qrecv[\kidstar]$ is only updated by \deckey{}$(\sid, \kidstar)$ if $\key[\kidstar] \neq \bot$,
and an \honest{} key will be deleted after the first call to \deckey{}$(\sid, \kidstar)$.
This proves that there is at most one completed matching session.

Each responder session calls \enckey{} once and increments $\kidctr$,
so there is only one honest responder session that contains $\kidstar$ in \sent.
Since the QKD key is \honest, it is only delivered to one initiator session.
Therefore there is at most one honest initiator session that
received a key from $\deckey(\sid, \kidstar)$,
implying there is at most one accepting initiator session with $\kidstar$ in \received.

Game \gamenqz{3,b} makes a guess $\sqsend'$ for $\sqsend$ and returns 0 if the guess was incorrect.
\Cref{line:sqsend-check} checks this by comparing the guess $\sqsend'$ against $\qsent[\kidstar]$,
which equals $\sqsend$ by definition.
Since $\Pr[\sqsend' = \sqsend] = 1/\numberSessions$, we have
\begin{equation}\label{eq:3h1}
  \Pr[\gamenqz{2,b} \Rightarrow 1] = \numberSessions \Pr[\gamenqz{3,b} \Rightarrow 1].
\end{equation}

Game \gamenqz{4,b} samples $\widetilde{\kqkd}$
and replaces $\kqkd$ as sampled by the QKD oracle
with $\widetilde{\kqkd}$ in sessions $\sqsend$ and (if it exists) in $\sqrec$.
The predicate on \cref{line:c3-checkinit} ($\sqsend' = \qsent[\kid]$)
is equivalent to the predicate $\sid = \sqrec$,
except it will be false for all sessions if $\sqrec$ does not exist.
The equivalence holds since $\flag[\kidstar] = \honest$,
and thus the QKD key is delivered to at most one initiator session.
For any other adversary queries \orSendMTwo{} containing $\kidstar$,
\deckey{} will return $\bot$ and those sessions will abort on \cref{line:c3-qkdabort}.
Since all we did was move the uniform sampling from the oracle to the game, we have
\begin{equation}\label{eq:3h2}
  \Pr[\gamenqz{3,b} \Rightarrow 1] = \Pr[\gamenqz{4,b} \Rightarrow 1].
\end{equation}

Game $\gamenqz{9,b}$ replaces $\ksess$ with the uniformly random sampled $\widetilde{\ksess}$.
Note that $\kqkds$ acts as a one-time pad on $\kpqcs$,
meaning that this gamehop is indistinguishable if $\kqkds^*$
(the value of $\kqkds$ used by the test session)
is independent of the adversary view (without the test session key).
We will prove this by case distinction.

\setcounter{pfcase}{0}
\heading{\Cref*{case3a}: \sqrec{} does not exist.}%
\refstepcounter{pfcase}\label{case3a}
In this case we assume $\qrecv[\kidstar] = [\ ]$:
no honest session received the test session QKD key.
Then $\kqkds^*$ only exists in session $\sqsend$
(and it must hold that $\sqsend = \sid^*)$, so that
\begin{equation}\label{eq:3ah7}
  \Pr\left [\gamenqz{4,b} \Rightarrow 1 \land \qrecv[\kidstar] = [\ ]\right]
  = \Pr[\gamenqz{9,b} \Rightarrow 1].
\end{equation}

\heading{\Cref*{case3b}: \sqrec{} exists.}%
\refstepcounter{pfcase}\label{case3b}
In this case we assume $\qrecv[\kidstar] = [\sqrec]$. This is encoded in game $\gamenqr{5,b}$, so that
\begin{equation}\label{eq:3bh3}
  \Pr\left[\gamenqz{4,b} \Rightarrow 1 \land \qrecv[\kidstar] = [\sqrec]\right]
  = \Pr[\gamenqr{5,b} \Rightarrow 1].
\end{equation}

We prove $\gamenqr{5,b}$ is indistinguishable from $\gamenqz{9,b}$ by looking at several subcases,
based on whether or not the transcripts and/or tags are equal.
Let $\tinit = ((\cbob, \pk_e), (\calice', \ceph', \kidstar))$ be the partial transcript of $\sqrec$,
and $\tresp = ((\cbob', \pk_e'), (\calice, \ceph, \kidstar))$ be the partial transcript of $\sqsend$.
Similarly let $\tau_1, \tau_2$ be the tags sent by $\sqsend$,
and let $\tau_1', \tau_2'$ be the tags as received by $\sqrec$.

\heading{\Cref*{case3b1}: transcripts differ.}%
\refstepcounter{pfsubcase}\label{case3b1}
This case assumes $\tinit \neq \tresp$. This is encoded in game $\gamenqnt{6,b}$,
so that
\begin{equation}\label{eq:3b1h4}
  \Pr[\gamenqr{5,b} \Rightarrow 1 \land \tinit \neq \tresp]
  = \Pr[\gamenqnt{6,b} \Rightarrow 1].
\end{equation}

Game $\gamenqnt{7,b}$ always aborts session $\sqrec$.
Note that $\gamenqnt{6,b}$ and $\gamenqnt{7,b}$ are identical unless
$\gamenqnt{6,b}$ does not abort session $\sqrec$ on \cref{line:c3-tauabort}.
Then by the difference lemma, we have
\[
  \left| \Pr[\gamenqnt{6,b} \Rightarrow 1]
  - \Pr[\gamenqnt{7,b} \Rightarrow 1]\right|
\leq \Pr[\gamenqnt{6,b} \text{ does not abort } \sqrec \text{ on \cref{line:c3-tauabort}}],
\]
This event would require at least $\tau_1 = \tau_1'$ to hold in $\sqrec$,
which lets us bound this by a reduction to the $\OTSUFCMA$ security of $\QKDMAC$,
using the following adversary $B_7 = (B_{7,0},B_{7,1})$:
$B_7$ runs game $\gamenqnt{6,b}$, with the following modifications.
On \cref{line:m1-tau}, if $\sid = \sqsend'$,
then instead of computing $\tau$ directly, $B_{7,0}$ outputs
$m = (\tresp, \peer[\sqsend], \owner[\sqsend])$
and uses the response $\tau$.
Just before \cref{line:c3-tauabort},
if $\sqsend = \qsent[\kid]$ (ie. if $\sid = \sqrec$),
$B_{7,1}$ outputs
$(m', \tau') = ((\tinit, \owner[\sqrec], \peer[\sqrec]), \tau_1')$.
Then $m \neq m'$, and $B_7$ created a successful forgery exactly when $\sqsend$ accepts $\tau_1'$.
Thus
\[
\Pr[\gamenqnt{6,b} \text{ does not abort } \sqrec \text{ on \cref{line:c3-tauabort}}]
  \leq \Adv_{\QKDMAC}^{\OTSUFCMA}(B_7)
\]
and therefore
\begin{equation}\label{eq:3b1h5}
  \left| \Pr[\gamenqnt{6,b} \Rightarrow 1]
  - \Pr[\gamenqnt{7,b} \Rightarrow 1]\right|
  \leq \Adv_{\QKDMAC}^{\OTSUFCMA}(B_7).
\end{equation}

Since \gamenqnt{7,b} always aborts $\sqrec$, $\kqkds^*$ only exists in $\sqsend$, so that:
\begin{equation}\label{eq:3b1h7}
  \Pr[\gamenqnt{7,b} \Rightarrow 1]
  = \Pr[\gamenqz{9,b} \Rightarrow 1].
\end{equation}

We combine \cref{eq:3b1h4,eq:3b1h5,eq:3b1h7} to conclude that for \cref{case3b1}:
\begin{equation}\label{eq:3b1}
  | \Pr[\gamenqr{5,b} \Rightarrow 1 \land \tinit \neq \tresp]
  - \Pr[\gamenqz{9,b} \Rightarrow 1] |
  \leq \Adv_{\QKDMAC}^{\OTSUFCMA}(B_7).
\end{equation}

\heading{\Cref*{case3b2}: transcripts are the same.}%
\refstepcounter{pfsubcase}\label{case3b2}
In this case: $\tinit = \tresp$. Game $\gamenqt{6,b}$ encodes this:
\begin{equation}\label{eq:3b2h4}
  \Pr[\gamenqr{5,b} \Rightarrow 1 \land \tinit = \tresp]
  = \Pr[\gamenqt{6,b} \Rightarrow 1].
\end{equation}

By definition of $\sqsend$ and $\sqrec$, both sessions use the same $\kqkd$,
and thus the same $\kqkdm$.
Note that $(\owner[\sqsend],\allowbreak \peer[\sqsend]) = (\peer[\sqrec], \owner[\sqrec])$,
because \deckey{} checked this.
We conclude that both sessions have the same key and message input to \QKDMAC,
so they compute the same tag $\tau_1$.
If the adversary changed $\tau_1' \neq \tau_1$, then the initiator will abort, therefore:
\begin{equation}\label{eq:3b2h7a}
  \Pr[\gamenqt{6,b} \Rightarrow 1 \land \tau_1' \neq \tau_1]
  = \Pr[\gamenqz{9,b} \Rightarrow 1].
\end{equation}

Next we prove that if $\tinit = \tresp$,
then almost certainly (by correctness of the \KEM{}s)
both sessions compute the same $\kpqc$.
Game $\gamenqt{7,b}$ makes a uniformly random guess $\sqrec'$
for $\sqrec$ and aborts if the guess was incorrect.
Since the guess is correct with probability $1/(\numberSessions-1)$, we have
\begin{equation}\label{eq:3b2h5}
  \Pr[\gamenqt{6,b} \Rightarrow 1 \land \tau_1' = \tau_1]
  = (\numberSessions-1) \Pr[\gamenqt{7,b} \Rightarrow 1 \land \tau_1' = \tau_1].
\end{equation}

Game $\gamenqt{8,b}$ records the encapsulated secrets in $\sqrec$
and replaces the decapsulation outputs in $\sqsend$ with the recorded values,
and vice versa. By the correctness of the two static \KEM{}s and the one
ephemeral \KEM{}, we have
\begin{equation}\label{eq:3b2h6}
  | \Pr[\gamenqt{7,b} \Rightarrow 1 \land \tau_1' = \tau_1]
  - \Pr[\gamenqt{8,b} \Rightarrow 1 \land \tau_1' = \tau_1] | \leq 2\deltastat + \deltaeph.
\end{equation}
We emphasize that the \KEM{} correctness is a statistical property and not a computational one,
since the probability in \cref{def:KemCorrectness} is independent of any adversary.

In $\gamenqt{8,b}$, both $\sqsend$ and $\sqrec$ compute the same $\kpqc$,
and thus the same $\kpqcm$.
Since $\tinit = \tresp$, $\tau_1' = \tau_1$, and
$(\peer[\sqsend], \owner[\sqsend]) = (\owner[\sqrec], \peer[\sqrec])$,
both key and message input to $\PQCMAC$ are the same in both sessions.
That means that if the adversary sends $\tau_2' \neq \tau_2$,
then $\sqrec$ will reject with certainty since \PQCMAC is assumed to be canonical.\footnote{
  Statistical strong unforgeability would suffice but induces as small loss,
  however \emph{computational} strong unforgeability is insufficient.
}
On the other hand, if $\tau_2' = \tau_2$, then $\sqsend$ and $\sqrec$ are matching sessions
and the adversary may not query \orReveal{} on either session.
In either case, $\kqkds$ does not leak through any session, so that
\begin{equation}\label{eq:3b2h7b}
  \Pr[\gamenqt{8,b} \Rightarrow 1 \land \tau_1' = \tau_1] = \Pr[\gamenqz{9,b} \Rightarrow 1].
\end{equation}

We combine \cref{eq:3b2h5,eq:3b2h6,eq:3b2h7b} to conclude that
\begin{equation}\label{eq:3b2th7}
  | \Pr[\gamenqr{6,b} \Rightarrow 1 \land \tau_1' = \tau_1]
  - \Pr[\gamenqz{9,b} \Rightarrow 1] |
  \leq \numberSessions(2\deltastat + \deltaeph),
\end{equation}
summing \cref{eq:3b2h7a,eq:3b2th7} we get
\begin{equation}\label{eq:3b2h7}
  | \Pr[\gamenqr{6,b} \Rightarrow 1]
  - \Pr[\gamenqz{9,b} \Rightarrow 1] |
  \leq \numberSessions(2\deltastat + \deltaeph),
\end{equation}
and combining \cref{eq:3b2h4,eq:3b2h7} we conclude \cref{case3b2}
\begin{equation}\label{eq:3b2}
  | \Pr[\gamenqr{5,b} \Rightarrow 1 \land \tinit = \tresp]
  - \Pr[\gamenqz{9,b} \Rightarrow 1] |
  \leq \numberSessions(2\deltastat + \deltaeph).
\end{equation}

Then we conclude \cref{case3b} by considering disjoint \cref{case3b1,case3b2}
by taking the sum of \cref{eq:3b1,eq:3b2} and combining with \cref{eq:3bh3}:
\begin{multline}\label{eq:3bh7}
  \left| \Pr\left[\gamenqz{4,b} \Rightarrow 1 \land \qrecv[\kidstar] = [\sqrec]\right]
  - \Pr[\gamenqz{9,b} \Rightarrow 1] \right| \\
  \leq \numberSessions(2\deltastat + \deltaeph) + \Adv_{\QKDMAC}^{\OTSUFCMA}(B_7).
\end{multline}
We add \cref{eq:3ah7,eq:3bh7} to combine disjoint \cref{case3a,case3b}:
\begin{equation}\label{eq:3h7}
  \left| \Pr[\gamenqz{4,b} \Rightarrow 1]
  - \Pr[\gamenqz{9,b} \Rightarrow 1] \right|
  \leq \numberSessions(2\deltastat + \deltaeph) + \Adv_{\QKDMAC}^{\OTSUFCMA}(B_7)
\end{equation}

Since $\ksess$ is uniformly random in \gamenqz{9,b}, we have
\begin{equation}\label{eq:3-7}
  \Pr[\gamenqz{9,0} \Rightarrow 1] = \Pr[\gamenqz{9,1} \Rightarrow 1],
\end{equation}
so that combining \cref{eq:3h0,eq:3h1a,eq:3h2a,eq:3h1,eq:3h2,eq:3-7} concludes \cref{case3}, thereby proving \cref{thm:qkd-security}
\begin{multline*}
  \big| \Pr[\INDAAQKD_0(\Pi, A) \Rightarrow 1]
  - \Pr[\INDAAQKD_1(\Pi, A) \Rightarrow 1] \big| \\
  \leq 2 \numberSessions \left( \numberSessions (2\deltastat + \deltaeph) + \Adv_{\QKDMAC}^{\OTSUFCMA}(B_7) \right) \, .
\end{multline*}
\end{proof}


\section{Conclusion and Future Work} \label{sec:conclusion}
  In this paper we identify some gaps within state-of-the-art modelling of combined
QKD and PQC authenticated key exchange,
most notably we consider the omission of QKD key IDs in the models.
We demonstrate that this omission allows an adversary to mix-and-match QKD keys of different sessions,
which could lead to full leakage of the session keys.
To enable the proper handling of key IDs in cryptographic contexts, we propose to model QKD as an oracle
that explicitly models key IDs as input and output of oracle queries.
The oracle provides queries for honest parties to get QKD keys,
and it provides adversarial queries to model both passive and active attacks on QKD.

To demonstrate the usefulness of the QKD oracle,
we construct a QKD/PQC-combined protocol that generates secure session keys
if either QKD is secure or the PQC (or PQC/classical-combined) primitives are.
We emphasize that our protocol preserves the information-theoretic security provided by QKD.
Our solution uses nested \MAC{}s to prevent \attackname attacks,
a construction that is inspired by the XOR-then-MAC \KEM{}-combiner,
but which avoids some of the pitfalls we identified in its security proof.

\heading{Future work.}
Our QKD oracle is modular and can be integrated models other than CK$^+$:
many variations on AKE models exist and (in principle) the oracle could
be integrated into any of them,
allowing formal modelling of other QKD/PQC-combined protocols.
For example, it could be integrated into HAKE,
so that the security for the Muckle family of protocols could be proven
when QKD key IDs are included into the protocol.

	One noteworthy advantage of the HAKE model over ours is that it captures Post-Compromise Security (PCS)~\cite{CSF:CohCreGar16},
	which allows some security even after session compromise.
	PCS seems a meaningful property to have in some contexts,
	we view it as an interesting future direction to combine our QKD oracle
  with AKE models that model PCS (such as HAKE).

We identified an open problem in \cref{ssec:xtm-combiner} about the construction
of robust one-time strong unforgeable \MAC{}s for the XtM \KEM combiner.
Future work could explore the existence of such a \MAC,
or it might be possible to fix the proof for the XtM combiner.

Our QKD oracle returns uniformly random bitstrings,
while real QKD can only guarantee statistical security:
the output key is statistically close to uniformly random.
The oracle could be adapted into a more accurate representation of reality
if this QKD security parameter is reflected in its output.

The QKD oracle presented in this paper is designed to model \emph{end-to-end QKD}.
Many deployments of QKD instead secure point-to-point links in a larger network,
and trusted repeater nodes are used to relay messages and/or keys.
While the QKD oracle could be used to output relayed keys,
we point out that this is a strong assumption on the security of both QKD and on the relaying protocol itself.
Instead for relaying protocols,
it would be more realistic to use the QKD oracle only to model point-to-point links,
and to precisely state the trust assumptions that are required
in order to guarantee security of the relaying protocol itself.

\subheading{Composability of the QKD interface.}
In this work we aim to prevent \attackname attacks \emph{within one} AKE protocol.
We emphasize, however, that attacks on (concurrently) \emph{composed} protocols are not captured in game-based AKE models --
in other words, our security model does not guarantee secure composability.
Even if QKD is secure and the AKE protocol is secure,
this does not always mean that its combination via the ETSI 014 QKD interface (described in \cref{sssec:etsi}) is secure.

To illustrate the concern, consider the following \attackname attack \emph{between different protocols}.
Imagine a scenario where Alice runs some (secure) protocol $\Pi$,
where she uses a (secure) QKD key retrieved via ``Get key''.
Bob runs two protocols, $\Pi$ and $\Pi'$, and in both he retrieves the QKD key via ``Get key with key IDs''.
The adversary can take the key ID sent by Alice in protocol $\Pi$,
and deliver it to Bob in protocol $\Pi'$.
If $\Pi'$ happens to be vulnerable, this may leak the QKD key to the adversary,
which could be sufficient to impersonate Bob in protocol $\Pi$.
Note that neither QKD nor $\Pi$ need to be broken by the adversary for this attack.

For most PQC protocols this issue does not result in real-world attacks,
because proper operational security practices (such as using a key only for a single protocol and protocol-specific key derivation) ensure that a key is only used in a single protocol.
In a typical QKD setup, however, these methods are often not deployed or not available. 

To protect against such attacks, some binding between the QKD key and the protocol (and session) in which it is used seems to be necessary, which might even require a change in the specification of the QKD interface.
It is essential future work to extend our model to capture attacks of this type in order to handle such composability issues.



\bibliographystyle{alphaurl}

\ifTightOnSpace
	\bibliography{additional,abbrev4,cryptobib/crypto}													
\else
	\bibliography{additional,cryptobib/abbrev0,cryptobib/crypto}

\newcommand{\etalchar}[1]{$^{#1}$}
\begin{thebibliography}{RDM{\etalchar{+}}24}

\bibitem[AAC{\etalchar{+}}22]{AAC+22}
Gorjan Alagic, Daniel Apon, David Cooper, Quynh Dang, Thinh Dang, John Kelsey, Jacob Lichtinger, Carl Miller, Dustin Moody, Rene Peralta, Ray Perlner, Angela Robinson, Daniel Smith-Tone, and Yi-Kai Liu.
\newblock Status {{Report}} on the {{Third Round}} of the {{NIST Post-Quantum Cryptography Standardization Process}}.
\newblock Technical Report NIST Internal or Interagency Report (NISTIR) 8413, {National Institute of Standards and Technology}, 9 2022.
\newblock \href {https://doi.org/10.6028/NIST.IR.8413-upd1} {\path{doi:10.6028/NIST.IR.8413-upd1}}.

\bibitem[ABC{\etalchar{+}}25]{ABC+25}
Gorjan Alagic, Maxime Bros, Pierre Ciadoux, David Cooper, Quynh Dang, Thinh Dang, John Kelsey, Jacob Lichtinger, Yi-Kai Liu, Carl Miller, Dustin Moody, Rene Peralta, Ray Perlner, Angela Robinson, Hamilton Silberg, Daniel Smith-Tone, and Noah Waller.
\newblock Status {{Report}} on the {{Fourth Round}} of the {{NIST Post-Quantum Cryptography Standardization Process}}.
\newblock Technical Report NIST Internal or Interagency Report (NISTIR) 8545, {National Institute of Standards and Technology}, 3 2025.
\newblock \href {https://doi.org/10.6028/NIST.IR.8545} {\path{doi:10.6028/NIST.IR.8545}}.

\bibitem[ABS14]{ACISP:AlaBoySte14}
Janaka Alawatugoda, Colin Boyd, and Douglas Stebila.
\newblock Continuous after-the-fact leakage-resilient key exchange.
\newblock In Willy Susilo and Yi~Mu, editors, {\em ACISP 14: 19th Australasian Conference on Information Security and Privacy}, volume 8544 of {\em Lecture Notes in Computer Science}, pages 258--273, Wollongong, NSW, Australia, July~7--9, 2014. Springer, Heidelberg, Germany.
\newblock \href {https://doi.org/10.1007/978-3-319-08344-5_17} {\path{doi:10.1007/978-3-319-08344-5_17}}.

\bibitem[ACD{\etalchar{+}}25]{ACD+25b}
Nick Aquina, Bruno Cimoli, Soumya Das, Kathrin Hövelmanns, Fiona~Johanna Weber, Chigo Okonkwo, Simon Rommel, Boris Škorić, Idelfonso {Tafur Monroy}, and Sebastian Verschoor.
\newblock A critical analysis of deployed use cases for quantum key distribution and comparison with post-quantum cryptography.
\newblock {\em EPJ Quantum Technology}, 12(51):1--42, 5 2025.
\newblock \href {https://doi.org/10.1140/epjqt/s40507-025-00350-5} {\path{doi:10.1140/epjqt/s40507-025-00350-5}}.

\bibitem[ADK{\etalchar{+}}22]{EPRINT:ADKPRY22}
Nimrod Aviram, Benjamin Dowling, Ilan Komargodski, Kenneth~G. Paterson, Eyal Ronen, and Eylon Yogev.
\newblock Practical (post-quantum) key combiners from one-wayness and applications to {TLS}.
\newblock Cryptology ePrint Archive, Report 2022/065, 2022.
\newblock \url{https://eprint.iacr.org/2022/065}.

\bibitem[ANS22]{ANSSIhybrid}
ANSSI.
\newblock {ANSSI views on the Post-Quantum Cryptography transition}, 2022.
\newblock URL: \url{https://cyber.gouv.fr/en/publications/anssi-views-post-quantum-cryptography-transition}.

\bibitem[ART24]{ART24}
Nick Aquina, Simon Rommel, and Idelfonso {Tafur Monroy}.
\newblock Quantum secure communication using hybrid post-quantum cryptography and quantum key distribution.
\newblock In {\em 2024 24th {{International Conference}} on {{Transparent Optical Networks}} ({{ICTON}})}, pages 1--4, 7 2024.
\newblock \href {https://doi.org/10.1109/ICTON62926.2024.10648124} {\path{doi:10.1109/ICTON62926.2024.10648124}}.

\bibitem[BB84]{BEN84}
C.~H. Bennett and G.~Brassard.
\newblock {Quantum cryptography: Public key distribution and coin tossing}.
\newblock In {\em Proceedings of IEEE International Conference on Computers, Systems, and Signal Processing}, page 175, India, 1984.
\newblock \href {https://doi.org/10.1016/j.tcs.2014.05.025} {\path{doi:10.1016/j.tcs.2014.05.025}}.

\bibitem[BBB{\etalchar{+}}25]{BBB+25a}
Jaime~S. Buruaga, Augustine Bugler, Juan~P. Brito, Vicente Martin, and Christoph Striecks.
\newblock Versatile quantum-safe hybrid key exchange and its application to {{MACsec}}.
\newblock {\em EPJ Quantum Technology}, 12(1):1--27, 7 2025.
\newblock \href {https://doi.org/10.1140/epjqt/s40507-025-00382-x} {\path{doi:10.1140/epjqt/s40507-025-00382-x}}.

\bibitem[BBF{\etalchar{+}}18]{EPRINT:BBFGS18}
Nina Bindel, Jacqueline Brendel, Marc Fischlin, Brian Goncalves, and Douglas Stebila.
\newblock Hybrid key encapsulation mechanisms and authenticated key exchange.
\newblock Cryptology ePrint Archive, Report 2018/903, 2018.
\newblock \url{https://eprint.iacr.org/2018/903}.

\bibitem[BBF{\etalchar{+}}19]{PQCRYPTO:BBFGS19}
Nina Bindel, Jacqueline Brendel, Marc Fischlin, Brian Goncalves, and Douglas Stebila.
\newblock Hybrid key encapsulation mechanisms and authenticated key exchange.
\newblock In Jintai Ding and Rainer Steinwandt, editors, {\em Post-Quantum Cryptography - 10th International Conference, PQCrypto 2019}, pages 206--226, Chongqing, China, May~8--10, 2019. Springer, Heidelberg, Germany.
\newblock \href {https://doi.org/10.1007/978-3-030-25510-7_12} {\path{doi:10.1007/978-3-030-25510-7_12}}.

\bibitem[BCD20]{sp800-56c}
Elaine Barker, Lily Chen, and Richard Davis.
\newblock Recommendation for key-derivation methods in key-establishment schemes.
\newblock Recommendation NIST SP 800-56C, National Institute of Standards and Technology, 8 2020.
\newblock \href {https://doi.org/10.6028/NIST.SP.800-56Cr2} {\path{doi:10.6028/NIST.SP.800-56Cr2}}.

\bibitem[BCK98]{STOC:BelCanKra98}
Mihir Bellare, Ran Canetti, and Hugo Krawczyk.
\newblock A modular approach to the design and analysis of authentication and key exchange protocols (extended abstract).
\newblock In {\em 30th Annual {ACM} Symposium on Theory of Computing}, pages 419--428, Dallas, TX, USA, May~23--26, 1998. {ACM} Press.
\newblock \href {https://doi.org/10.1145/276698.276854} {\path{doi:10.1145/276698.276854}}.

\bibitem[BCNP08]{ACISP:BCNP08}
Colin Boyd, Yvonne Cliff, Juan~Gonzalez Nieto, and Kenneth~G. Paterson.
\newblock {Efficient One-Round Key Exchange in the Standard Model}.
\newblock ACISP 08: 13th Australasian Conference on Information Security and Privacy, 2008.
\newblock \href {https://doi.org/10.1007/978-3-540-70500-0_6} {\path{doi:10.1007/978-3-540-70500-0_6}}.

\bibitem[BDF{\etalchar{+}}11]{AC:BDFLSZ11}
Dan Boneh, {\"O}zg{\"u}r Dagdelen, Marc Fischlin, Anja Lehmann, Christian Schaffner, and Mark Zhandry.
\newblock Random oracles in a quantum world.
\newblock In Dong~Hoon Lee and Xiaoyun Wang, editors, {\em Advances in Cryptology -- {ASIACRYPT}~2011}, volume 7073 of {\em Lecture Notes in Computer Science}, pages 41--69, Seoul, South Korea, December~4--8, 2011. Springer, Heidelberg, Germany.
\newblock \href {https://doi.org/10.1007/978-3-642-25385-0_3} {\path{doi:10.1007/978-3-642-25385-0_3}}.

\bibitem[BDK{\etalchar{+}}18]{BDK+18}
Joppe Bos, Leo Ducas, Eike Kiltz, T~Lepoint, Vadim Lyubashevsky, John~M. Schanck, Peter Schwabe, Gregor Seiler, and Damien Stehle.
\newblock {{CRYSTALS}} - {{Kyber}}: {{A CCA-Secure Module-Lattice-Based KEM}}.
\newblock In {\em 2018 {{IEEE European Symposium}} on {{Security}} and {{Privacy}} ({{EuroS}}\&{{P}})}, pages 353--367, 4 2018.
\newblock \href {https://doi.org/10.1109/EuroSP.2018.00032} {\path{doi:10.1109/EuroSP.2018.00032}}.

\bibitem[BL17]{BL17}
Daniel~J. Bernstein and Tanja Lange.
\newblock Post-quantum cryptography.
\newblock {\em Nature}, 549(7671):188--194, 9 2017.
\newblock \href {https://doi.org/10.1038/nature23461} {\path{doi:10.1038/nature23461}}.

\bibitem[BR94]{C:BelRog93}
Mihir Bellare and Phillip Rogaway.
\newblock Entity authentication and key distribution.
\newblock In Douglas~R. Stinson, editor, {\em Advances in Cryptology -- {CRYPTO}'93}, volume 773 of {\em Lecture Notes in Computer Science}, pages 232--249, Santa Barbara, CA, USA, August~22--26, 1994. Springer, Heidelberg, Germany.
\newblock \href {https://doi.org/10.1007/3-540-48329-2_21} {\path{doi:10.1007/3-540-48329-2_21}}.

\bibitem[BRS23]{MucklePlus}
Sonja Bruckner, Sebastian Ramacher, and Christoph Striecks.
\newblock {Muckle}+: {End}-to-{End} {Hybrid} {Authenticated} {Key} {Exchanges}.
\newblock In Thomas Johansson and Daniel Smith-Tone, editors, {\em Post-{Quantum} {Cryptography}}, pages 601--633, Cham, 2023. Springer Nature Switzerland.
\newblock \href {https://doi.org/10.1007/978-3-031-40003-2_22} {\path{doi:10.1007/978-3-031-40003-2_22}}.

\bibitem[BSP{\etalchar{+}}24]{BSP+24}
Christopher Battarbee, Christoph Striecks, Ludovic Perret, Sebastian Ramacher, and Kevin Verhaeghe.
\newblock Quantum-{{Safe Hybrid Key Exchanges}} with {{KEM-Based Authentication}}.
\newblock arXiv:2411.04030 [cs.CR], 11 2024.
\newblock \href {https://doi.org/10.48550/arXiv.2411.04030} {\path{doi:10.48550/arXiv.2411.04030}}.

\bibitem[CCG16]{CSF:CohCreGar16}
Katriel {Cohn-Gordon}, Cas J.~F. Cremers, and Luke Garratt.
\newblock On post-compromise security.
\newblock In Michael Hicks and Boris Köpf, editors, {\em CSF 2016: IEEE 29th Computer Security Foundations Symposium}, pages 164--178, Lisbon, Portugal, June~27--1, 2016. {IEEE} Computer Society Press.
\newblock \href {https://doi.org/10.1109/CSF.2016.19} {\path{doi:10.1109/CSF.2016.19}}.

\bibitem[CK01]{EC:CanKra01}
Ran Canetti and Hugo Krawczyk.
\newblock Analysis of key-exchange protocols and their use for building secure channels.
\newblock In Birgit Pfitzmann, editor, {\em Advances in Cryptology -- {EUROCRYPT}~2001}, volume 2045 of {\em Lecture Notes in Computer Science}, pages 453--474, Innsbruck, Austria, May~6--10, 2001. Springer, Heidelberg, Germany.
\newblock \href {https://doi.org/10.1007/3-540-44987-6_28} {\path{doi:10.1007/3-540-44987-6_28}}.

\bibitem[CK02]{EC:CanKra02}
Ran Canetti and Hugo Krawczyk.
\newblock Universally composable notions of key exchange and secure channels.
\newblock In Lars~R. Knudsen, editor, {\em Advances in Cryptology -- {EUROCRYPT}~2002}, volume 2332 of {\em Lecture Notes in Computer Science}, pages 337--351, Amsterdam, The Netherlands, April~28~--~May~2, 2002. Springer, Heidelberg, Germany.
\newblock \href {https://doi.org/10.1007/3-540-46035-7_22} {\path{doi:10.1007/3-540-46035-7_22}}.

\bibitem[DHP20a]{PQCRYPTO:DowHanPat20}
Benjamin Dowling, Torben~Brandt Hansen, and Kenneth~G. Paterson.
\newblock Many a mickle makes a muckle: {A} framework for provably quantum-secure hybrid key exchange.
\newblock In Jintai Ding and Jean-Pierre Tillich, editors, {\em Post-Quantum Cryptography - 11th International Conference, PQCrypto 2020}, pages 483--502, Paris, France, April~15--17, 2020. Springer, Heidelberg, Germany.
\newblock \href {https://doi.org/10.1007/978-3-030-44223-1_26} {\path{doi:10.1007/978-3-030-44223-1_26}}.

\bibitem[DHP20b]{EPRINT:DowHanPat20}
Benjamin Dowling, Torben~Brandt Hansen, and Kenneth~G. Paterson.
\newblock Many a mickle makes a muckle: {A} framework for provably quantum-secure hybrid key exchange.
\newblock Cryptology ePrint Archive, Report 2020/099, 2020.
\newblock \url{https://eprint.iacr.org/2020/099}.

\bibitem[DSL{\etalchar{+}}22]{Dev+22}
Oskar~van Deventer, Nicolas Spethmann, Marius Loeffler, Momtchil Peev, Andreas Poppe, Stefan Rass, Michael Sauerwein, Peter Schindler, Damien Stucki, Nino Walenta, Helmut Weier, Christiane Wichmann, and Anton Zeilinger.
\newblock Towards {{European Standards}} for {{Quantum Technologies}}.
\newblock {\em EPJ Quantum Technology}, 9(1):33, 11 2022.
\newblock \href {https://doi.org/10.1140/epjqt/s40507-022-00150-1} {\path{doi:10.1140/epjqt/s40507-022-00150-1}}.

\bibitem[Eke91]{Ekert91}
Artur~K. Ekert.
\newblock Quantum cryptography based on bell's theorem.
\newblock {\em Phys. Rev. Lett.}, 67:661--663, Aug 1991.
\newblock \href {https://doi.org/10.1103/PhysRevLett.67.661} {\path{doi:10.1103/PhysRevLett.67.661}}.

\bibitem[ETS19]{etsi014}
ETSI.
\newblock {Quantum Key Distribution (QKD); Protocol and data format of REST-based key delivery API}.
\newblock Standard GS QKD 014, European Telecommunications Standards Institute, 2 2019.
\newblock URL: \url{https://www.etsi.org/deliver/etsi_gs/QKD/001_099/014/01.01.01_60/gs_qkd014v010101p.pdf}.

\bibitem[ETS20]{etsi004}
ETSI.
\newblock Quantum {{Key Distribution}} ({{QKD}}); {{Application Interface}}.
\newblock Standard GS QKD 0004, European Telecommunications Standards Institute, 8 2020.
\newblock URL: \url{https://www.etsi.org/deliver/etsi_gs/QKD/001_099/004/02.01.01_60/gs_qkd004v020101p.pdf}.

\bibitem[{Fed}21]{BSIhybrid}
{Federal Office for Information Security}.
\newblock Quantum-safe cryptography – fundamentals, current developments and recommendations, 2021.
\newblock URL: \url{https://www.bsi.bund.de/SharedDocs/Downloads/EN/BSI/Publications/Brochure/quantum-safe-cryptography.html}.

\bibitem[FSXY12]{PKC:FSXY12}
Atsushi Fujioka, Koutarou Suzuki, Keita Xagawa, and Kazuki Yoneyama.
\newblock Strongly secure authenticated key exchange from factoring, codes, and lattices.
\newblock In Marc Fischlin, Johannes Buchmann, and Mark Manulis, editors, {\em PKC~2012: 15th International Conference on Theory and Practice of Public Key Cryptography}, volume 7293 of {\em Lecture Notes in Computer Science}, pages 467--484, Darmstadt, Germany, May~21--23, 2012. Springer, Heidelberg, Germany.
\newblock \href {https://doi.org/10.1007/978-3-642-30057-8_28} {\path{doi:10.1007/978-3-642-30057-8_28}}.

\bibitem[FSXY13]{ASIACCS:FSXY13}
Atsushi Fujioka, Koutarou Suzuki, Keita Xagawa, and Kazuki Yoneyama.
\newblock Practical and post-quantum authenticated key exchange from one-way secure key encapsulation mechanism.
\newblock In Kefei Chen, Qi~Xie, Weidong Qiu, Ninghui Li, and Wen-Guey Tzeng, editors, {\em ASIACCS 13: 8th ACM Symposium on Information, Computer and Communications Security}, pages 83--94, Hangzhou, China, May~8--10, 2013. {ACM} Press.
\newblock \href {https://doi.org/10.1145/2484313.2484323} {\path{doi:10.1145/2484313.2484323}}.

\bibitem[GHP18]{PKC:GiaHeuPoe18}
Federico Giacon, Felix Heuer, and Bertram Poettering.
\newblock {KEM} combiners.
\newblock In Michel Abdalla and Ricardo Dahab, editors, {\em PKC~2018: 21st International Conference on Theory and Practice of Public Key Cryptography, Part~I}, volume 10769 of {\em Lecture Notes in Computer Science}, pages 190--218, Rio de Janeiro, Brazil, March~25--29, 2018. Springer, Heidelberg, Germany.
\newblock \href {https://doi.org/10.1007/978-3-319-76578-5_7} {\path{doi:10.1007/978-3-319-76578-5_7}}.

\bibitem[GPH{\etalchar{+}}24]{GPH+24}
Lydia Garms, Taofiq~K. Paraïso, Neil Hanley, Ayesha Khalid, Ciara Rafferty, James Grant, James Newman, Andrew~J. Shields, Carlos Cid, and Maire O'Neill.
\newblock Experimental {{Integration}} of {{Quantum Key Distribution}} and {{Post-Quantum Cryptography}} in a {{Hybrid Quantum-Safe Cryptosystem}}.
\newblock {\em Advanced Quantum Technologies}, 7(4):2300304, 2 2024.
\newblock \href {https://doi.org/10.1002/qute.202300304} {\path{doi:10.1002/qute.202300304}}.

\bibitem[HKSU20]{PKC:HKSU20}
Kathrin H{\"o}velmanns, Eike Kiltz, Sven Sch{\"a}ge, and Dominique Unruh.
\newblock Generic authenticated key exchange in the quantum random oracle model.
\newblock In Aggelos Kiayias, Markulf Kohlweiss, Petros Wallden, and Vassilis Zikas, editors, {\em PKC~2020: 23rd International Conference on Theory and Practice of Public Key Cryptography, Part~II}, volume 12111 of {\em Lecture Notes in Computer Science}, pages 389--422, Edinburgh, UK, May~4--7, 2020. Springer, Heidelberg, Germany.
\newblock \href {https://doi.org/10.1007/978-3-030-45388-6_14} {\path{doi:10.1007/978-3-030-45388-6_14}}.

\bibitem[{ITU}20]{X1714}
{ITU-T}.
\newblock Key combination and confidential key supply for quantum key distribution networks.
\newblock Recommendation ITU-T X.1714, International Telecommunication Union, 10 2020.
\newblock URL: \url{http://handle.itu.int/11.1002/1000/14453}.

\bibitem[Kra03]{C:Krawczyk03}
Hugo Krawczyk.
\newblock {SIGMA}: The ``{SIGn-and-MAc}'' approach to authenticated {Diffie}-{Hellman} and its use in the {IKE} protocols.
\newblock In Dan Boneh, editor, {\em Advances in Cryptology -- {CRYPTO}~2003}, volume 2729 of {\em Lecture Notes in Computer Science}, pages 400--425, Santa Barbara, CA, USA, August~17--21, 2003. Springer, Heidelberg, Germany.
\newblock \href {https://doi.org/10.1007/978-3-540-45146-4_24} {\path{doi:10.1007/978-3-540-45146-4_24}}.

\bibitem[Kra05]{C:Krawczyk05}
Hugo Krawczyk.
\newblock {HMQV}: A high-performance secure {Diffie}-{Hellman} protocol.
\newblock In Victor Shoup, editor, {\em Advances in Cryptology -- {CRYPTO}~2005}, volume 3621 of {\em Lecture Notes in Computer Science}, pages 546--566, Santa Barbara, CA, USA, August~14--18, 2005. Springer, Heidelberg, Germany.
\newblock \href {https://doi.org/10.1007/11535218_33} {\path{doi:10.1007/11535218_33}}.

\bibitem[Kra10]{C:Krawczyk10}
Hugo Krawczyk.
\newblock Cryptographic extraction and key derivation: The {HKDF} scheme.
\newblock In Tal Rabin, editor, {\em Advances in Cryptology -- {CRYPTO}~2010}, volume 6223 of {\em Lecture Notes in Computer Science}, pages 631--648, Santa Barbara, CA, USA, August~15--19, 2010. Springer, Heidelberg, Germany.
\newblock \href {https://doi.org/10.1007/978-3-642-14623-7_34} {\path{doi:10.1007/978-3-642-14623-7_34}}.

\bibitem[LLM07]{PROVSEC:LaMLauMit07}
Brian~A. LaMacchia, Kristin Lauter, and Anton Mityagin.
\newblock Stronger security of authenticated key exchange.
\newblock In Willy Susilo, Joseph~K. Liu, and Yi~Mu, editors, {\em ProvSec 2007: 1st International Conference on Provable Security}, volume 4784 of {\em Lecture Notes in Computer Science}, pages 1--16, Wollongong, Australia, November~1--2, 2007. Springer, Heidelberg, Germany.
\newblock \href {https://doi.org/10.1007/978-3-540-75670-5_1} {\path{doi:10.1007/978-3-540-75670-5_1}}.

\bibitem[MSU13]{PQCRYPTO:MosSteUst13}
Michele Mosca, Douglas Stebila, and Berkant Ustaoglu.
\newblock Quantum key distribution in the classical authenticated key exchange framework.
\newblock In Philippe Gaborit, editor, {\em Post-Quantum Cryptography - 5th International Workshop, PQCrypto 2013}, pages 136--154, Limoges, France, June~4--7, 2013. Springer, Heidelberg, Germany.
\newblock \href {https://doi.org/10.1007/978-3-642-38616-9_9} {\path{doi:10.1007/978-3-642-38616-9_9}}.

\bibitem[RCV{\etalchar{+}}23]{GAO+23}
Carlos {Rubio Garc\'ia}, Abraham {Cano Aguilera}, Juan~Jose {Vegas Olmos}, Idelfonso {Tafur Monroy}, and Simon Rommel.
\newblock Quantum-{{Resistant TLS}} 1.3: {{A Hybrid Solution Combining Classical}}, {{Quantum}} and {{Post-Quantum Cryptography}}.
\newblock In {\em 2023 {{IEEE}} 28th {{International Workshop}} on {{Computer Aided Modeling}} and {{Design}} of {{Communication Links}} and {{Networks}} ({{CAMAD}})}, pages 246--251, 11 2023.
\newblock \href {https://doi.org/10.1109/CAMAD59638.2023.10478407} {\path{doi:10.1109/CAMAD59638.2023.10478407}}.

\bibitem[RDM{\etalchar{+}}24]{RDM+24}
Sara Ricci, Patrik Dobias, Lukas Malina, Jan Hajny, and Petr Jedlicka.
\newblock Hybrid {{Keys}} in {{Practice}}: {{Combining Classical}}, {{Quantum}} and {{Post-Quantum Cryptography}}.
\newblock {\em IEEE Access}, 12:23206--23219, 2024.
\newblock \href {https://doi.org/10.1109/ACCESS.2024.3364520} {\path{doi:10.1109/ACCESS.2024.3364520}}.

\bibitem[SCN{\etalchar{+}}24]{SCN+24}
{Secure Information Technology Center Austria}, {Centre for Cybersecurity Belgium}, {National Cyber and Information Security Agency Czech Republic}, {Centre for Cyber Security Denmark}, {Information System Authority Estonia}, {Finnish transport and Communication Agency}, {French National Agency for the Security of Information Systems}, {Federal Office for Information Security Germany}, {National Cyber Security Authority Hellenic Republic}, {National Cyber Security Centre Ireland}, {National Cybersecurity Agency Italy}, {Ministry of Defense Latvia}, {National Cyber Security Centre Ministry of Defense Lithuania}, {High Commission for National Protection Luxemburg}, {Netherlands National Communication Security Agency}, {Ministry of Interior and Kingdom Relations Netherlands}, {National Cyber Security Centre Ministry of Security and Justice Netherlands}, {Research and Academic Research Center Poland}, {Government Information Security Office Slovenia}, and {National Cryptologic Center Spain}.
\newblock Securing {{Tomorrow}}, {{Today}}: {{Transitioning}} to {{Post-Quantum Cryptography}}, 11 2024.
\newblock URL: \url{https://www.bsi.bund.de/SharedDocs/Downloads/EN/BSI/Crypto/PQC-joint-statement.pdf}.

\bibitem[Sho94]{FOCS:Shor94}
Peter~W. Shor.
\newblock Algorithms for quantum computation: Discrete logarithms and factoring.
\newblock In {\em 35th Annual Symposium on Foundations of Computer Science}, pages 124--134, Santa Fe, NM, USA, November~20--22, 1994. {IEEE} Computer Society Press.
\newblock \href {https://doi.org/10.1109/SFCS.1994.365700} {\path{doi:10.1109/SFCS.1994.365700}}.

\bibitem[Sho04]{shoup2004sequences}
Victor Shoup.
\newblock Sequences of games: a tool for taming complexity in security proofs.
\newblock Cryptology {ePrint} Archive, Paper 2004/332, 2004.
\newblock URL: \url{https://eprint.iacr.org/2004/332}.

\bibitem[SPP{\etalchar{+}}24]{Saez2024}
Juan~Morales Sáez, Antonio~Pastor Perales, Rafael~Cantó Palancar, Diego~R. Lopez, Jesús~Folgueira Chavarria, Vicente~Martin Ayuso, and Juan~Pedro Brito~Mendez.
\newblock Current status, gaps, and future directions in quantum key distribution standards: Implications for industry.
\newblock In {\em 2024 International Conference on Quantum Communications, Networking, and Computing (QCNC)}, pages 341--345, 2024.
\newblock \href {https://doi.org/10.1109/QCNC62729.2024.00059} {\path{doi:10.1109/QCNC62729.2024.00059}}.

\bibitem[SSW20]{CCS:SchSteWig20}
Peter Schwabe, Douglas Stebila, and Thom Wiggers.
\newblock Post-quantum {TLS} without handshake signatures.
\newblock In Jay Ligatti, Xinming Ou, Jonathan Katz, and Giovanni Vigna, editors, {\em ACM CCS 2020: 27th Conference on Computer and Communications Security}, pages 1461--1480, Virtual Event, USA, November~9--13, 2020. {ACM} Press.
\newblock \href {https://doi.org/10.1145/3372297.3423350} {\path{doi:10.1145/3372297.3423350}}.

\bibitem[TLLM18]{TLLM18}
Piotr~K. Tysowski, Xinhua Ling, Norbert Lütkenhaus, and Michele Mosca.
\newblock The engineering of a scalable multi-site communications system utilizing quantum key distribution ({{QKD}}).
\newblock {\em Quantum Science and Technology}, 3(2):024001, 1 2018.
\newblock \href {https://doi.org/10.1088/2058-9565/aa9a5d} {\path{doi:10.1088/2058-9565/aa9a5d}}.

\bibitem[Unr15]{Unr15}
Dominique Unruh.
\newblock Revocable {{Quantum Timed-Release Encryption}}.
\newblock {\em Journal of the ACM}, 62(6):49:1--49:76, 12 2015.
\newblock \href {https://doi.org/10.1145/2817206} {\path{doi:10.1145/2817206}}.

\bibitem[WC81]{WegCar81}
Mark~N. Wegman and Larry Carter.
\newblock New hash functions and their use in authentication and set equality.
\newblock {\em Journal of Computer and System Sciences}, 22:265--279, 1981.

\end{thebibliography}
\fi

\ifCameraReady \else	
	
	\ifSupplementaryMaterial
		\newpage
		\bigskip\noindent{\Huge\textbf{Supplementary material}}\vspace{1cm}					
	\fi
	
	\appendix
	
\fi

\end{document}